\documentclass{pasa}%

\usepackage{graphicx}
\usepackage[para]{threeparttable}
\usepackage{subfigure}
\usepackage{wrapfig,lipsum}
\usepackage{setspace}
\usepackage[toc,page]{appendix}
\usepackage{adjustbox}
\usepackage{rotating,tabularx,siunitx}
\def\Robbie{{\sc Robbie}}
\title[Short GRB survey]{Early-time Searches for Coherent Radio Emission from Short GRBs with the Murchison Widefield Array}

\author[J. Tian et al.]{J. Tian$^{1}$\thanks{E-mail: jun.tian@postgrad.curtin.edu.au}, G. E. Anderson$^1$, P. J. Hancock$^1$, J. C. A. Miller-Jones$^1$, M. Sokolowski$^1$, A. Rowlinson$^{2,3}$, A. Williams$^1$, J. Morgan$^1$, N. Hurley-Walker$^1$, D. L. Kaplan$^4$, Tara Murphy$^{5,9}$, S.J. Tingay$^1$, M. Johnston-Hollitt$^1$, K. W. Bannister$^6$, M. E. Bell$^7$, B. W. Meyers$^8$
\affil{$^1$International Centre for Radio Astronomy Research, Curtin University, GPO Box U1987, Perth, WA 6845, Australia}%
\affil{$^2$Anton Pannekoek Institute, University of Amsterdam, Postbus 94249, 1090 GE, Amsterdam, The Netherlands}
\affil{$^3$ASTRON, the Netherlands Institute for Radio Astronomy, Oude Hoogeveensedijk 4, 7991 PD, Dwingeloo, The Netherlands}
\affil{$^4$Department of Physics, University of Wisconsin-Milwaukee, 1900 E. Kenwood Boulevard, Milwaukee, WI 53211, USA}
\affil{$^5$Sydney Institute for Astronomy, School of Physics, The University of Sydney, NSW 2006, Australia}
\affil{$^6$Australia Telescope National Facility, CSIRO Astronomy and Space Science, PO Box 76, Epping, NSW 1710, Australia}
\affil{$^7$University of Technology Sydney, 15 Broadway, Ultimo NSW 2007, Australia}
\affil{$^8$Department of Physics and Astronomy, University of British Columbia, Vancouver, British Columbia, Canada}
\affil{$^9$ARC Centre of Excellence for Gravitational Wave Discovery (OzGrav), Hawthorn, Victoria, Australia}
}%

\jid{PASA}
\doi{10.1017/pas.\the\year.xxx}
\jyear{\the\year}

\usepackage{aas_macros}
\usepackage{hyperref} 
\hypersetup{colorlinks,citecolor=blue,linkcolor=blue,urlcolor=blue}

\begin{document}

\begin{frontmatter}
\maketitle

\begin{abstract}

Many short gamma-ray bursts (GRBs) originate from binary neutron star mergers, and there are several theories that predict the production of coherent, prompt radio signals either prior, during, or shortly following the merger, as well as persistent pulsar-like emission from the spin-down of a magnetar remnant. Here we present a low frequency (170\textendash200\,MHz) search for coherent radio emission associated with
nine short GRBs detected by the \textit{Swift} and/or \textit{Fermi} satellites using the Murchison Widefield Array (MWA) rapid-response observing mode. 
The MWA began observing these events within 30 to 60\,s of their high-energy detection, enabling us to capture any dispersion delayed signals emitted by short GRBs for a typical range of redshifts. We conducted transient searches at the GRB positions on timescales of 5\,s, 30\,s and 2\,min, resulting in the most constraining flux density limits on any associated transient of 0.42, 0.29, and 0.084\,Jy, respectively. 
We also searched for dispersed signals at a temporal and spectral resolution of 0.5\,s and 1.28\,MHz but none were detected. 
However, the fluence limit of $80-100$\,Jy\,ms derived for GRB 190627A is the most stringent to date for a short GRB. Assuming the formation of a stable magnetar for this GRB, we compared the fluence and persistent emission limits to short GRB coherent emission models, placing constraints on key parameters including the radio emission efficiency of the nearly merged neutron stars ($\epsilon_r\lesssim10^{-4}$), the fraction of magnetic energy in the GRB jet ($\epsilon_B\lesssim2\times10^{-4}$), and the radio emission efficiency of the magnetar remnant ($\epsilon_r\lesssim10^{-3}$). Comparing the limits derived for our full GRB sample (along with those in the literature) to the same emission models, we demonstrate that our fluence limits only place weak constraints on the prompt emission predicted from the interaction between the relativistic GRB jet and the interstellar medium for a subset of magnetar parameters. However, the 30-min flux density limits were sensitive enough to theoretically detect the persistent radio emission from magnetar remnants up to a redshift of $z\sim0.6$. 
Our non-detection of this emission could imply that some GRBs in the sample were not genuinely short or did not result from a binary neutron star merger, the GRBs were at high redshifts, these mergers formed atypical magnetars, the radiation beams of the magnetar remnants were pointing away from Earth, or the majority did not form magnetars but rather collapse directly into black holes.
\end{abstract}

\begin{keywords}
surveys -- radio continuum: transients -- gamma-ray bursts
\end{keywords}
\end{frontmatter}

\section{INTRODUCTION }
\label{sec:intro}

Short Gamma-ray bursts (GRBs) originate from binary neutron star (BNS) or neutron star - black hole (NS-BH) mergers at cosmological distances \citep{Narayan92}. These compact binary mergers are gravitational wave (GW) emitters, as confirmed by the near-coincident detection of GRB 170817A \citep{Goldstein17} with peculiarly low luminosity \citep{Begue17} and GW170817 \citep{Abbott17}, increasing interest in a multi-messenger era for short GRB observations.

In addition to a short-duration ($\lesssim\,$2\,s) sudden increase in gamma-ray flux, short GRBs have been detected at other wavelengths. These include the detection of X-ray afterglows from many short GRBs by the \textit{Neil Gehrels Swift Observatory} (hereafter referred to as \textit{Swift}; \citealt{Gehrels04}), with a considerable fraction of the resulting light curves showing a plateau phase during the X-ray decay \citep{Evans09}, which suggests continued energy injection after the prompt emission phase of the GRB \citep{Burrows06, Soderberg06}. Either a BH or a quasi-stable, highly magnetised, rapidly rotating NS (magnetar) may be formed via the merger depending on the NS equation of state \citep{Duncan92, Usov92, Lattimer12}, and both of them can supply energy for the extended emission through different channels. While the accretion onto the black hole is likely to finish within a short time~\citep{Rezzolla11}, the magnetar model can account for on-going energy injection, and has been fitted to the X-ray light curves observed in a significant fraction of short GRBs~\citep{Rowlinson13}. In this paper, we consider the magnetar model to explain the origin of the X-ray emission after the BNS merger.

Apart from the X-ray emission, radio synchrotron afterglows that result from ejecta (likely from the relativistic jets) interacting with the circum-merger media that last $\sim$\,1\textendash10\,days have been observed among a few short GRBs at GHz frequencies (e.g. \citealt{Fong15}, \citealt{Fong21}, \citealt{Anderson21b}). However, until recently, low radio frequency studies of short GRBs have been limited, particularly as the fireball model \citep{Cavallo78, Rees92} does not predict synchrotron radio emission to be very bright below 300\,MHz (e.g. \citealt{Horst08}). 
However, there are several models that predict short GRBs may produce coherent low-frequency radio emission (e.g.,\,\,\citealt{Lyutikov13},~\citealt{Zhang14},~\citealt{Usov00},~\citealt{Totani13}). At least four potential mechanisms that predict either prompt, fast radio burst (FRB) like signals or persistent pulsar-like emission during different phases of the compact binary merger have been explored in the context of current low-frequency radio facilities \citep{Chu16, Rowlinson19}. The earliest prompt FRB-like signals could come from the inspiral phase, where GWs may excite the surrounding plasma, or through interactions of the NS magnetic fields just preceding the merger \citep{Lyutikov13}. During the merger, an extremely relativistic jet may be launched that may then produce an FRB-like signal when it interacts with the interstellar medium (ISM; \citealt{Usov00}). If the merger remnant is a magnetar, the coherent radio emission powered by dipole magnetic braking may appear as persistent or pulsed emission during the lifetime of the magnetar \citep{Totani13}. Finally, if the magnetar cannot support its high mass due to spinning down, a final FRB-like event may be produced as it collapses into a BH and ejects its magnetosphere \citep{Zhang14}.

There have been many searches for prompt radio emission associated with GRBs but so far none have yielded a detection. One of the earliest searches was at 151\,MHz between 1970 and 1973 but no signals were observed over a sensitivity limit $\sim10^{5}$\,Jy from GRBs detected by the \textit{Vela} satellites \citep{Baird75}. Later observations have also showed no prompt radio emission from GRBs (see \citealt{Dessenne96}, \citealt{Obenberger14}, \citealt{Bannister12}, \citealt{Palaniswamy14} and \citealt{Kaplan15}). Recently, \citet{AndersonM18} performed a low-frequency search (below 100\,MHz) using the Owens Valley Radio Observatory Long Wavelength Array (OVRO-LWA), which continuously monitored GRBs detected by the \textit{Swift} Burst Alert Telescope (BAT; \citealt{Barthelmy05}) and the \textit{Fermi} Gamma-ray Burst Monitor (GBM; \citealt{Meegan09}). They performed an image de-dispersion analysis and found no simultaneous radio emission associated with short GRB 170112A above 4.5\,Jy on 13\,s timescales. Another telescope, the LOw Frequency ARray (LOFAR; \citealt{Haarlem13}), was used to trigger rapid-response observations on \textit{Swift} GRB 180706A (long) and 181123B (short), resulting in deep limits of 1.7\,mJy and 153\,mJy, respectively over a 2\,hr timescale on associated coherent, persistent radio emission from a magnetar remnant~\citep{Rowlinson19b, Rowlinson20}. 
In addition, the neutron star merger origin has also been investigated from the FRB context, where the properties of two non-repeating FRBs and their host environments were used to constrain merger models, make electromagnetic light curve predictions, and were searched for evidence of temporally coincident, sub-threshold gamma-ray counterparts \citep{gourdji20}.  

We have been using the Murchison Widefield Array (MWA; \citealt{Tingay13}, \citealt{Wayth18}) to perform triggered observations of GRBs since 2015 (e.g. \citealt{Kaplan15}). In 2018, the MWA triggering system was upgraded to enable it to trigger on VOEvents (Virtual Observatory Events, which is a standard information packet for the communication of transient celestial events; \citealt{Seaman11}), allowing the MWA to point to a GRB position and begin observations within 20\,s of receiving an alert \citep{Hancock19}. The first triggered MWA observation on a short GRB was performed by \citet{Kaplan15}, and yielded an upper limit of 3\,Jy on 4\,s timescales. \citet{Anderson20} reported the first short GRB trigger with the upgraded MWA triggering system, and obtained a flux density upper limit of 270\textendash630\,mJy on 5\,s timescales and a fluence upper-limit range from 570\,Jy\,ms at a dispersion measure  (DM) of $3000$\,pc\,cm$^{-3}$ ($z\sim 2.5$) to 1750\,Jy\,ms at a DM of $200$\,pc\,cm$^{-3}$ ($z\sim 0.1)$, corresponding to the known redshift range of short GRBs \citep{Rowlinson13}.

There are several possible reasons for the non-detections in these previous efforts. While all-sky instruments can continuously monitor for GRB occurrences, they usually have lower sensitivity. The more sensitive pointed observations are prone to miss the earliest signal because they often take a few minutes to slew/repoint and begin observing the event. Moreover most previous searches have focused on the more common long GRBs. Although long GRBs have some of the same expected mechanisms to short GRBs for producing coherent radio emission (e.g. the impact of the gamma-ray jets into the ISM and the formation of a magnetar; \citealt{Usov00}, \citealt{Evans09}), such signals may not penetrate through the dense medium surrounding core-collapse supernovae \citep{Zhang14} and would therefore be difficult to detect.

In this work, we use MWA rapid-response observations to search for coherent low-frequency radio emission from a sample of nine short GRBs.
In Section~\ref{sec:data}, we describe the MWA rapid-response mode and the processing pipeline and analysis we used to search for prompt radio emission on 2\,min, 30\,s, and 5\,s timescales. We also describe our search for dispersed prompt emission using image de-dispersion techniques with an image temporal and spectral resolution of 0.5\,s and 1.28\,MHz. Our results are then presented in Section~\ref{sec:results}. We compare the flux density and fluence upper limits derived from different GRBs and select the best ones to constrain the models of BNS mergers, as shown in Section~\ref{sec:EMmod}. We focus on GRB 190627A, which is the only GRB in our sample with a redshift measurement and therefore represents the most sensitive low frequency, short timescale limit among the population of short GRBs included in this paper.
In Section~\ref{sec:discussion}, we discuss how our sample along with other low frequency radio limits on prompt and persistent coherent emission from short GRBs constrain some of the explored emission models, and how these studies can be improved in the context of the MWA.

We assume a cosmology with $H_0 = 71\,\text{km}\,\text{s}^{-1}\,\text{Mpc}^{-1}$, $\Omega_{\text{m}} = 0.27$ and $\Omega_{\lambda} = 0.73$ throughout this work \citep{Spergel03}.

\section{observations and analysis}
\label{sec:data}
In this section, we describe the observations and analysis of nine short GRBs for which we obtained triggered observations using the MWA rapid-response mode.

\begin{table*}
\begin{threeparttable}
\resizebox{2.2\columnwidth}{!}{\hspace{-1.3cm}\begin{tabular}{l c l l l c l l l l c c}
\hline
GRB & Detector\tnote{1} & Trigger No.\tnote{2} & Start Time\tnote{3} & Time (sec) & Config.\tnote{5} & RA  & Dec  & Uncertainty\tnote{6}  & $\text{T}_{90}\tnote{7}$  & Localisation\tnote{8} & Calibrators\tnote{9}\\
 & & & (UT) & post-burst\tnote{4} & & (deg) & (deg) &   & (sec) & instrument & \\
\hline
170827B & \textit{Fermi} & 525555489 & 19:38:38 & 34 & I & 30.39 & -47.86 & $1.29^{\circ}$ & 0.18\tnote{a} & IPN\tnote{g} & PicA \\

190420.98 & \textit{Fermi} & 577495949 & 23:33:18 & 53 & IIE & 319.29 & -66.41 & $2.04^{\circ}$ & 1.47\tnote{b} & \textit{Fermi}\textendash GBM\tnote{h} & IF \\

190627A & \textit{Swift} & 911609 & 11:19:26 & 55 & IIE & 244.828 & -5.289 & 1.7'' & 1.6\tnote{c} & \textit{Swift}\textendash XRT\tnote{i} & HerA+3C353\\

190712.02 & \textit{Fermi} & 584583925 & 00:26:06 & 46 & IIE & 341.06 & -38.41 & $9.21^{\circ}$ & 1.79\tnote{b} & \textit{Fermi}\textendash GBM\tnote{h} & IF \\

190804A & \textit{Fermi} & 586574612 & 01:32:54 & 567 & IIE & 108.02 & -64.86 & $12.02^{\circ}$ & 2\tnote{d} & \textit{Fermi}\textendash GBM\tnote{h} & IF \\

190903A & \textit{Fermi} & 589223981 & 17:20:30 & 54 & IIC & 63.95 & -59.485 & $0.46^{\circ}$ & 0.27\tnote{e} & IPN\tnote{g} & PicA \\

191004A & \textit{Swift} & 927825 & 18:13:34 & 392 & IIC & 31.668 & -36.933 & 1.7'' & 2.44\tnote{f} & \textit{Swift}\textendash XRT\tnote{i} & PicA \\

200325A & \textit{Fermi} & 606799116 & 03:19:26 & 54 & IIE & 31.72 & -31.816 & 4' & 0.96\tnote{b} & \textit{Swift}\textendash BAT\tnote{i} & IF \\

200327A & \textit{Fermi} & 607035413 & 20:57:50 & 62 & IIE & 236.574 & -4.217 & 13.7' & 0.64\tnote{b} & IPN\tnote{g} & IF \\
\hline
\end{tabular}}
\caption{Short GRBs that triggered the MWA rapid-response mode. GRB 191004A had a burst duration slightly longer than 2\,s but is assumed to be short (see the text in Section 2.2.1). \\
1: The GRB detector that triggered the MWA rapid-response mode; \\
2: The unique number assigned to each GRB detected by \textit{Fermi} or \textit{Swift}. \\
3: The start time in UT of the first MWA observation that contained the position of the GRB within 50\% of the MWA primary beam. Note that the date is given by the GRB name in the first column with the convention of YYMMDD;\\ 
4: The delay of the MWA observation with respect to the GRB detection by \textit{Fermi} or \textit{Swift};\\
5: The MWA array configuration of the GRB observation, including phase I (`I'), or phase II extended (`IIE') or compact (`IIC');\\
6: The uncertainty corresponds to the $1\sigma$ and 90\% positional confidence for \textit{Fermi}- and \textit{Swift}-detected GRBs, respectively;\\
7: The time taken to accumulate 90\% of the burst fluence starting at the 5\% fluence level (short GRBs are usually considered to have a $\text{T}_{90}\leq2\,s$; \citealt{Kouveliotou93}): a: \citet{Svinkin17}; b: Fermi\textendash GBM burst catalog at HEASARC: \url{https://heasarc.gsfc.nasa.gov/W3Browse/fermi/fermigbrst.html}; c: \citet{Barthelmy19}; d: \citet{Ghumatkar19}; e: \citet{Mailyan19}; f: \citet{Sakamoto19}; \\
8: The instrument that provides the best localisation for the GRB: g: the IPN GRB database table: \url{https://heasarc.gsfc.nasa.gov/w3browse/all/ipngrb.html}; h: \textit{Fermi} trigger information: \url{https://gcn.gsfc.nasa.gov/fermi_grbs.html}; i: \textit{Swift} trigger information: \url{https://gcn.gsfc.nasa.gov/swift_grbs.html}; \\
9: The calibration of the MWA observation was either performed using an external calibrator (named) or via an infield calibration (IF) using the GLEAM survey \citep{Hurley17} as a sky model.
}
\label{SGRBs}
\end{threeparttable}
\end{table*}

\subsection{MWA rapid-response observations}\label{sec:MWA}
The MWA operational frequency range is between 80 and 300\,MHz, with an instantaneous bandwidth of 30.72\,MHz, and a field-of-view ranging from $\sim300-1000$\,deg$^2$ \citep{Tingay13}. The MWA observations of GRBs have a central frequency of 185\,MHz, and are taken in the standard correlator mode, which has a time and frequency resolution of 0.5\,s and 10\,kHz. 
The MWA has been triggering on GRBs since 2017, resulting in rapid-response observations taken during Phase I and Phase II, which differ in array configuration, and baseline length and distribution. The MWA phase I has a maximum baseline of 2864\,m, corresponding to an angular resolution of $\sim2$\,arcmin at 185\,MHz \citep{Tingay13}. The MWA phase II has two configurations: extended and compact configurations with an angular resolution of $\sim1$ and $\sim10$\,arcmin, respectively \citep{Wayth18}.

The rapid-response mode of the MWA can respond to a GRB trigger within 20-30\,s of receiving an alert (see \citealt{Kaplan15} and \citealt{Anderson20}). We trigger on both \textit{Swift}-BAT and \textit{Fermi}-GBM GRBs, which have typical positional uncertainties of $1-4$\,arcmin \citep{Gehrels04} and 1\textendash10\,deg \citep{Meegan09}, respectively. There were several upgrades to the system in 2018 (for the old rapid-response mode see \citealt{Kaplan15}), including triggering on VOEvents, allowing the MWA to repoint in the case of a position update from \textit{Fermi}, and a Sun suppression algorithm (for details see \citealt{Hancock19}). As we are mostly searching for short timescale transients, we trigger a GRB observation regardless of the array configuration as we are less affected by classical confusion noise. The rapid-response observation of a GRB lasts for 30\,min ($15\times2\,\text{min}$ snapshot observations) immediately following the transient alert. We chose 185\,MHz as the central frequency of our rapid-response observations as the dispersion delay $\tau$ of any prompt signal 
emitted between a redshift of $0.1 <z<  2.5$ \citep[the observed redshift range of short GRBs;][]{Gompertz20} is enough ($18\,\text{s}\lesssim\tau\lesssim6\,\text{min}$) to allow the MWA to be on target in-time to detect it (see figure 1 in \citealt{Hancock19}).

\subsection{Sample selection}\label{sec:sample}

There were 22 short GRBs detected by \textit{Swift} or \textit{Fermi} during the period from April 2017 to September 2020 that triggered MWA rapid-response observations. After inspecting their images individually, we selected nine GRBs with good image quality to be included in this paper as listed in Table \ref{SGRBs} (for their image quality see Figure~\ref{images_swift} and \ref{images_fermi} in Appendix~\ref{appendix:images}). Most of the GRBs were discovered by \textit{Fermi}\textendash GBM as it monitors 50\% of the sky at any one time \citep{Meegan09}, much larger than the 1.4\,sr field of view of \textit{Swift}\textendash BAT \citep{Gehrels04}.
Among the GRBs not chosen for analysis, six were contaminated by the Sun \citep[these observations were taken before we implemented the Sun suppression algorithm;][]{Hancock19}, three were close to the Galactic plane (these may be analysed in the future when we have a good MWA Galactic plane sky model), three \textit{Fermi} GRBs had final positions outside the MWA primary beam \citep[before we implemented real-time pointing updates as the \textit{Fermi} position is improved;][]{Hancock19}, and GRB 180805A was published in a previous paper \citep{Anderson20}. The MWA configuration in which each GRB was observed is listed in Table \ref{SGRBs}.

\subsubsection{\textit{Swift} triggers}\label{sec:swift_trig}

MWA rapid-response observations were triggered on \textit{Swift}\textendash BAT events GRB 190627A \citep{Sonbas19b} and GRB 191004A \citep{Cenko19b}. The advantage of \textit{Swift}\textendash BAT detected GRBs over \textit{Fermi} detected events is that they are often localised by the \textit{Swift} X-ray Telescope (XRT; \citealt{Burrows05}), which provides arcsec position localisations and makes afterglow detections/redshift determinations much more likely. The XRT also performs these observations rapidly following the BAT trigger (109.8\,s and 81.5\,s for GRB 190627A and GRB 191004A, respectively; \citealt{Sonbas19}, \citealt{Cenko19}), with the resulting X-ray light curves allowing us to constrain coherent radio emission models (see Section 4.1). 

GRB 190627A has a brief duration of $1.6$\,s \citep{Barthelmy19}, placing it in the short GRB category based on the usual criterion $\text{T}_{90}\leq2$\,s~\citep{Kouveliotou93}. Nonetheless, there are a few caveats about this identification. Despite the short duration of GRB 190627A, the relative softness of its spectrum makes it intermediate between most short and long bursts detected by \textit{Swift}\textendash BAT \citep{Barthelmy19}. GRB 190627A is the only event with a spectroscopic redshift of $z=1.942$ \citep{Japelj19} in our sample, which helps us to constrain its DM when searching for associated dispersed signals as described in Section~\ref{sec:disp_search}. However, this redshift is unusually high for a short GRB, requiring a very high efficiency to produce observable emission given the limited energy reservoir of binary mergers \citep{Berger14}. Additionally, GRB 190627A has a bright optical afterglow \citep{Japelj19}, making it unusual in the population of short GRBs \citep{Kann13}. Therefore, there is ambiguity in classifying GRB 190627A as being short.

GRB 191004A has been included in our sample as one of our rare triggers on \textit{Swift} short GRBs, though it had a duration of 2.44\,s \citep{Sakamoto19}. We calculated its hardness ratio (i.e., fluence in 50–100\,keV over fluence in 25–50\,keV) to be $\approx1.6$, intermediate between the short and long population of \textit{Swift} GRBs (see figure 8 in \citealt{Lien16}). There is no other information available for determining whether it is short or long, such as rest-frame duration \citep{Zhang09, Belczynski10} or the isotropic gamma-ray energy and spectral peak \citep{Lu10}. A few high-redshift long GRBs with rest-frame durations shorter than 2\,s have been found to possess the properties of short GRBs, such as a hard spectrum and a large offset from the host galaxy centre (e.g. \citealt{Ahumada21}). Given there is no clear distinction in the durations of long and short GRBs \citep{Berger14} and the rest-frame duration of GRB 191004A would be $<2$\,s should it occur at $z>0.2$, we treated GRB 191004A as if it were short. Note that the first three 2\,min observations following the \textit{Swift} trigger were corrupted so our first MWA observation of this source was delayed 6.35\,min with respect to the GRB  detection as shown in Table \ref{SGRBs}.

\subsubsection{\textit{Fermi} triggers}\label{sec:fermi_trig}

Seven GRBs were triggered by the \textit{Fermi}\textendash GBM and detected only in the gamma-ray band. For five out of the seven events, we were able to obtain 15 continuous MWA snapshot observations, covering 30\,min post-burst. During our MWA observation of GRB 190804A, the source coordinates were updated by \textit{Fermi} by more than 20\,deg, resulting in the first few MWA pointing centres being well separated from the final GRB position, and subsequently discarded. Note that in Table \ref{SGRBs}, the quoted start time is the earliest time that the MWA was actually pointed at the GRB, with the time post-burst corresponding to this delay with respect to the GRB detection by the \textit{Fermi}\textendash GBM. We also discarded the last few observations of GRB 190420.98, of which the pointing centres were driven away from the GRB location (beyond 50\% of the primary beam) by the Sun suppression algorithm. Among the seven \textit{Fermi} GRBs, three were further localised by the Interplanetary Network (IPN; \citealt{Hurley13}), one was later localised by \textit{Swift}\textendash BAT, whereas for the other four events, the only position was provided by the \textit{Fermi}\textendash GBM. The positional information contained in the GBM Final Position Notice usually have uncertainties of about a  1\textendash10\,deg radius, which are less accurate than the IPN and BAT localisations.

\subsection{Data Processing}
For our data processing we used the MWA-fast-image-transients pipeline\footnote{\mbox{\url{https://github.com/PaulHancock/MWA-fast-image-transients}}}, which automates the reduction of MWA data of transient events, including downloading, calibration and imaging on different timescales (for details about the pipeline see \citealt{Anderson20}). The final data products are images on timescales of 2\,m, 30\,s, and 5\,s as well as 0.5\,s/1.28\,MHz (coarse channel, i.e. splitting the 30.72\,MHz bandwidth into 24) images. It should be noted that every 2\,min observation actually means 112\,s of data due to the flagging of data at the beginning and end of each observation (see \citealt{Anderson20} for details). Here we present specific details regarding the image calibration and cleaning of the nine short GRBs in our sample.

\subsubsection{Calibration}

We used two calibration methods depending on the location of the GRB relative to bright sources in the field. Our preferred method is in-field calibration, which uses the GaLactic and Extragalactic All-sky MWA (GLEAM) survey \citep{Hurley17} as a source model within the field of view, and avoids the bulk refractive offset resulting from transferring solutions from dedicated calibration observations. However, for the GRBs with bright sources such as PicA or CenA within the primary beams, the in-field calibration was not possible as these sources were not included in the GLEAM catalogue. Instead, we derived our calibration solution from a nearby calibrator that had been observed within 12 hours of the GRB observation.

The calibration methods adopted for each of the nine GRBs are listed in the last column of Table \ref{SGRBs}. We applied the in-field calibration to five GRBs. 
Since there was a significant amount of extended emission in the region of GRB 190712.02, we discarded the baselines shorter than 500\,m before applying the in-field calibration. Self-calibration was applied on GRB 190420.98 to improve the image quality. For the other four GRBs, we derived calibration solutions using calibrator observations of a single bright source including PicA, HerA, and 3C353. 

\subsubsection{Imaging}
The MWA-fast-image-transients pipeline, which incorporates the {\sc WSClean} algorithm \citep{Offringa14, Offringa17}, was used to image and deconvolve the 2\,min observations of each GRB. We adopted a pixel scale of 32\,arcsec for the observations taken in the MWA phase I configuration, 16\,arcsec for the phase II extended configuration, and 1.6\,arcmin for the phase II compact configuration. The default image size was $4096\times4096$ pixels. For the \textit{Fermi} GRBs in our sample with poor localisations, we increased the image size to as large as $8000\times8000$ pixels. For the GRBs observed with the lower angular resolution compact configuration, we made smaller images of $1000\times1000$ pixels. See Figure~\ref{images_swift} and \ref{images_fermi} in Appendix~\ref{appendix:images} for the first 2\,min snapshots of the nine short GRBs in our sample.

Next we made images on 30\,s and 5\,s timescales, which accommodate the expected dispersive smearing of a prompt radio signal across the MWA observing bandwidth \textbf{for a redshift of $z\sim0.1$ (DM$\sim120\,\text{pc}\,\text{cm}^{-3}$) and $z\sim0.7$ (DM$\sim840\,\text{pc}\,\text{cm}^{-3}$)}, the lowest and average redshift known for short GRBs \citep{Rowlinson13}. This was done by splitting each 2\,min observation into 4 intervals of 30\,s, which were cleaned and imaged separately. We then split the 2\,min observation into 22 intervals of 5\,s starting from the first timestamp with real data (i.e. abandoning the first 4\,s and last 6\,s of data, see \citealt{Anderson20}), and created the 5\,s images without cleaning. Instrumental XX and YY polarisation images were converted into primary beam corrected Stokes I images using the fully embedded element MWA beam model \citep{Sokolowski17}.

To allow for a de-dispersion search for prompt signals, we created sub-band images by splitting the 2\,min observation into 0.5\,s intervals and the $30.72\,\text{MHz}$ bandwidth into 24 1.28\,MHz channels,
\textbf{which is the native coarse channel resolution of the MWA correlator (for the MWA system design see \citealt{Tingay13}). Note that we estimate our largest sensitivity loss due to intrachannel smearing within a 1.28\,MHz channel at the maximum DM search value of 2500\,$\text{pc}\,\text{cm}^{-3}$ (see Section~\ref{sec:disp_search}) to be $\sim60$\%, which we consider acceptable when weighed against the massive disk space and computational resources required to process images with a higher frequency resolution.}
We also created 0.5\,s full band images, which could be used to correct source positions in the 0.5\,s sub-band images as discussed below. We performed no cleaning on the 0.5\,s timescale images. 

Source position offsets caused by ionospheric effects and the errors in the absolute flux density calibration were corrected using {\sc fits\_warp}\footnote{\url{https://github.com/nhurleywalker/fits\_warp}} \citep{Hurley18} and {\sc flux\_warp}\footnote{\url{https://gitlab.com/Sunmish/flux\_warp}} \citep{Duchesne20}, which apply corrections to MWA images via comparisons to the GLEAM catalogue. Ionospheric corrections using {\sc fits\_warp} were applied to each individual 2\,min, 30\,s, and 5\,s image. There were too few sources found in the 0.5\,s\,/\,1.28\,MHz images to perform a reliable position correction, so instead we generated a solution from each 0.5\,s\,/\,30.72\,MHz image and applied it to the sub-band images in the same time bin. \textbf{We did not apply a chromatic correction to the frequency dependent ionospheric position offsets ($\propto1/\nu^2$). At our central observing frequency of 185\,MHz, the relative difference in the position offset within the 30.72\,MHz bandwidth is expected to be $\sim18\%$. Given the ionospheric position offset has a typical value of $\sim$arcmin \citep{Hurley18}, this difference would be smaller than the MWA synthesized beam size (see Section~\ref{sec:MWA}) and thus negligible.} However, the flux density calibration was computed by running {\sc flux\_warp} on the ionospherically corrected high signal-to-noise 2\,min images, with the resulting solutions then being transferred to the 30\,s, 5\,s and 0.5\,s images.

We did not correct the source positions and flux scales for the observations of GRB 190903A and GRB 191004A taken in the compact configuration as their low angular resolution meant that few sources could be matched to the GLEAM catalogue. We expected the ionospheric correction, typically a few tens of arcsec, to be much smaller than the $\sim\,10$\,arcmin resolution of the compact configuration and thus should not distort our images or analysis. We also expected a consistent flux calibration for the different timescale images across the 30\,min compact configuration observations of GRB 190903A and GRB 191004A as all 15 snapshots were calibrated using a single solution derived from an external calibrator.

\subsection{Data Analysis}

In this section we first describe the software used to search for transient and variable candidates in the MWA images, followed by the criteria we set to remove 
invalid candidates. 
We consider transient candidates as sources that appear in individual epochs and variable candidates as sources that remain detectable in multiple epochs but with a variable flux density, both of which may be coherent radio emission associated with GRBs (see the model descriptions in Section 4.2). 

\subsubsection{Transient and variable search}\label{sec:tran_var_search}

We adopted the {\sc Robbie}~\citep{Hancock19b} work-flow, which was further updated in \citet{Anderson20} to process the MWA images and search for variable and transient events within the positional error regions of each GRB.
For each GRB dataset, {\sc Robbie} first runs {\sc fits\_warp} to correct for ionospheric positional shifts in the individual images and then creates a mean image, which is then used to extract a persistent source catalogue and corresponding light curves. Comparisons between the mean and individual images, and a statistical analysis of the catalogue, are then used to identify variable and transient candidates. 

Provided the GRB position is known to within the synthesised beam of the images, we can add a monitoring position into the catalogue of persistent sources, which forces {\sc Robbie} to perform priorized fitting and extract a light curve at the best known position of the GRB to search for associated radio emission \citep{Hancock18}. 
We performed this analysis on the two \textit{Swift} GRBs in our sample as they were localised by XRT, which resulted in smaller position errors than the angular resolution of MWA \citep{Wayth18}.

We characterised the variability of all light curves output by {\sc Robbie} through the derivation of three parameters: the modulation index ($m$); the de-biased modulation index (\textit{$\text{m}_d$}), which takes into account the errors on the flux densities (see eq. 4 in \citealt{Hancock19b}); and the probability of being a non-variable source (p\_val).
The value p\_val is calculated from the light curves after being normalised by the uncertainty on each data point to account for the effect of varying uncertainties caused by the telescope changing its pointing centre throughout the observation (see eq. 3 in \citealt{Anderson20}).

For the less well localised \textit{Fermi} GRBs, we assumed that any associated variable or transient candidates were located within the \textit{Fermi}\textendash GBM $1\sigma$ error region \citep{Narayana16} if they were not further localised by \textit{Swift}\textendash BAT or the IPN. In addition, since the noise level increases towards the edge of the MWA primary beam that can create spurious signals, we restricted our source search to the inner part of the MWA primary beam, within 50\% of the maximum sensitivity. Among the different MWA pointings in the $15\times2$\,min snapshot observations of a GRB, we picked the first pointing to determine the MWA primary beam as the prompt radio emission we are targeting is most likely to appear in the first few snapshots (see Section 2.1). 
We therefore only searched for candidates within the overlap between these two regions. An example of the overlap region between the IPN position of GRB 170827B and 50\% of the MWA primary beam within which we searched for transients and variables is shown in Figure \ref{image_170827B} (for other \textit{Fermi} GRBs see Figures~\ref{images_swift} and \ref{images_fermi} in Appendix~\ref{appendix:images}). The final list of variable and transient candidates within the region of interest (ROI) of all \textit{Fermi} GRBs were retained for further inspection.

\subsubsection{Transient and variable selection} \label{sec:tran_var_sel}

For the \textit{Swift} GRBs, we looked for any transients or variables detected by {\sc Robbie} at their known positions. We also inspected the light curves and corresponding variability statistics as output by {\sc Robbie} that were generated via priorized fitting at the GRB positions for evidence of transient or variable behaviour. 

For the \textit{Fermi} GRBs, we inspected the {\sc Robbie} transient and variable candidates found within the ROI.  
In order to make a first cut on transient and variable candidate selection, we devised a set of tests to filter out false positives (such as noise fluctuations or imaging artefacts) as described in the following.

\begin{figure}
\centering
\includegraphics[width=\linewidth]{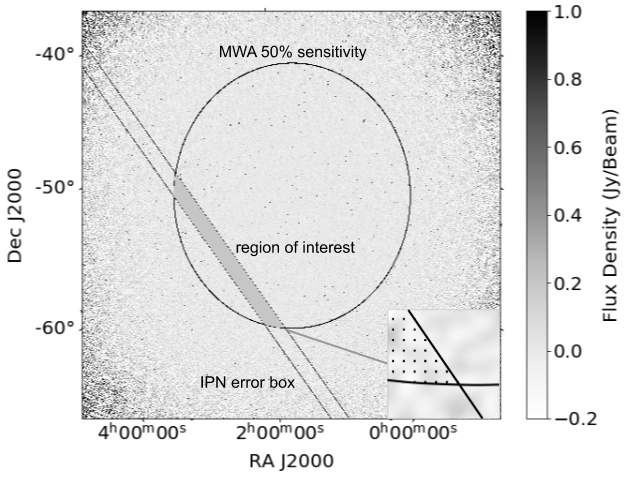}
\caption{MWA image of the field of GRB 170827B. The image size is $30^\circ\times30^\circ$, and the integration time is 2\,min beginning 34s post-burst. The boundaries of the MWA primary beam and the IPN error box are shown with black lines, where the overlap (grey shaded area) shows the region of interest (ROI) we searched for transients and variables. The inset at the bottom right corner is a zoomed in view to display the black dots that illustrate the independent pixels selected for the de-dispersion analysis (see section 2.4.3).}
\label{image_170827B}
\end{figure}

All transient candidates had to have a signal-to-noise ratio (SNR) $\geq6$, which corresponds to a false positive rate of $\sim10^{-9}$ under the assumption of Gaussian noise that is independent in both the space and time dimensions. The number of trials was estimated by the number of synthesised beams in the ROI. 
For example, for GRB 190420.98 there were $\sim2\times10^5$ synthesised beams in the ROI, and there were ten 2\,min snapshots, resulting in a final trial number of $\sim2\times10^6$. Therefore, we expected a false positive transient rate of $2\times10^{-3}$ at a $6\,\sigma$ level in the ROI for GRB 190420.98 for the full 20\,min observation.
See Table \ref{transient} in Appendix~\ref{sec:appendix}, which lists the number of synthesised beams, the transient false positive rate $\geq6\sigma$ within each ROI for each transient timescale (2\,min, 30\,s, and 5\,s) assuming Gaussian statistics, and the total number of transient candidates detected by {\sc Robbie} for each \textit{Fermi} GRB. 

As a sanity check, we used another method to estimate the expected transient false positive rate in each GRB ROI.
Again assuming the image noise conforms to Gaussian statistics, we estimated the expected number of false positive events in the ROI by taking the number of candidates found in a larger image region defined by the 50\% MWA primary beam and then multiplying by the ratio of the region areas.
In the case where more transient candidates were detected by {\sc Robbie} in the ROI than predicted via this method, we inspected the individual images and removed any candidates that were consistent with sidelobes from bright sources or other imaging artefacts.
For example, while we detected a transient candidate in the ROI for GRB 190420.98, there were significant sidelobe artefacts from the nearby radio galaxy PKS 2153-69, indicating the candidate was unlikely to be real.
The expected transient false positive rate within the ROI of each of the seven \textit{Fermi} GRBs based on comparisons to the number of false events in the MWA 50\% primary beam can be found in Table \ref{transient}.

For variable candidates, we followed the threshold set in \citet{Hancock19b}, i.e. p\_val $< 10^{-3}$ and \textit{$m_d$} > 0.05, to distinguish variables from non-variables.
As a sanity check, we again computed the number of variables in the MWA 50\% primary beam to estimate the expected variable false positive rate in the ROI as we did for transient candidates.
In the case of an excess of variables in the ROI, we compared their light curves with the light curves of nearby sources. If a variable candidate showed the same trend of variation as the nearby sources, the variation was probably caused by short-timescale calibration, measurement, or instrumental errors rather than being intrinsic to the source \citep{Bell19}.
All variable candidates found by {\sc Robbie} and the expected variable false positive rate within each \textit{Fermi} GRB ROI are listed in Table \ref{transient}.

Following this analysis, no transients or variables were identified in the ROIs of the \textit{Fermi} GRBs. Based on minimising the false positive rate assumed from Gaussian statistics, we quote $6\,\sigma$ flux density upper limits (6 times the average RMS within the ROI) for each of the \textit{Fermi} GRBs on timescales of 2\,min, 30\,s, and 5\,s, as shown in Table \ref{limit}. 
As the light curves generated at the position of the \textit{Swift} GRBs on different timescales are consistent with the noise, we quote  $3\,\sigma$ flux density upper limits.
We also include a $3\,\sigma$ deep limit derived from the 30\,min full observation for each GRB in Table \ref{limit}, which can be used to constrain persistent emission models (see Section~\ref{sec:model_pers}).

\subsubsection{Search for dispersed signals} \label{sec:disp_search}

Our observations are specifically targeting prompt, coherent radio signals predicted to be associated with short GRBs, which will be dispersed in time by the intervening medium between their origin and the Earth. We therefore perform a search for dispersed signals across a wide range of DMs over the entire $\sim$\,30\,min triggered observation.
A technique has been developed to search for dispersed signals in short timescale, sub-band radio images of a transient event, which generates a de-dispersed time series in units of SNR at a single sky position (for details see \citealt{Anderson20}). 
The de-dispersion code has four main steps for creating de-dispersed time series, which was run on all relevant sky positions within the ROI of each short GRB. 

\begin{enumerate}
\item \textit{Create a dynamic spectrum at each relevant sky position using the 0.5\,s sub-band images.} For \textit{Swift} GRBs, we created a dynamic spectrum at the pixel coincident with the GRB position. For \textit{Fermi} GRBs with large positional errors, we only processed the independent sky/field positions within the ROI, which were essentially one position per synthesised beam as illustrated in Figure \ref{image_170827B}. The search areas of GRB 190712.02 and GRB 190804A were prohibitively large (see Figure~\ref{images_fermi} in Appendix~\ref{appendix:images}) so we did not perform this analysis on these two \textit{Fermi} GRBs. 

\item \textit{Create a de-dispersed time series from each dynamic spectrum.} For every 0.5\,s time step and $12\,\text{pc}\,\text{cm}^{-3}$ DM trial across the whole 30\,min observation (see details in the next paragraph) we calculated the average de-dispersed flux density over the dynamic spectrum pixels crossed by the dispersive sweep (see \citealt{Anderson20} for a visualisation of the dynamic spectrum and de-dispersed time series).

\item \textit{Estimate the noise levels of the de-dispersed time series.} For \textit{Swift} GRBs, this was calculated by running the de-dispersion code on 100 nearby position pixels (in addition to the signal pixel) to create 100 de-dispersed time series. To remove any persistent emission that may be at that position, we averaged each of the 101 de-dispersed timeseries in time and subtracted this mean from their corresponding parent timeseries. 
We then calculated the standard deviation of the 100 mean subtracted timeseries (not including the signal pixel), which we defined as the de-dispersed time series of the noise.
For \textit{Fermi} GRBs, the noise was estimated in the same way but by averaging the de-dispersed time series created at each independent pixel in the ROI.

\item \textit{Create the final de-dispersed time series in SNR units.} This was done by dividing the de-dispersed time series at the signal pixel (\textit{Swift}) or independent pixels in the ROI (\textit{Fermi}) by the de-dispersed time series of the noise derived in the previous step.
\end{enumerate}

The DM resolution of our search for dispersed signals ($12\,\text{pc}\,\text{cm}^{-3}$) was chosen by equating the expected dispersion smearing across the full 30.72\,MHz bandwidth to the temporal resolution of 0.5\,s (see eq. 1 in \citealt{Anderson18}). All but one of the GRBs in our sample have no redshift measurement so we searched for dispersed signals across the DM space that covers the known redshift range of short GRBs ($0.1<z<2.5$; \citealt{Rowlinson13}).
The contribution of the intergalactic medium to the DM of a short GRB can be estimated from the redshift for the cosmological paradigm of a flat universe. We adopted the method described by \citet{JP20} to calculate $\text{DM}_{\text{IGM}}$, taking into account the redshift evolution of the fraction of cosmic baryons in diffuse ionized gas. The redshift range corresponded to a range of $90 < \text{DM}_{\text{IGM}} < 2400$ $\text{pc}\,\text{cm}^{-3}$. Considering the typically large offset of short GRBs from the centers of their host galaxies \citep{Fong13}, we assumed the DM contribution from their host galaxies to be small  ($\text{DM}_{\text{host}}\sim30\,\text{pc}\,\text{cm}^{-3}$; \citealt{Cordes02}). Assuming a similarly small contribution from the Milky Way based on the YMW16 DM model ($\text{DM}_{\text{MW}}\sim30\,\text{pc}\,\text{cm}^{-3}$; \citealt{YMW16}), we adopted a DM range of \mbox{150\textendash2500 $\text{pc}\,\text{cm}^{-3}$} for our search for dispersed signals associated with GRBs without a known redshift. 

The redshift of GRB 190627A corresponded to a $\text{DM}_{\text{IGM}}$ of $\sim1800\,\text{pc}\,\text{cm}^{-3}$. However, this value can vary depending on the number of galactic halos intersected by the line of sight, corresponding to a possible DM range between 1400 and 2400\,$\text{pc}\,\text{cm}^{-3}$, which encompasses 90\% of the expected values \citep{JP20}.
The DM contribution from the Milky Way in the direction of GRB 190627A ($l=8.17^{\circ}$, $b=30.26^{\circ}$) is estimated to be $\text{DM}_{\text{MW}}\sim50\,\text{pc}\,\text{cm}^{-3}$ based on the YMW16 electron-density model \citep{YMW16}. We therefore estimate a DM of $\text{DM}_{\text{host}}+\text{DM}_{\text{MW}}+\text{DM}_{\text{IGM}} = 1900^{+600}_{-400}\,\text{pc}\,\text{cm}^{-3}$ for GRB 190627A.

In order to set a threshold for selecting dispersed signal candidates for further investigation in the resulting de-dispersed time series, we first considered the \textit{Swift} GRBs and determined the number of trials based on the time and DM steps used in our analysis.
Given that there are $\sim3\times10^6$ trials for GRB 191004A and $\sim10^{6}$ for GRB 190627A, we set a threshold of $5\sigma$, corresponding to less than one false positive for each \textit{Swift} GRB.
In the case of \textit{Fermi} GRBs, which are not localised to a single pixel, we assessed the noise statistics of the de-dispersed time series in the ROI for each event by creating a set of time series from the same dataset using (nonphysical) negative DM values. If the GRB error region contains only noise, a de-dispersion analysis with positive and negative DMs (same range of absolute values) should give similar SNR distributions.
Table \ref{dedispersion_Fermi} shows a comparison of the maximum SNR and number of high SNR events above $5\sigma$ for the set of positive and negative DMs for each \textit{Fermi} GRB. We also include the expected number of false positive events $>5\,\sigma$, along with the maximum SNR for which we expect there to be only one false positive event assuming a Gaussian distribution.
From Table \ref{dedispersion_Fermi}, one can see that the high SNR events observed in our dataset are consistent with noise. There are fewer detected events above $5\sigma$ than the expected number of false positive events, which may be caused by an overestimation of the noise. The noise calculated in our data using the standard deviation of a population of pixels was affected by the sensitivity changing across the image, which is higher than the value expected in the case of an unchanged sensitivity. If any signals are detected with a higher SNR in the positive de-dispersed time series than the maximum measured in the negative de-dispersed time series, then it is possible that they are real signals.

In the case of no dispersed signal detections in the de-dispersed time series, we derived an upper limit for each \textit{Swift} GRB using signal simulations (see Section~\ref{sec:disp_sim}) and adopted a $7\sigma$ upper limit for each \textit{Fermi} GRB given less than one event above $7\sigma$ is expected from Gaussian statistics (see Table \ref{dedispersion_Fermi}).
We present our dispersed signal search results in units of fluence (Jy\,ms, the integrated flux density over the pulse width), which is common in the fields of FRB and pulsar astrophysics (see Tables \ref{dedispersion_Fermi} and \ref{dedispersion_Swift}).

\begin{table}
\centering
\resizebox{.75\columnwidth}{!}{\vspace{-3cm}\begin{tabular}{l l l l l}
\hline
GRB & \multicolumn{4}{c}{Upper limit (Jy/beam)} \\
\cline{2-5}
 & 30\,min & 2\,min & 30\,s & 5\,s \\
\hline
170827B & 0.41 & 0.94 & 2.1 & 2.1\\
190420.98 & 0.29 & 0.27 & 0.58 & 1.1 \\
190627A & 0.027 & 0.084 & 0.29 & 0.42 \\
190712.02 & 0.33 & 0.86 & 2.1 & 4.6 \\
190804A & 0.21 & 0.58 & 2.1 & 4.1 \\
190903A & 2.9 & 11 & 15 & 20 \\
191004A & 1.1 & 2.0 & 1.9 & 1.9 \\
200325A & 0.19 & 0.39 & 0.81 & 1.7 \\
200327A & 0.2 & 0.71 & 1.4 & 3.2 \\
\hline
\end{tabular}}
\caption{Upper limits on the radio flux density of 
transient and variable emission associated with the nine short GRBs in our sample. We quote $6\,\sigma$ upper limits (6 times the average RMS within the ROI) for \textit{Fermi} GRBs and $3\,\sigma$ for \textit{Swift} GRBs. Given the noise evolves with time for each GRB, we quote the maximum value for the upper limit. We also include the $3\sigma$ deep limit derived from the 30\,min full observation for each GRB.}
\label{limit}
\end{table}

\begin{table*}
\begin{threeparttable}
\centering
\resizebox{2\columnwidth}{!}{\begin{tabular}{c c c c c c c c}
\hline
 GRB & DM\tnote{1} & Trial number\tnote{2} & Maximum SNR\tnote{3} & Critical SNR & Detected events & Expected false events & Fluence limit\\
 & &  &  & for 1 event\tnote{4} & above $5\sigma$\tnote{5} & above $5\sigma$\tnote{6} & (Jy\,ms)\\
\hline
170827B & P & $\sim2\times10^{10}$ & 6.564 & 6.5 & 1920 & 5733 & 610\textendash1616 \\
& N & & 6.662 & & 1395 & & \\
\hline
190420.98 & P & $\sim10^{11}$ & 6.936 & 6.9 & 15195 & 28665 & 609\textendash1609 \\
& N & & 6.750 & & 15375 & & \\
\hline
190903A & P & $\sim10^9$ & 5.237 & 6 & 40 & 287 & 2067\textendash12110 \\
& N &  & 5.351 &  & 50 &  & \\
\hline
200327A & P & $\sim10^{10}$ & 6.190 & 6.4 & 2415 & 2866 & 166\textendash2064 \\
& N & & 6.177 & & 2324 & & \\
\hline
\end{tabular}}
\caption{Results following the search for dispersed signals associated with \textit{Fermi} GRBs. This includes a comparison of high SNR events arising from the positive and negative DM time series analysis described in Section~\ref{sec:disp_search}. We also list the $7\,\sigma$ upper limits on the fluence for each GRB in the last column. Given the noise varies with DM, time and the GRB position within the ROI due to the MWA primary beam, we present a range for the fluence upper limits (see details in Section~\ref{sec:disp_lim_f}). GRB 190712.02 and GRB 190804A are not included in this analysis due to their poor localisations (see Section~\ref{sec:disp_search}), and GRB 200325A is analyzed along with the \textit{Swift} GRBs (see Section~\ref{sec:disp_sim}). \\
1: The positive (P) or negative (N) DMs used to create the de-dispersed time series; \\
2: The trial number estimated from the number of time steps, DM trials and synthesised beams (for which a de-dispersed time series was generated) within the ROI;\\
3: The maximum SNR event detected; \\
4: The critical SNR beyond which we expect there to be just one event;\\
5: The number of events detected with SNRs above $5\sigma$ in the positive and negative DM de-dispersed time series;\\ 
6: The expected false positive event rate above $5\sigma$  assuming a Gaussian distribution.\\
7: The $7\sigma$ fluence limits on dispersed signals associated with each GRB.
}
\label{dedispersion_Fermi}
\end{threeparttable}
\end{table*}

\subsubsection{Dispersed signal simulations for well localised GRBs} \label{sec:disp_sim}

\begin{figure}[ht!]
\subfigure[GRB 190627A]{
  \label{efficiency_1}
  \includegraphics[width=.89\linewidth]{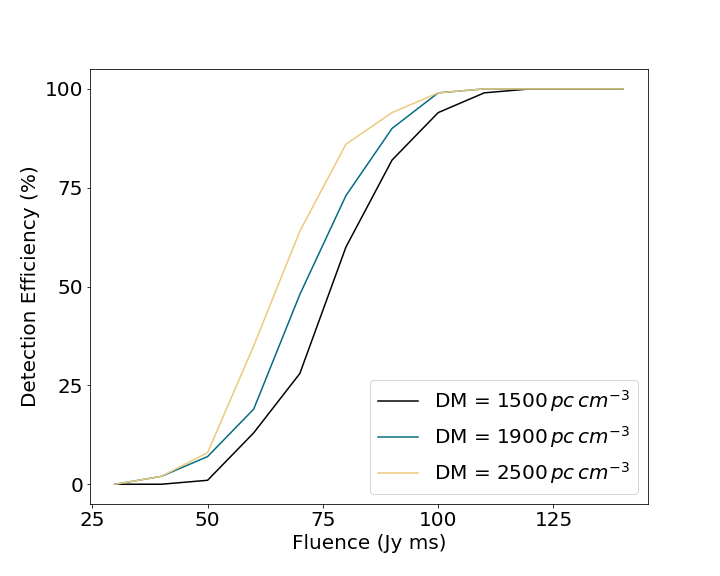}} \\
\subfigure[GRB 191004A]{
  \label{efficiency_2}
  \includegraphics[width=.89\linewidth]{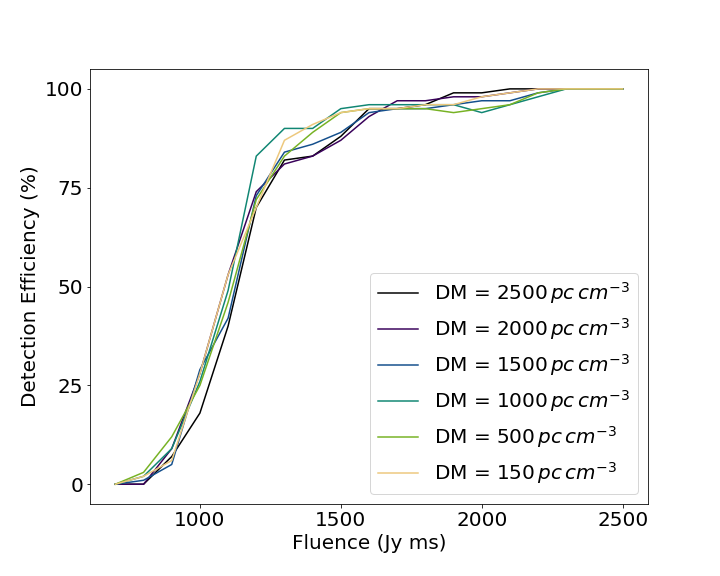}} \\
\subfigure[GRB 200325A]{
  \label{efficiency_3}
  \includegraphics[width=.89\linewidth]{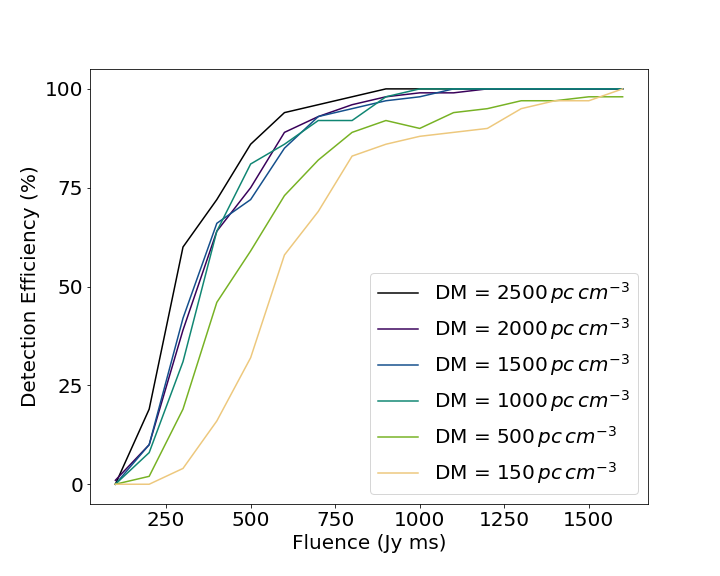}}
\caption{Detection efficiency of dispersed signals as a function of fluence for GRB 190627A, GRB 191004A and GRB 200325A calculated through signal injection (see Section~\ref{sec:disp_sim}). The DM ranges of the simulated signals were based on the known redshift of GRB 190627A (see Section~\ref{sec:disp_search}) or in the case of GRB 191004A and GRB 200325A, the known redshift range of short GRBs.
}
\label{efficiency}
\end{figure}

As GRBs 190627A, 191004A, and GRB 200325A were well localised by \textit{Swift} (\textit{Fermi} GRB 200325A was localised by \textit{Swift}\textendash BAT to within 50 synthesised beams), 
we were able to inject simulated pulses into the de-dispersed time series to determine our sensitivity to such signals in each MWA observation, thereby allowing us to derive fluence limits as a function of DM (as done by \citealt{Anderson20}). 
As GRB 191004A and GRB 200325A have no known redshift, we simulated pulses over a large DM range, including 150, 500, 1000, 1500, 2000 and 2500\,$\text{pc}\,\text{cm}^{-3}$. 
As the redshift is known for GRB 190627A, we simulated pulses over a smaller DM range between 1500 and 2500\,$\text{pc}\,\text{cm}^{-3}$. 
By injecting signals over a wide range of fluence values, we were able to test the efficiency of our detection algorithm in the three GRB datasets, which are plotted in Figure \ref{efficiency}.
The fluence limits quoted in Table \ref{dedispersion_Swift} are the signal fluence corresponding to the 90\% detection efficiency of our algorithm.
The performance of our algorithm was different for each MWA observation depending on many factors such as the presence of bright sources in the field, the GRB location within the primary beam, and the elevation of the observation.

\subsubsection{Fluence limits for \textit{Fermi} GRBs} \label{sec:disp_lim_f}

Signal injection was not a viable method for calculating the fluence limits of the \textit{Fermi} GRBs as their poorer localisations means that the sensitivity changes significantly across the ROI, and the performance of our algorithm is dependent on the sky position we choose to inject the signals. 
Although we could provide a fluence limit as a function of DM and signal position in the ROI using the signal injection technique for a given \textit{Fermi} GRB, it would be hugely computationally expensive. Instead we created a de-dispersed time series for each of the independent positions in the ROI and derived a $7\,\sigma$ fluence upper limit for each of the \textit{Fermi} GRBs using the noise calculated from these de-dispersed time series, corresponding to a false positive rate of $\sim10^{-12}$ under the assumption of Gaussian noise (see Table \ref{dedispersion_Fermi}). Given the noise varies with DM, time, and the position in the MWA primary beam, we present a range for the fluence upper limits for the \textit{Fermi} GRBs.

\section{results}
\label{sec:results}
\subsection{\textit{Swift} GRBs}
The light curves derived at the position of the two \textit{Swift} GRBs using priorised fitting showed no evidence of variable or transient radio emission over any of the timescales investigated (see Table \ref{var} in Section~\ref{sec:appendix} for the corresponding variability parameters described in Section~\ref{sec:tran_var_search}) and were consistent with the local rms noise.
For both events we quote the $3\sigma$ upper limits on the flux density of an associated radio transient on timescales of 30\,min, 2\,min, 30\,s, and 5\,s in Table \ref{limit}.

\begin{table}
\begin{threeparttable}
\centering
\resizebox{.6\columnwidth}{!}{\hspace{-0.0cm}\begin{tabular}{c c}
\hline
GRB & fluence limit (Jy\,ms) \\
\hline
190627A & 80\textendash100 \\

191004A & 1300\textendash1600 \\

200325A & 600\textendash1200 \\
\hline
\end{tabular}}
\caption{The fluence limit on dispersed signals for the two \textit{Swift} GRBs and one \textit{Fermi} GRB. These limits correspond to the 90\% detection efficiency of our detection algorithm to simulated signals injected into the dedispersed timeseries of these three events (see Section~\ref{sec:disp_sim}).}
\label{dedispersion_Swift}
\end{threeparttable}
\end{table}

We also performed a search for dispersed signals at the position of the two \textit{Swift} GRBs but none were detected above $5\,\sigma$ (which is well below our detection threshold of $7\,\sigma$). 
In order to calculate the efficiency of our detection algorithm to dispersed signals, we injected simulated pulses covering a fluence range of 30\textendash140\,Jy\,ms and a DM range of 1500\textendash2500\,$\text{pc}\,\text{cm}^{-3}$ 
into the dedispersed timeseries of GRB 190627A (see Sections~\ref{sec:disp_search} and \ref{sec:disp_sim}). 
The variation of detection efficiency as a function of fluence for this GRB is shown in Figure \ref{efficiency_1}, which increases with increasing DM. For this DM range, we found a 90\% detection efficiency for signals with a fluence of 80\textendash100\,Jy\,ms.
\textbf{Note that the 90\% detection efficiency is commonly used as a threshold in FRB simulations to validate FRB search pipelines (e.g. \citealt{Farah19, Gupta21}).}
As GRB 191004A has no known redshift, we performed this simulation over a much broader DM range of 150\textendash2500\,$\text{pc}\,\text{cm}^{-3}$ (see Section~\ref{sec:disp_search}). 
The detection efficiency of dispersed signals in the GRB 191004A dataset as a function of fluence for various DM values are also shown in Figure \ref{efficiency_2}, which indicates a 90\% detection efficiency between 1300 and 1600\,Jy\,ms within the DM range explored. 
The GRB 191004A dataset is therefore less sensitive than that of GRB 190627A, which reflects the different noise levels in the images of these two GRBs (see Figure~\ref{images_swift} in Appendix~\ref{appendix:images}). The much higher noise in the field of GRB 191004A is caused by the sidelobes from a nearby bright source (Fornax A).
We use the 90\% detection efficiency as the fluence limit in our analysis in Section~\ref{sec:EMmod} for both \textit{Swift} GRBs, which are also listed in Table \ref{dedispersion_Swift}.

\subsection{\textit{Fermi} GRBs}
We inspected the transient candidates detected on timescales of 2\,min, 30\,s, and 5\,s from the seven \textit{Fermi} GRBs (see Table \ref{transient}).
We used two different methods to estimate the expected false positive detection rate (see Section~\ref{sec:tran_var_sel}).
Given that there was no transient candidate passing our inspection, we quote a $6\sigma$ upper limit on the flux density for the \textit{Fermi} GRBs, as shown in Table \ref{limit}.

We inspected the variable candidates detected from the seven \textit{Fermi} GRBs (see Table \ref{transient}), and found the number of variables detected in each GRB ROI to be consistent with the expected false positive event rate.
Further inspection revealed that all candidates showed similar light curve variations to nearby sources, 
with larger flux density variations likely due to errors associated with ionospheric positional corrections (GRB 190712.02 and GRB 190804A) or errors associated with the flux density scale correction (GRB 190420.98). No variable candidates passed our inspection.

We searched for dispersed signals from the \textit{Fermi} GRBs (excluding GRB 190712.02 and GRB 190804A for the reason given in Section~\ref{sec:disp_search}) following the procedure described in Sections~\ref{sec:disp_search}, \ref{sec:disp_sim} and \ref{sec:disp_lim_f}.
\textit{Fermi} GRB 200325A was localised by \textit{Swift}-BAT to within 4\,arcmin \citep{DeLaunay20} so we can assume that the image noise did not change over this small region (see Figure~\ref{images_swift} in Appendix~\ref{appendix:images}). We therefore performed the same simulation analysis as for \textit{Swift} GRBs at the best known GRB position. We found no signal above $5\,\sigma$ in the de-dispersed time series. 
Our detection efficiency of the simulated signals as a function of fluence is shown in Figure \ref{efficiency_3}, which yields a fluence upper limit of 600\textendash1200\,Jy\,ms for the DM range of \mbox{150\textendash2500 $\text{pc}\,\text{cm}^{-3}$}  
(90\% detection efficiency, see Table \ref{dedispersion_Swift}).
The four events GRB 170827B, GRB 190420.98, GRB 190903A and GRB 200327A were localised by \textit{Fermi}-GBM or IPN, and had large positional errors (see Figure~\ref{images_fermi} in Appendix~\ref{appendix:images}). We compared the high SNR events observed in the real data to the SNRs produced by negative DMs. 
The SNR of the brightest dispersed signal detected by our algorithm for these four \textit{Fermi} GRBs is given in Table \ref{dedispersion_Fermi}, which are all SNR$<7$.
Given the similar maximum SNR values resulting from the processing of both the positive and negative DM datasets, we conclude there is no compelling evidence of any dispersed signals. The range in $7\,\sigma$ fluence upper limits derived from the de-dispersed time series created for each independent pointing in the ROI for the above four \textit{Fermi} GRBs are given in Table \ref{dedispersion_Fermi}.

In summary, no associated radio emission was observed for any of the nine short GRBs.
Table \ref{limit} shows the flux density upper limits derived from the transient/variable analysis of the nine GRBs over different timescales, and Table \ref{dedispersion_Fermi} and \ref{dedispersion_Swift} show the fluence upper limits derived from a search for dispersed signals via an image dedispersion analysis and signal simulations.
The most stringent limits are from GRB 190627A as it was located close to the MWA pointing centre with no nearby bright sources. All these upper limits can now be used to constrain the theoretical coherent radio emission models applicable to BNS mergers (e.g. \citealt{Rowlinson19}).

\section{constraints on emission models}\label{sec:EMmod}

We discuss the implications of our fluence and flux density upper-limits with respect to
three models that predict coherent radio emission arising from BNS mergers within the context of our sample of short GRBs. 
Two of the models described in the following predict emission that results from the production of a (quasi-)stable magnetar so we assume that such a remnant was formed by each of the short GRBs studied in this paper. 
In this section, we first discuss the magnetar parameters derived or assumed for each event before describing each model. We then discuss the coherent emission constraints placed by individual short GRBs and by the whole sample. 

\subsection{Central engine activity}\label{sec:cen_eng}

The positions of the two \textit{Swift} GRBs are well known due to their XRT detections, with the resulting X-ray light curves allowing us to derive additional parameters for any remnant.
The plateau phase observed in many GRB X-ray light curves is often interpreted as energy injection from a (quasi-)stable magnetar remnant formed from the BNS merger \citep{Zhang01}. We consider two different magnetar models: stable and unstable. If the magnetar remnant possesses a mass less than the maximum mass allowed for a NS (depending on the equation of state; \citealt{Lasky14}), the merger product would be a stable magnetar. In the other case, the magnetar is unstable as its mass is initially  
supported by its rapid spin.
However, it soon collapses into a BH as it spins down, resulting in a steep decay in the X-ray light curve.

\begin{figure*}
\centering
\subfigure[GRB 190627A]{
  \label{GRB190627A_4}
  \includegraphics[width=.47\linewidth]{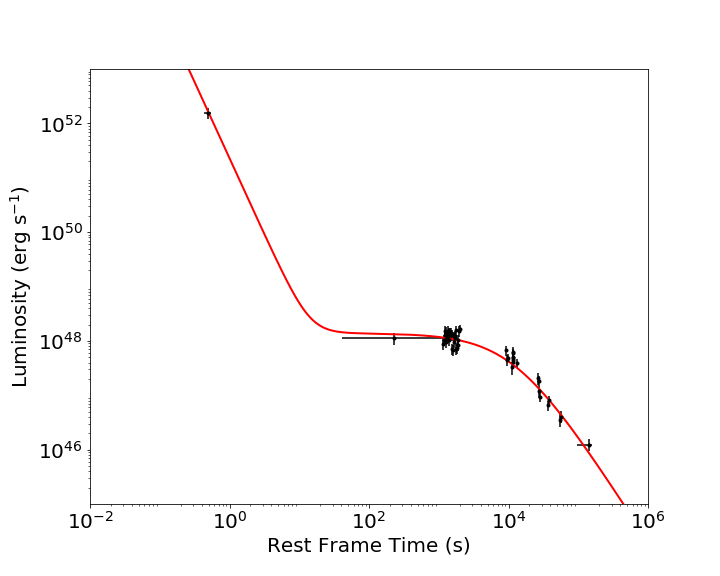}}
\qquad  
\subfigure[GRB 191004A]{
  \label{GRB191004A_4}
  \includegraphics[width=.47\linewidth]{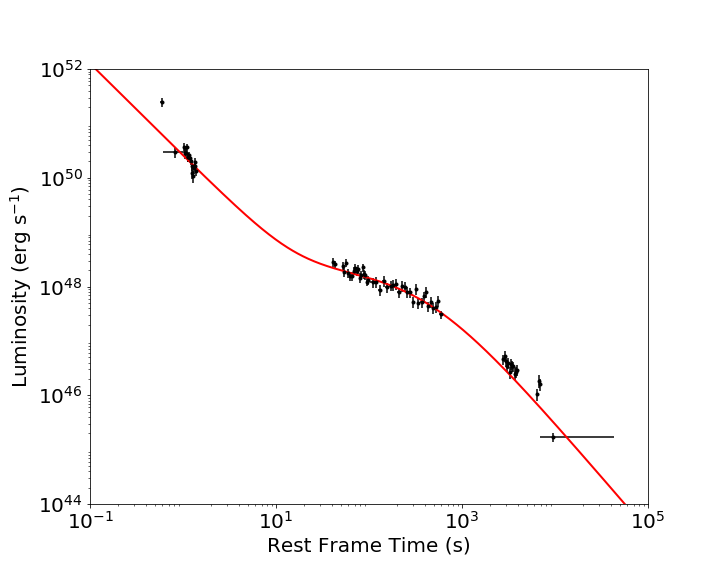}}
\caption{The rest-frame \textit{Swift}\textendash BAT and \textendash XRT light curves of GRB 190627A and GRB 191004A. The black points represent the BAT and XRT data, and the red lines show the fit to the magnetar central engine powering the plateau phase (see Section~\ref{sec:cen_eng}). We used the redshift $z=1.942$ for GRB 190627A and assumed a typical short GRB redshift of $z=0.7$ for GRB 191004A.}
\label{X-ray}
\end{figure*}

In order to determine the magnetar parameters (for the method see \citealt{Rowlinson13}), we created the rest-frame X-ray light curves for GRB 190627A and GRB 191004A by combining their \textit{Swift}-BAT and XRT data from the \textit{Swift} Burst Analyser \citep{Evans10}. Although the \textit{Fermi} GRB 200325A was also detected by \textit{Swift}-BAT, this observation lasted only $\sim150$\,s \citep{James20}, insufficient for fitting the magnetar model.
Since we do not know the redshift of GRB 191004A, we assumed a value of 0.7, the average redshift for short GRBs. We also applied a $k$-correction~\citep{Bloom01} to obtain the 1\textendash10000\,keV rest-frame luminosity light curves, as shown in Figure \ref{X-ray}.
Considering that the decay phase following the plateau in both X-ray light curves does not agree with the simple curvature effect expected for the collapse of an unstable magnetar into a BH \citep{Rowlinson10, Rowlinson13}, we fitted a stable magnetar model to these data
to determine the bolometric luminosity of the magnetar (see eq. 8 in \citealt{Rowlinson19}).
This bolometric luminosity and the duration of the X-ray plateau are used to calculate the magnetar magnetic field strength and the initial spin period, assuming a magnetar mass of $2.1\,\text{M}_\odot$, a radius of $10^6$\,cm, and an efficiency factor of $f=3.45$
(see eq. 6, 7 and section 2.2.2 in \citealt{Rowlinson19}).
Table \ref{magnetar} shows the fitted magnetar parameters for the two \textit{Swift} GRBs. Compared to the typical range of magnetar parameters proposed by \citet{Rowlinson19}, the fitted magnetar remnant for GRB 191004A may be considered typical, while for GRB 190627A it has a much faster spin period and weaker magnetic field. Note that in our modelling we assumed the magnetar spins down solely through magnetic dipole radiation. Incorporating other radiative losses such as gravitational wave emission (e.g. \citealt{Sarin20}) may further improve our fitting results.

As the \textit{Fermi} GRBs in our sample do not have X-ray data for deriving the magnetar parameters, we assumed the formation of a `typical' stable magnetar remnant, which, from the distribution of the magnetar parameters fitted to known X-ray plateaus (see figure 8 in \citealt{Rowlinson19}), may be defined as possessing a magnetic field of $2.4^{+4.6}_{-1.6}\times10^{16}$\,G and a spin period of $9.7^{+20.8}_{-6.6}$\,ms.
In addition, given the lack of X-ray data, we cannot make any assumptions regarding the stability of the magnetar remnant resulting from the \textit{Fermi} GRBs, so we do not consider coherent emission produced by the collapse of the magnetar into a BH (e.g. \citealt{Zhang14}).

\begin{table}
\begin{threeparttable}
\centering
\resizebox{0.7\columnwidth}{!}{\hspace{0cm}\begin{tabular}{c c c}
\hline
GRB & P (ms) & B ($10^{15}$\,G) \\ [0.2cm]
\hline
  190627A      & $1.353^{+0.047}_{-0.042}$ & $0.668^{+0.052}_{-0.045}$ \\ [0.2cm]
  191004A      & $6.16^{+0.32}_{-0.25}$ & $17.4^{+1}_{-0.9}$ \\ [0.2cm]
\hline
\end{tabular}}
\caption{Magnetar parameters derived from magnetar model fitting to the X-ray light curves of GRB 190627A and GRB 191004A, assuming a NS mass of $2.1\,\text{M}_\odot$ (as described in Section~\ref{sec:cen_eng}). P and B represent the spin period and magnetic field with $1\,\sigma$ uncertainties.}
\label{magnetar}
\end{threeparttable}
\end{table}

\subsection{Coherent emission models}\label{sec:models}

Various models have predicted the production of coherent radio emission associated with short GRBs (for a review see \citealt{Rowlinson19}). The emission predicted to be produced by the alignment of the merging NS magnetic fields is independent of the magnetar remnant (see Section~\ref{sec:model_NSB}), while the other models require us to derive magnetar parameters from X-ray data (see Sections~\ref{sec:model_ism} and \ref{sec:model_pers}).
Here we briefly review the three emission mechanisms we have chosen to test.

\subsubsection{Interactions of NS magnetic fields}\label{sec:model_NSB}

The earliest coherent radio emission may come from the inspiral phase, where interactions between the NS magnetospheres just preceding the merger may spin them up to millisecond spin periods and lead to a revival of the pulsar emission mechanism \citep{Lipunov96,Metzger16}. The emission starts from a few seconds prior to the merger and peaks at the first contact of the NS surfaces, giving rise to a short duration radio flash. Its flux density is predicted to be

\begin{equation}\label{eq:ns}
F_{\nu} \sim 2\times10^8\,(1 + z)\,\frac{B_{15}^{2}M_{1.4}^{3}}{R_6\nu_{9,\text{obs}}D^2}\,\epsilon_{\text{r}}\,\text{Jy},
\end{equation}

\noindent where $\nu_{9,\text{obs}}$ is the observing frequency in units of $10^9$\,Hz, $D$ is the distance to the binary system in Gpc, the radio emission efficiency ($\epsilon_{\text{r}}$) is the fraction of the wind luminosity being converted into coherent radio emission \citep[typically  $10^{-4}$ for known pulsars;][]{Taylor93}, 
and 
we assume typical NS parameters for the magnetic field of $B_{15}=10^{-3}$ in units of $10^{15}$\,G, 
a mass of $M_{1.4}=1$ in units of $1.4\,\text{M}_\odot$, 
and a radius of $R_6=1$ in units of $10^6$\,cm \citep{Rowlinson19}. 
Note that the parameters $B_{15}$, $M_{1.4}$ and $R_6$ are assumed properties of the merging NSs, not of the magnetar remnant. The distance can be estimated from the redshift $z$ using the predefined cosmology parameters~\citep{Wright06}.

\subsubsection{Interaction of relativistic jets with the ISM}\label{sec:model_ism}

If a Poynting flux dominated wind is produced by the magnetar remnant formed following the binary merger \citep{Usov94,Thompson94}, its interaction with the ISM would generate a low frequency, coherent radio pulse at the highly magnetised shock front \citep{Usov00}. As this emission is linked to the relativistic gamma-ray jet and only propagates through the pre-existing low density surrounding medium, it should be detectable within the first few minutes of a GRB trigger. The radio fluence is given by

\begin{equation}\label{eq:jet_ism}
    \Phi_{\nu}\sim\frac{0.1\epsilon_B(\beta-1)}{\nu_{\text{max}}}\Big(\frac{\nu}{\nu_{\text{max}}}\Big)^{-\beta}\Phi_{\gamma}\,\text{erg}\,\text{cm}^{-2}\,\text{Hz}^{-1}
\end{equation}

\noindent where $\epsilon_B$ is the fraction of magnetic energy in the relativistic jet, and $\beta\simeq1.6$ is the spectral index \citep{Usov00}. The peak frequency of the coherent radio emission ($\nu_{\text{max}}$ in GHz) is determined by the magnetic field at the shock front \citep[see equations 11-13 in section 2.4 of][]{Rowlinson19}, and $\Phi_{\gamma}$ is the observed gamma-ray fluence in $\text{erg}\,\text{cm}^{-2}$.

\subsubsection{Persistent emission following the formation of a magnetar}\label{sec:model_pers}

If the merger remnant is a magnetar, the coherent radio emission powered by dipole magnetic braking may be detectable during the lifetime of the magnetar \citep{Totani13,Metzger17}. If the beam of the magnetar aligns with the GRB jet and points towards the Earth, the coherent emission will be detected as persistent emission. The flux density of this emission is given by

\begin{equation}\label{eq:pers}
    F_{\nu}\sim8\times10^{7}\,\nu^{-1}_{\text{obs}}\,\epsilon_{\text{r}}\,D^{-2}\,B^{2}_{15}\,R^6_6\,P^{-4}_{-3}\,\text{Jy},
\end{equation}

\noindent where $P_{-3}$ is the magnetar spin period in units of $10^{-3}$\,s~\citep{Totani13,Rowlinson19}. 
This persistent radio emission may only be detectable for up to a few hours following the merger 
as its rapid spin-down rate would cause the dipole radiation to weaken very quickly or (if unstable) it collapses into a BH (which is $\lesssim2$\,h, depending on the magnetar parameters $B$ and $P$ derived in Section~\ref{sec:cen_eng}).

\subsection{Constraints from individual GRBs}
\subsubsection{GRB 190627A}\label{sec:1906}

GRB 190627A is the only event amid our nine short GRBs that has a known redshift ($z=1.942$).
Thanks to the redshift measurement, we can calculate a luminosity distance of 15.2\,Gpc and directly constrain the fundamental parameters of the emission models, 
such as the efficiency of radio emission $\epsilon_{\text{r}}$, which directly scales with the predicted flux densities from the interaction of the merging NS magnetic fields, and that of the persistent pulsar emission from the magnetar remnant (Eq. 1 and 3). We can also constrain the fraction of magnetic energy $\epsilon_B$ that directly scales with the predicted fluence from the interaction between the relativistic jet and ISM (Eq. 2). 
This has not been possible in previous works as low frequency radio upper limits of coherent radio emission have only poorly constrained these emission parameters under a presumed redshift (e.g. \citealt{Rowlinson19b}, \citealt{Rowlinson20}, \citealt{Anderson20}).
In addition, as this \textit{Swift} GRB has also been localised to within an MWA synthesised beam, we are able to derive a constraining dispersed signal (fluence) limit.

\begin{figure}
\centering
\includegraphics[width=.45\textwidth]{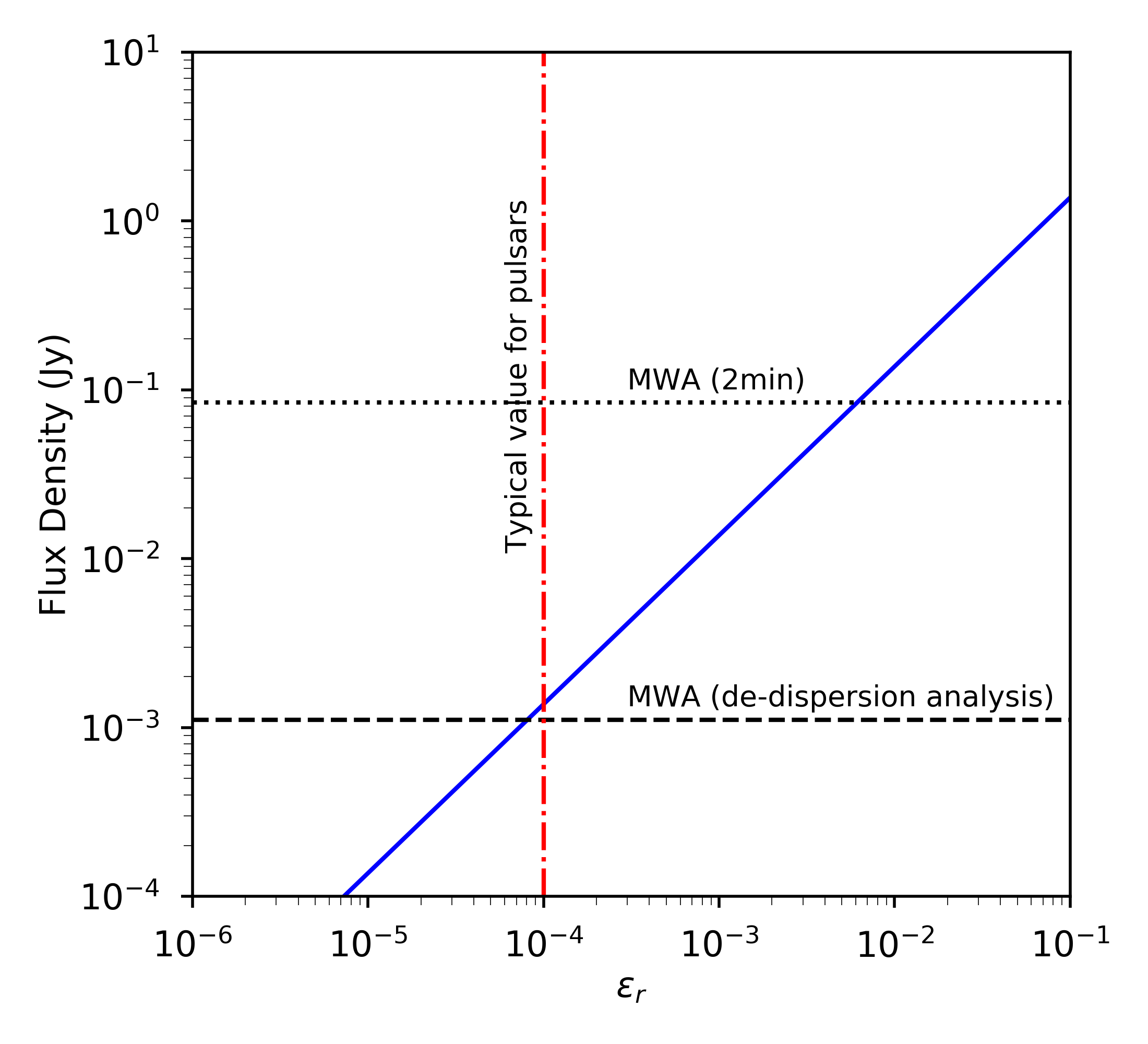}
\caption{The predicted 185\,MHz flux density (blue line) of the prompt signal emitted by the alignment of the merging NS magnetic fields (Section~\ref{sec:model_NSB}) in GRB 190627A as a function of the radio emission efficiency ($\epsilon_r$). The horizontal dotted line shows the least constraining flux density upper limit derived from the 2\,min snapshots of GRB 190627A (Table~\ref{limit}). 
The horizontal dashed line shows the flux density upper limit converted from the least constraining fluence limit derived from the image de-dispersion analysis (Table~\ref{dedispersion_Swift}), and the vertical line shows the typical efficiency observed for known pulsars ($\epsilon_r\sim10^{-4}$).} 
\label{GRB190627A_1}
\end{figure}

In Figure \ref{GRB190627A_1} we plot the predicted flux density of a signal produced by the alignment of the merging NS magnetic fields in GRB 190627A, as a function of the radio emission efficiency $\epsilon_r$ (see Eq. 1 in Section~\ref{sec:model_NSB}). 
We compare this prediction to the upper limit derived from the 2\,min snapshots of GRB 190627A as this timescale is comparable to the dispersive smearing of a prompt, coherent signal across the MWA observing band for a redshift of 1.942 ($\sim90$\,s).
We can see from Figure \ref{GRB190627A_1} that this coherent emission would only be detectable on a 2\,min timescale if it has an efficiency $\gtrsim10^{-2}$. We also plot the deeper limit derived from the dispersed signal simulations (see Section~\ref{sec:disp_sim}) in Figure \ref{GRB190627A_1}. In this case, the emission would be marginally detectable for a typical pulsar efficiency.

\begin{figure}
\centering
\includegraphics[width=.43\textwidth]{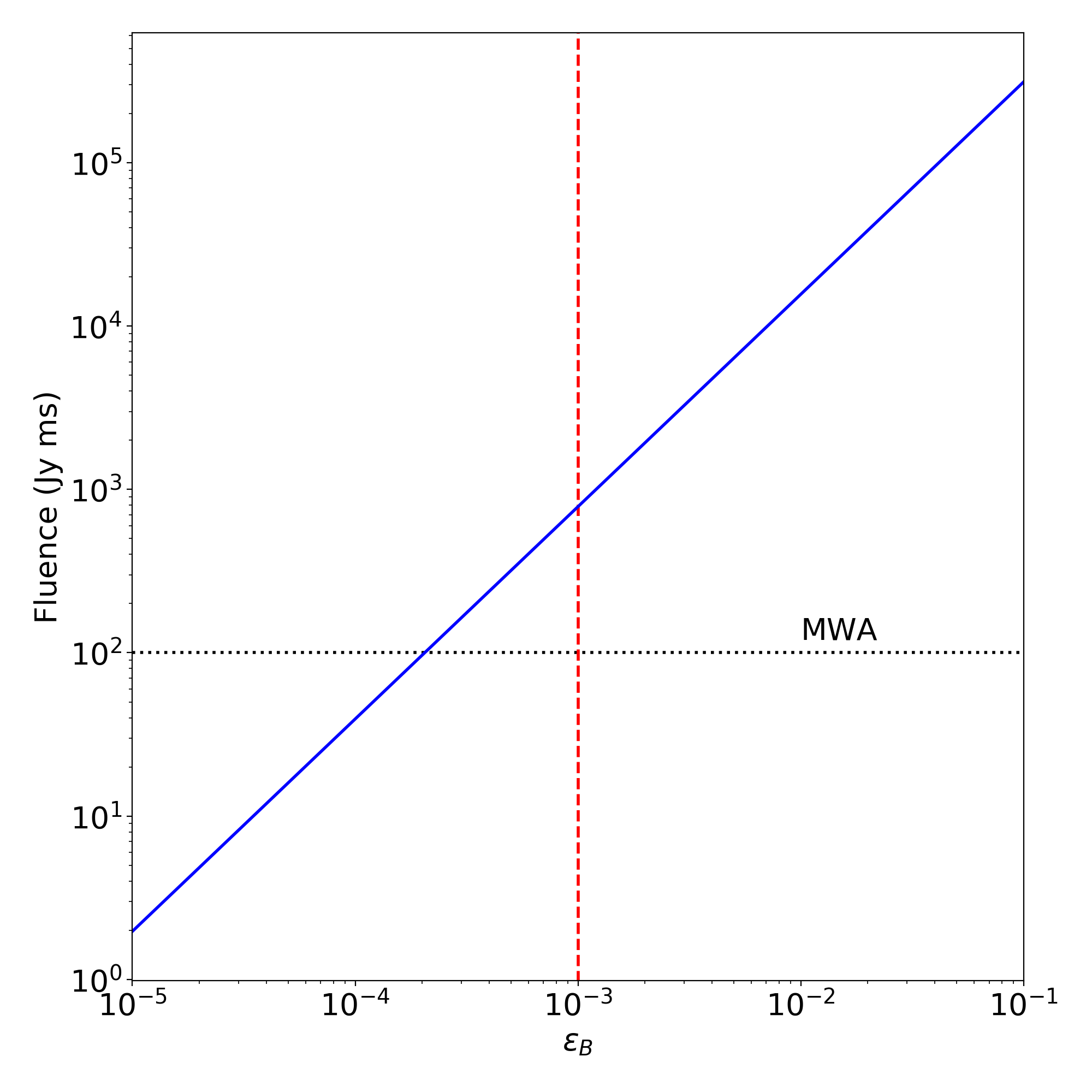}
\caption{The predicted fluence (blue line) of a prompt signal produced by the relativistic jet and ISM interaction (Section~\ref{sec:model_ism}) for GRB 190627A as a function of the fraction of magnetic energy in the GRB jet ($\epsilon_B$). The horizontal dotted line shows the least constraining fluence upper limit derived from our image de-dispersion analysis (Table~\ref{dedispersion_Swift}), and the vertical dashed line shows a typical value for the magnetic energy fraction of $\epsilon_B=10^{-3}$ \citep{Katz97}.} 
\label{GRB190627A_2}
\end{figure}

In Figure \ref{GRB190627A_2} we plot the predicted fluence produced by the relativistic jet and ISM interaction as a function of the fraction of magnetic energy $\epsilon_B$ (see Eq. 2 in Section~\ref{sec:model_ism}).
The gamma-ray fluence of GRB 190627A was measured to be $(9.9\pm2.2)\times10^{-8}\,\text{erg}\,\text{cm}^{-2}$ by \textit{Swift}\textendash BAT in the 15\textendash150\,keV energy band~\citep{Barthelmy19}. 
In this case, the jetted outflow is assumed to be a magnetised wind that is powered by a magnetar central engine \citep{Usov00} so is dependent on the magnetar parameters derived for GRB 190627A (see Table~\ref{magnetar}).
By also adopting a typical value of $10^{49}$\,erg for the kinetic energy from short GRBs~\citep{Fong15}, $\Gamma=1000$ for the Lorentz factor of the relativistic wind~\citep{Ackermann10}, and $n=10^{-2}$ $\text{cm}^{-3}$ for the poorly known electron density of the surrounding medium \citep{Fong15} we were able to derive $\nu_{max}$ \citep[see Section~\ref{sec:model_ism} and][]{Rowlinson19}. 
We then compared the fluence upper limit we derived through our de-dispersion analysis (Table~\ref{dedispersion_Swift}; horizontal dotted line in Figure~\ref{GRB190627A_2}) to the fluence prediction (solid line), constraining the fraction of magnetic energy in the relativistic jet launched by GRB 190627A to $\epsilon_B<2\times10^{-4}$, which is consistent with the limit $\epsilon_B\lesssim10^{-3}$ given by the requirement that the magnetic stress at the shock front should not disrupt the thin colliding shells in internal shock models (vertical dashed line in Figure~\ref{GRB190627A_2}; \citealt{Katz97}).

\begin{figure}
\centering
\includegraphics[width=.45\textwidth]{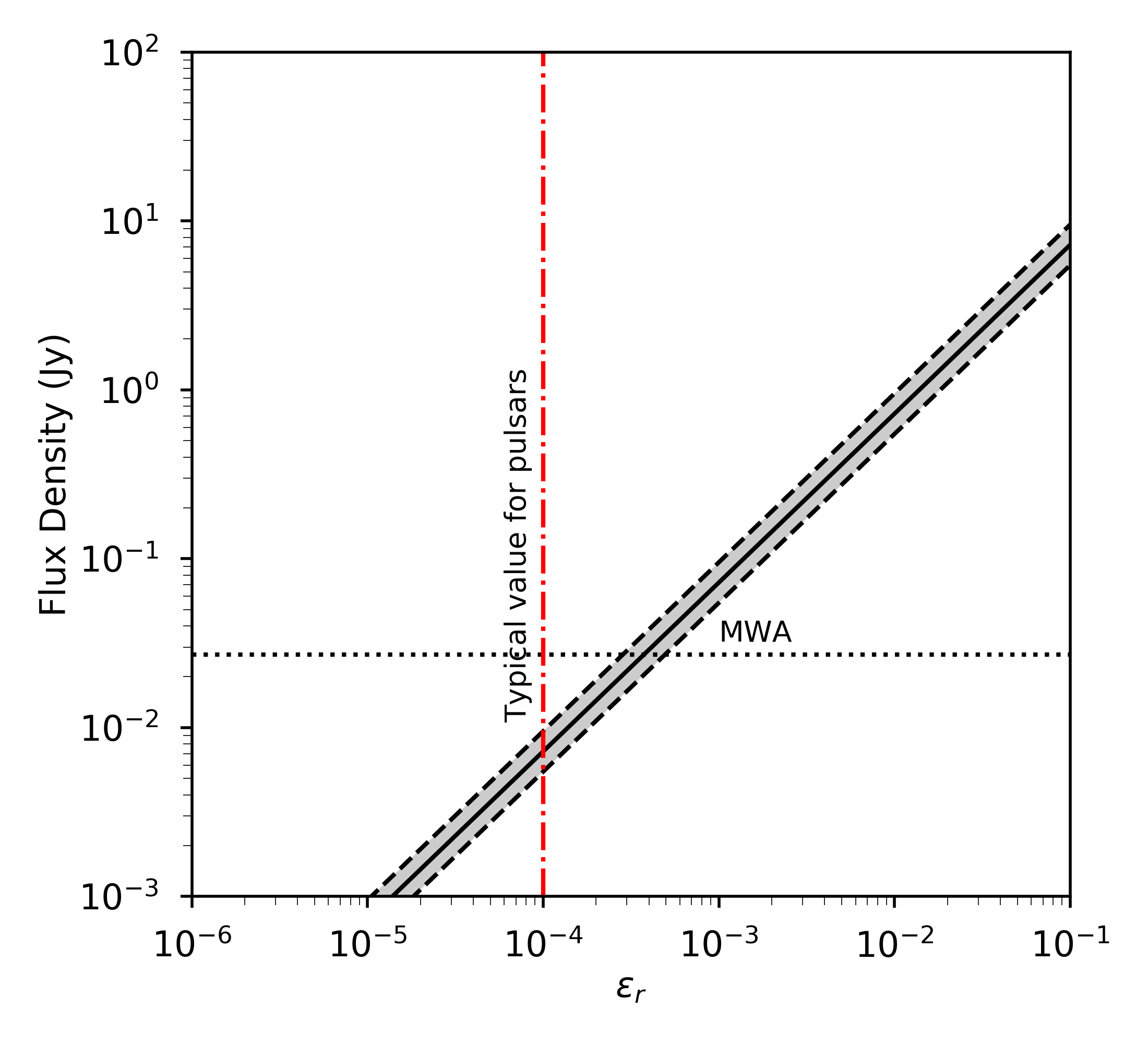}
\caption{The predicted flux density of the persistent emission from a magnetar remnant (Section~\ref{sec:model_pers}) resulting from GRB 190627A as a function of the radio emission efficiency ($\epsilon_r$). The shaded region corresponds to the $1\,\sigma$ uncertainty on the fitted magnetar remnant parameters listed in Table~\ref{magnetar} (see Section~\ref{sec:cen_eng}). The horizontal line shows the flux density upper limit obtained from the 30\,min integration of GRB 190627A (Table~\ref{limit}), and the vertical line shows the typical efficiency observed for known pulsars.} 
\label{GRB190627A_3}
\end{figure}

In Figure \ref{GRB190627A_3} we plot the predicted flux density of persistent pulsar emission from the magnetar remnant as a function of the radio emission efficiency $\epsilon_{r}$ using the magnetar parameters listed in Table~\ref{magnetar} and Eq. 3 in Section~\ref{sec:model_pers}.
We use the upper limit obtained from the 30\,min integration of GRB 190627A to constrain our detection of the persistent emission (Table~\ref{limit}). Figure \ref{GRB190627A_3} shows that for our limit, the magnetar remnant likely has a radio emission efficiency of $\epsilon_r\lesssim10^{-3}$.

From the above comparison between the upper limits and the theoretical predictions, we obtained the following constraints on the model parameters for GRB 190627A: the radio emission efficiency of the nearly merged NSs $\epsilon_r\lesssim10^{-4}$; the fraction of magnetic energy in the GRB jet $\epsilon_B\lesssim2\times10^{-4}$; and the radio emission efficiency of the magnetar remnant $\epsilon_r\lesssim10^{-3}$. 
While the merging NSs are predicted to have a lower radio emission efficiency than typical pulsars, the magnetar remnant may have a higher efficiency. The constraint on the fraction of magnetic energy in the GRB jet is consistent with the range $10^{-6} \lesssim \epsilon_B \lesssim 10^{-3}$ resulting from a systematic study of GRB magnetic fields \citep{Santana14}.
GRB afterglow analyses also show that $\epsilon_B$ downstream of the shock is much larger than $10^{-9}$, which is the typical value in the surrounding medium of short GRBs, assuming a density of $1\,\text{cm}^{-3}$ and a magnetic field of $\sim \mu G$ similar to the Milky Way (e.g., \citealt{Panaitescu02}, \citealt{Yost03} and \citealt{Panaitescu05}). If the magnetic field in the surrounding medium of GRB 190627A has a similar value to that of the Milky Way, the amplification factor of the magnetic energy fraction for this GRB would be $\lesssim2\times10^5$. Several magnetic field amplification mechanisms have been proposed, including the Weibel instability, the cosmic-ray streaming instability, and the dynamo effect (e.g., \citealt{Lucek00}, \citealt{Medvedev05}, \citealt{Milosavljevi06}, \citealt{Inoue11}, \citealt{Mizuno11}).

\subsubsection{GRB 191004A}\label{sec:1910}

GRB 191004A is one of the two events detected by \textit{Swift}.
Given that the first three 2\,min snapshots were corrupted, the delay between the MWA first being on-target with respect to the GRB detection 
(see Section~\ref{sec:swift_trig}) has meant we cannot test coherent emission models that predict the production of prompt radio signals either just prior to, or concurrent with, the merger (see Section~\ref{sec:model_NSB} and \ref{sec:model_ism}) for redshifts $z\lesssim2.7$.
We are therefore only able to constrain the
persistent emission from a magnetar remnant for this GRB, which is the model presented in Section~\ref{sec:model_pers}.
Given we don't know the redshift of this event, we need to explore how the predicted flux density changes between redshifts of $0.1<z<2$.
The magnetar parameters were derived by fitting the magnetar model to the rest frame X-ray light curve (assuming $z=0.7$; see Section~\ref{sec:cen_eng}).
These parameters therefore vary with redshift \citep{Rowlinson19}, with the magnetic field and spin period scaling according to

\begin{equation}\label{eq:B_z}
    B_{15}\propto D^{-1}\,(1+z),
\end{equation}

\begin{equation}\label{eq:P_z}
    P_{-3}\propto D^{-1}\,(1+z)^{1/2}.
\end{equation}

\begin{figure}
\centering
\includegraphics[width=.5\textwidth]{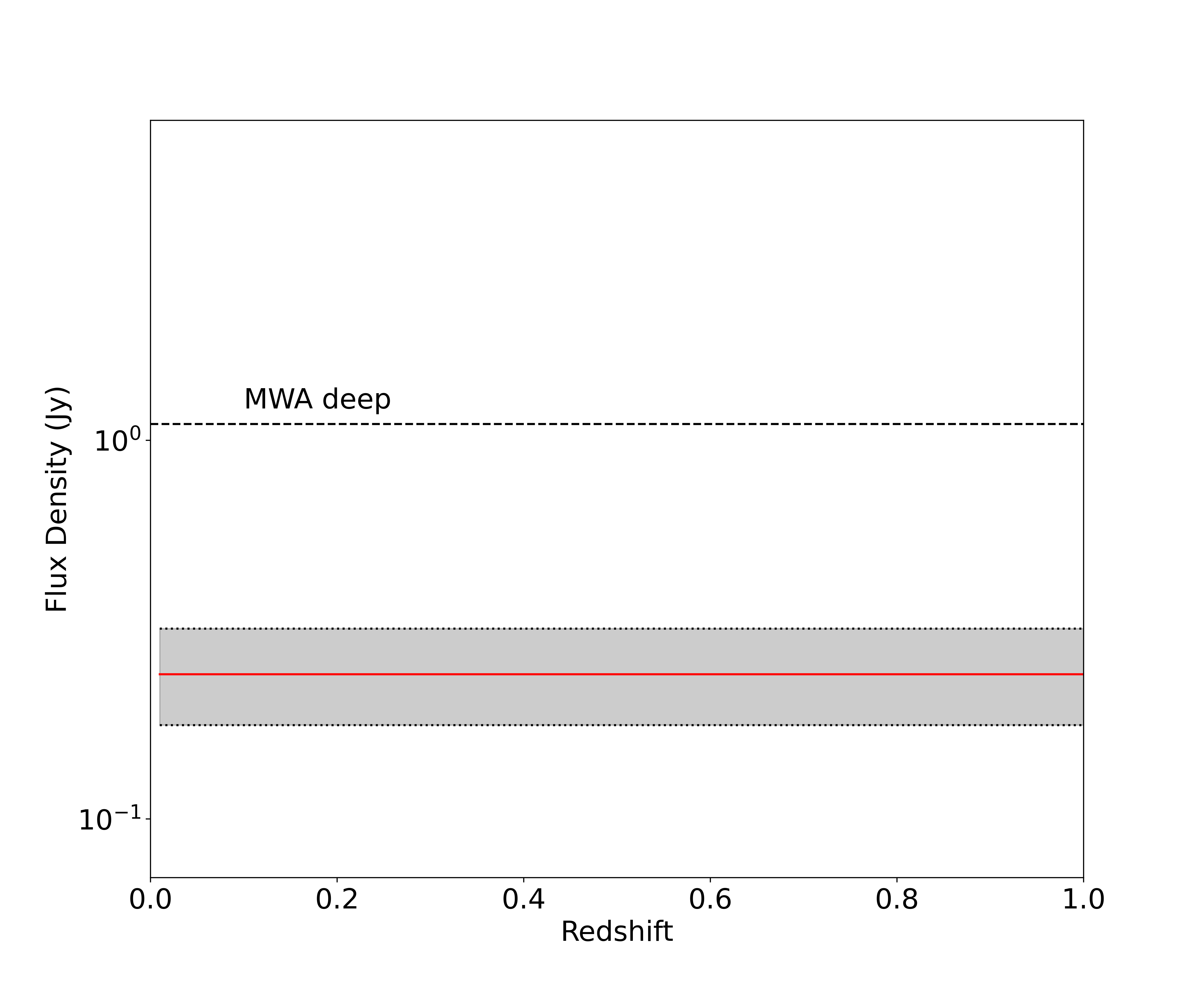}
\caption{The flux density of persistent emission (solid red line) predicted to be produced by a remnant magnetar resulting from GRB 191004A as a function of redshift (Section~\ref{sec:model_pers}). The shaded region corresponds to the $1\,\sigma$ uncertainties on the fitted magnetar parameters (see Section~\ref{sec:cen_eng} and Table \ref{magnetar}). 
The radio emission efficiency is assumed to be $\epsilon_r=10^{-4}$, which is the typical value for pulsars.
The horizontal dashed line indicates the flux density upper limit of 1.104\,Jy derived from the 30\,min integration of GRB 191004A.}
\label{GRB191004A_1}
\end{figure}

Figure \ref{GRB191004A_1} shows the predicted persistent pulsar emission from a magnetar remnant produced by GRB 191004A as a function of redshift.
We used the fitted magnetar parameters derived for GRB 191004A (listed in Table \ref{magnetar}) and the above scaling relations (Eq.~\ref{eq:B_z} and \ref{eq:P_z}) to calculate the predicted flux density. Otherwise, we assumed the same parameters as adopted for GRB 190627A in Section~\ref{sec:1906}. 
Since the observations of GRB 191004A were conducted with the MWA compact configuration, the confusion noise caused the upper limit on the 30\,min integration to be less constraining compared to other GRB observations taken in the extended configuration. Our MWA flux density limit for GRB 191004A is therefore insufficient for constraining this model.

\subsubsection{\textit{Fermi} GRBs}\label{sec:Fermi}

\begin{figure}[t]
\centering
\includegraphics[width=.5\textwidth]{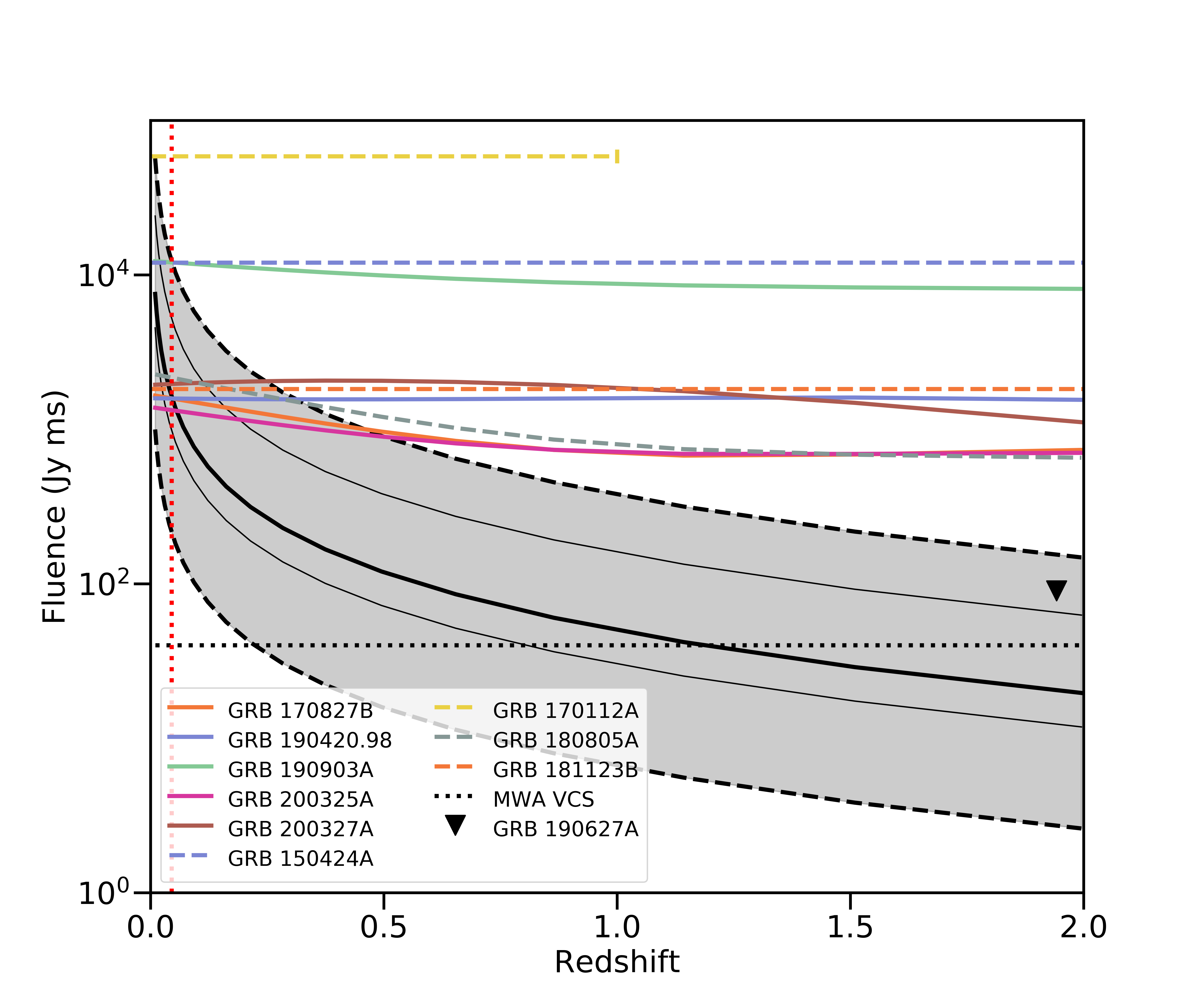}
\caption{The fluence of the prompt radio signal predicted to be produced by the relativistic jet and ISM interaction using the mean values of the magnetic field and spin period of known magnetar remnants (see figure 8 in \citealt{Rowlinson19}) and assuming the median value of the gamma-ray fluences measured for different \textit{Fermi} GRBs in Table \ref{gamma_fluence} (thick black curve). 
The two thin black curves show the radio fluence predictions corresponding to the minimum and maximum gamma-ray fluence measured for the \textit{Fermi} GRBs, and the shaded region corresponds to the $1\,\sigma$ scattering in the distribution of the parameters of typical magnetars. Different from Figure \ref{GRB191004A_1}, there is no rescaling of magnetic field and spin period with redshift. 
The fluence limit for GRB 190627A is plotted as a black triangle. 
The solid coloured curves represent the fluence upper limits as a function of DM (redshift) derived from the de-dispersion image analysis performed on the \textit{Fermi} GRBs.
We also include the fluence upper limits published for individual short GRBs (dashed coloured curves), including GRB 150424A (132\,MHz; \citealt{Kaplan15}), GRB 170112A (56\,MHz; \citealt{AndersonM18}), GRB 180805A (185\,MHz; \citealt{Anderson20}) and GRB 181123B (144\,MHz; \citealt{Rowlinson20}).
The dotted black line indicates a potential fluence limit we could achieve if we instead trigger observations using the MWA Voltage Capture System (VCS; see further details in Section~\ref{sec:disc_jetism}.)
}
\label{GRB190420.98_2}
\end{figure}

The other GRBs in our sample were detected by \textit{Fermi}\textendash GBM so no follow-up X-ray data are available to derive their magnetar remnant parameters. 
Assuming a typical magnetar remnant for all the \textit{Fermi} GRBs (see Section~\ref{sec:cen_eng}), we now compare the radio emission upper limits derived from our MWA observations to theoretically predicted values described by models presented in Section~\ref{sec:model_ism} and \ref{sec:model_pers}.
Note that since none of the \textit{Fermi} GRB fluence upper limits place constraints on the NS magnetic field interactions described in Section~\ref{sec:model_NSB}, we do not consider this model further (see discussion in Section~\ref{sec:disc_NS}).

Figure \ref{GRB190420.98_2} shows the predicted prompt radio emission produced by the GRB jet\textendash ISM interaction as a function of redshift assuming a typical magnetar remnant was formed.
The large uncertainties (the shaded region) come from the scatter in the distribution of the spin period and magnetic field strength of magnetar remnants from previously studied short GRBs \citep{Rowlinson13,Rowlinson19}.
The predicted radio fluence directly scales with the gamma-ray fluence (see Eq.~\ref{eq:jet_ism}), which we list in Table \ref{gamma_fluence} for each GRB as measured by \textit{Fermi}\textendash GBM in the 10\textendash1000\,keV energy band.
For this comparison, we adopt the median gamma-ray fluence value, i.e. $7.8\times10^{-7}\,\text{erg}\,\text{cm}^{-2}$, for predicting the  
prompt emission fluence (thick black curve in Figure \ref{GRB190420.98_2}). 
We also plot the model radio fluence predictions corresponding to the minimum and maximum gamma-ray fluence in our sample (Table~\ref{gamma_fluence}) in Figure \ref{GRB190420.98_2} (thin black curves).
Note that the uncertainty in the predicted emission due to the magnetar parameters 
encompasses the range in predictions caused by different gamma-ray fluence measurements. 
The other model parameters are assumed to be the same as for GRB 190627A.

We overplot all the fluence limits derived from our GRB sample (Tables~\ref{dedispersion_Fermi} and \ref{dedispersion_Swift}) on Figure \ref{GRB190420.98_2} to constrain the GRB jet-ISM interaction model (Section~\ref{sec:model_ism}).
Only those GRBs for which MWA was on-target $\lesssim1$\,min post-burst were included in this Figure as any prompt signals emitted at cosmological distances at the time of burst would have been dispersion delayed by up to $\sim2$\,min at MWA frequencies.
We incorporated the dependence of the fluence upper limits on the DM value and redshift, as was done in \citet{Anderson20}. 
In each case, we plot the maximum fluence limit in the range quoted in Tables~\ref{dedispersion_Fermi} and \ref{dedispersion_Swift}. 
It can be seen that the majority of our fluence limits in this paper are constraining for GRBs at low redshifts $z \lesssim 0.5$ for a subset of magnetar parameters.
In Section~\ref{sec:disc_jetism}, we further explore the implications of our upper limits on this emission model.

\begin{figure}[t]
\centering
\includegraphics[width=0.5\textwidth]{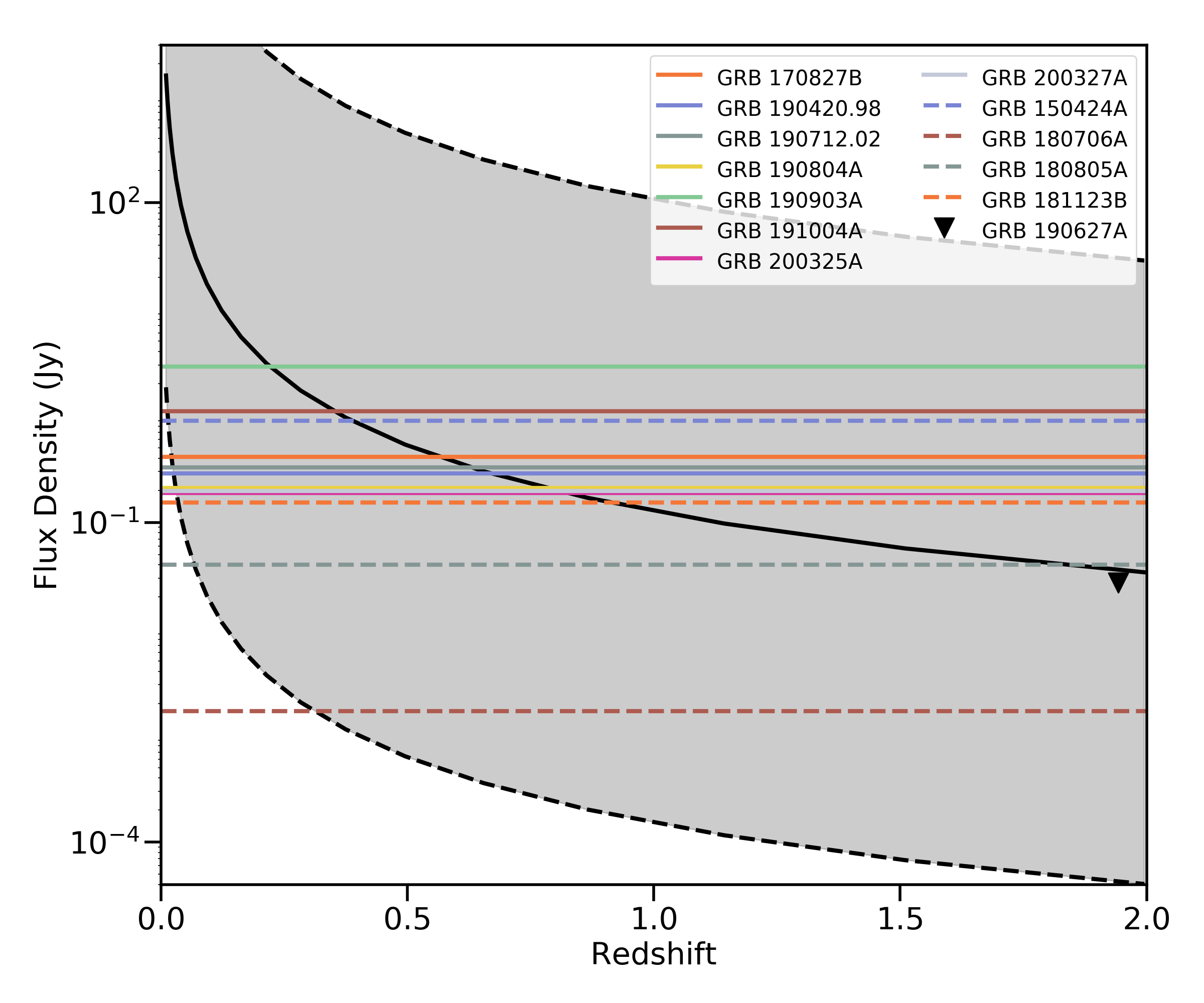}
\caption{Similar to Figure \ref{GRB190420.98_2}, here we plot the predicted flux density for the persistent radio emission from the dipole radiation of a magnetar remnant (see Section~\ref{sec:model_pers}). The solid black curve represents the predicted emission from a typical magnetar with the shaded region corresponding to the $1\sigma$ scatter in the distribution of magnetar parameters. The solid coloured curves represent the flux density upper limits derived from the 30\,min integration of our sample of short GRBs. We also plot the flux density upper limits from observations of other individual GRBs (dashed coloured curves), including GRB 150424A \citep{Kaplan15}, GRB 180706A \citep[a long GRB;][]{Rowlinson19b}, 
GRB 180805A \citep{Anderson20} and GRB 181123B \citep{Rowlinson20}.}
\label{GRB190420.98_3}
\end{figure}

Figure \ref{GRB190420.98_3} shows the predicted persistent dipole radiation from a typical magnetar as a function of redshift (thick black curve).
Again, the shaded region illustrates the uncertainty in the prediction due to the scatter in the distribution of known magnetar parameters \citep{Rowlinson19}, which spans six orders of magnitude.  
We plot the flux density upper limits derived from the 30\,min integrations of the GRBs as listed in Table~\ref{limit}.
For a typical magnetar remnant, the majority of our MWA observations could have detected persistent dipole radiation up to a redshift of $z\sim0.6$. 
We explore this further in Section~\ref{sec:disc_pers}.

\begin{table}
\begin{threeparttable}
\centering
\resizebox{.8\columnwidth}{!}{\hspace{-0.0cm}\begin{tabular}{c c}
\hline
GRB & $\gamma$-ray fluence ($10^{-7}\,\text{erg}\,\text{cm}^{-2}$) \\
\hline
170827B &  $4.7\pm0.1$\tnote{a}   \\
190420.98  & $6.5\pm0.3$\tnote{a} \\
190903A & $7.8\pm0.6$\tnote{b} \\
200325A & $24.9\pm1.2$\tnote{c} \\
200327A & $15.3\pm0.6$\tnote{d} \\
\hline
\end{tabular}}
\caption{The gamma-ray fluences (10\textendash1000\,keV) measured by \textit{Fermi}\textendash GBM for those \textit{Fermi} 
events for which we derived radio fluence limits.
References include: 
\\
a: Fermi\textendash GBM burst catalog at HEASARC: \url{https://heasarc.gsfc.nasa.gov/W3Browse/fermi/fermigbrst.html}; \\
b: \citet{Mailyan19};\\
c: \citet{Veres20};\\
d: \citet{Veres20b}.
}
\label{gamma_fluence}
\end{threeparttable}
\end{table}

\section{discussion}\label{sec:discussion}

We have performed a search for coherent radio emission associated with BNS mergers on the biggest sample of short GRBs at low frequencies using the MWA rapid-response system, obtaining radio fluence and flux density upper limits on this predicted emission (see Tables~\ref{limit} \ref{dedispersion_Swift}, and \ref{dedispersion_Fermi}). The de-dispersion analysis in the image space (see Section~\ref{sec:disp_search}) accommodates the dispersive smearing of the prompt signals across the MWA observing band for a range of short GRB redshifts, and thus is more sensitive to these signals than just imaging on different timescales (5\,s, 30\,s, 2\,min).
Thanks to the rapid response time of the MWA (for a comparison of response times of different low frequency facilities see figure 10 in \citealt{Anderson20}), we are able to use these fluence upper limits to constrain the prompt radio emission models in the early stages of BNS mergers. With the flux density upper limits obtained from the 30\,min observations, we are able to constrain the persistent emission model (see Section~\ref{sec:models} for a summary of these models).

\subsection{Constraints on the tested coherent emission models}\label{sec:disc_mod}

In this section, we consider the implications of the constraints the nine short GRBs in our sample place on the emission models described in Section~\ref{sec:models}. 
The upper limits we obtained for the \textit{Swift} GRB 190627A demonstrates the best sensitivity MWA can achieve with triggered observations using the standard correlator mode, as can be seen from Figure \ref{GRB190420.98_2} and \ref{GRB190420.98_3} (black triangle).
As part of this investigation, 
we also compare our results to the low frequency fluence and flux density upper limits obtained from investigations of individual short GRBs.
In Figure \ref{GRB190420.98_2}, we include the fluence upper limits on the prompt radio emission from individual short GRBs, including $1.2\times10^4$\,Jy\,ms for GRB 150424A observed by the MWA \citep{Kaplan15}, $5.85\times10^4$\,Jy\,ms for GRB 170112A observed by OVRO-LWA \citep{Anderson18}, 570\textendash1750\,Jy\,ms for GRB 180805A observed by the MWA \citep{Anderson20} and 1824\,Jy\,ms for GRB 181123B observed by LOFAR \citep{Rowlinson20}.
In Figure \ref{GRB190420.98_3}, we chose to include the flux density upper limits on the persistent radio emission from individual GRBs, including 0.9\,Jy on a 30\,min timescale for GRB 150424A observed by MWA \citep{Kaplan15}, 
1.7\,mJy on a 2\,h timescale for GRB 180706A observed by LOFAR \citep[a long GRB;][]{Rowlinson19b}, 
40.2\,mJy on a 30\,min timescale for GRB 180805A observed by MWA \citep{Anderson20}, and 153\,mJy on a 2\,h timescale for GRB 181123B observed by LOFAR \citep{Rowlinson20}.
Even compared to previous surveys and triggered observations, our triggered MWA observations of GRB 190627A obtained the most stringent limits on both prompt and early-time persistent coherent emission to date from a short GRB within a few hours post-burst at low frequencies.

\subsubsection{Interaction of NS magnetic fields}\label{sec:disc_NS}

In Section~\ref{sec:model_NSB}, we described 
that one of the earliest prompt, coherent signals predicted from a short GRB may be produced by the alignment of the NS magnetic fields prior to the merger.
Here we compare the model prediction to the fluence upper limits derived from our short GRB sample. From Table \ref{dedispersion_Fermi} and \ref{dedispersion_Swift}, we can see the fluence upper limits range from $80$ to $12,110$\,Jy\,ms.
According to Eq.~\ref{eq:ns}, assuming typical NS properties, we predict the fluence to be $\lesssim10$\,Jy\,ms at a reasonable redshift. Therefore, none of our observations are sensitive enough to detect this predicted emission.

\subsubsection{Relativistic jet and ISM interaction}\label{sec:disc_jetism}

As outlined in Section~\ref{sec:model_ism}, we expect the collision of a relativistic jet with the ISM to produce prompt, coherent radio emission. This model is dependent on the properties, i.e. the spin period and the magnetic field strength of the newly formed magnetar remnant. Considering a typical magnetar as proposed by \citet{Rowlinson19}, we compared the fluence upper limits obtained from our sample of short GRBs (solid curves) and those from previous triggered observations (dashed curves) to the model prediction in Figure \ref{GRB190420.98_2}. Note that the de-dispersion analysis performed for GRB 170112A by \citet{Anderson18} covers a limited range of DM values so the corresponding dashed lines spans a limited range of redshift.

Among all these searches at low frequencies, our observation of GRB 190627A is the most sensitive, and even at a redshift of $z=1.942$, we were sensitive enough to detect the predicted prompt signal from the jet-ISM interaction for some range of typical magnetar remnant parameters. The majority of fluence limits obtained from our sample and other triggered observations of short GRBs are sensitive enough to search for prompt signals up to a redshift of $z\sim0.5$ for a subset of typical magnetar remnant parameters. \textbf{Considering the magnetic energy fraction was assumed to be $\epsilon_\text{B}=10^{-3}$ for the predicted emission in Figure~\ref{GRB190420.98_2}, our non-detection might suggest a constraint of $\epsilon_\text{B}\lesssim10^{-3}$, comparable to the theoretical limit on $\epsilon_\text{B}$ for GRBs with narrow subpulses in internal shock models \citep{Usov00}. Compared to previous constraints on $\epsilon_\text{B}$ using radio observations of short GRBs, our limit is comparable to the $\epsilon_\text{B}\lesssim[10^{-4}\text{--}10^{-2}]$ (depending on pulse widths) derived for GRB 150424A \citep{Rowlinson19}, but less constraining than the $\epsilon_\text{B}\lesssim[3\times10^{-5}\text{--}2\times10^{-4}]$ derived for GRB 181123B \citep{Rowlinson21}.}
While it is unlikely that the majority of short GRBs are at $z>0.5$ \citep{Gompertz20},
overall, our non-detection of prompt radio emission from such a big sample of short GRBs is consistent with model predictions
when considering the full parameter space covered by the potential diveristy in both magnetar parameters and gamma-ray fluences and (see shaded region in Figure~\ref{GRB190420.98_2}).

In Figure~\ref{GRB190420.98_2} we also compared the expected sensitivity of the MWA Voltage Capture System \citep[VCS;][]{Tremblay15} to the model prediction and short GRB fluence limits. 
The VCS mode has a temporal and spectral resolution of $100\,\mu$s and 10\,kHz, respectively, making it specifically sensitive to narrow ($\sim10$\,ms) pulsed and therefore dispersed signals. 
Rapid-response MWA observations of short GRBs using the VCS would therefore be sensitive enough to detect prompt emission from a large subset of magnetar remnant parameters over a large redshift range (see further discussions in Section~\ref{sec:disc_future}).

\subsubsection{Persistent pulsar emission}\label{sec:disc_pers}

If a magnetar is formed via the BNS merger, it is predicted to emit in the same way as pulsars (see Section~\ref{sec:model_pers}). Again, this emission is dependent on the magnetar remnant properties. In Figure \ref{GRB190420.98_3}, we compared the flux density upper limits obtained from our sample of short GRBs (solid curves) and those from previous works (dashed curves) to the predicted flux densities assuming the formation of a typical magnetar remnant.

Our observation of \textit{Swift} GRB 190627A provides the second most sensitive limit on the persistent emission from a magnetar remnant. With this sensitivity, we are able to detect the predicted persistent emission from a typical magnetar up to a redshift of $\sim2$. Note that when plotting the limits for this GRB and the other \textit{Swift} GRB 191004A, we did not assume typical magnetar parameters but used those derived from their X-ray light curves in Section~\ref{sec:cen_eng}.

One can see that \citet{Rowlinson19b} performed the most sensitive search for persistent emission from the long GRB 180706A using LOFAR. A LOFAR observation with over a 2\,h integration can reach mJy sensitivities thanks to its large number of antennas and long interferometric baselines~\citep{Haarlem13}. This sensitivity is sufficient to detect the predicted persistent emission over a broad range of redshifts and a large magnetar parameter space. However, as GRB 180706A is a long GRB, the dense surrounding medium may prevent the transmission of low-frequency radio signals \citep{Zhang14}.

For the 11 short GRBs in Figure \ref{GRB190420.98_3} (excluding the long GRB 180706A), it is somewhat surprising to see no detection 
of any persistent emission given that all the flux density upper limits are comparable to the model predictions for a large subset of typical magnetar parameters over a wide range of redshifts, at least up to $z\sim0.6$.
The non-detections from such a big sample provide new implications for this persistent emission model. It is possible that these short GRBs did not produce any persistent radio emission, given that the probability of a BNS merger forming a magnetar remnant is between 5 and 97\%, depending on the NS equation of state \citep{Ravi14}.
However, if all the 11 short GRBs in Figure \ref{GRB190420.98_3} actually formed typical magnetars, then either their pulsar beams were pointed away from the Earth and/or they were located at redshifts of $z\gtrsim0.6$. In the population of short GRBs with known redshifts (see table 1 in \citealt{Gompertz20}), only one third have redshifts of $z>0.6$. Therefore, it is unlikely that all the 11 short GRBs are located at high redshifts, which suggests that either some of the short GRBs did not form magnetars or their radio emission is not aligned with our line of sight. There is still another possibility that the short GRBs formed magnetars that deviate from the assumed typical parameters, given the typical parameters are drawn from the distribution of a small population of fitted magnetars (see figure 8 in \citealt{Rowlinson19}). 
For an ideal case, we might expect two thirds of the 11 short GRBs to be located at $z<0.6$, half of which formed magnetars with persistent emission brighter than what we are considering a `typical' magnetar remnant, and all of which have their emission beams pointed towards us. If that were the case, then we might have predicted to detect dipole radiation from the magnetar remnants of approximately four short GRBs.

We therefore conclude that the non-detection of persistent emission from these 11 short GRBs could be related to the simplicity of the model, our assumption that the majority of short GRBs form a magnetar remnant is invalid, and/or our assumption that the magnetar's radio beam is aligned with the rotational axis and remains pointed towards Earth following the merger is invalid. It is also possible that some GRBs in the sample, such as GRB 190627A and 191004A (see Section~\ref{sec:swift_trig}) may actually be long GRBs, which means their higher density environments could prevent the radio emission from escaping.

\subsection{Future improvements}\label{sec:disc_future}

The biggest improvement to this experiment will be to perform triggered observations on short GRBs using the VCS as its high temporal resolution ($100\,\mu$s) will increase our sensitivity to FRB-like signals ($\sim10$ms width), which would otherwise be diluted by the 0.5\,s coarse sampling of the standard correlator.
Given that the data rate of VCS observations is high ($\sim$28\,TB/hr), we would not be able to continuously observe one GRB for more than $\sim$100\,min \citep{Tremblay15},
making it difficult for us to detect prompt signals predicted to be produced at late times (e.g. by the collapse of an unstable magnetar remnant into a BH, which may not occur for up to 2\,hr post-merger; \citealt{Zhang14}).
However, a 15\,min integration with the VCS will allow us to search for prompt, coherent emission predicted to occur either just prior, during, or shortly following the merger (see the models described in Sections~\ref{sec:model_NSB} and \ref{sec:model_ism}) 
for a wide range of DMs (150-2500\,pc\,cm$^{-3}$) to much deeper sensitivities while still creating a manageable data volume. 
This is demonstrated in Figure~\ref{GRB190420.98_2} where we include the estimated VCS sensitivity limit calculated by \citet{Rowlinson19}.
While the standard correlator (0.5\,s temporal resolution) observations can probe the prompt radio emission produced by the jet-ISM interaction up to a redshift of $z\sim0.5$ for a typical magnetar remnant, the more sensitive VCS observation can probe this emission up to redshifts of $z>1$ for a much wider range of magnetar parameters.
Therefore, VCS triggered observations of short GRBs is the most promising method for searching for associated prompt, coherent emission.

The poorly known magnetar remnant properties result in uncertainties in the predicted persistent emission that spans six orders of magnitude, as shown in Figure \ref{GRB190420.98_3}. As the population of short GRBs detected by \textit{Swift} keeps growing, we expect more of their X-ray light curves to be fitted by the magnetar model in the future. This will help narrow down the model parameters and better constrain the detectability of the emission. Additionally, more redshift measurements on short GRBs would allow us to constrain the key parameters of the emission models, as what we have done for GRB 190627A in Figure \ref{GRB190627A_1}, \ref{GRB190627A_2} and \ref{GRB190627A_3} (see also Section~\ref{sec:1906}).

\subsubsection{Implications for GW follow-up}

\textbf{As BNS mergers are plausible GW emitters, we plan to search for the predicted coherent radio emission using MWA triggered observations of GW events. Compared to short GRBs detected by \textit{Swift} and/or \textit{Fermi}, GW events detected by aLIGO/Virgo are much closer ($\lesssim190$\,Mpc in observing run O4 likely commencing mid-2022; \citealt{Abbott20}), which means their associated radio emission would be much brighter. In Figure~\ref{GRB190420.98_2} we plot the expected maximum redshift of GWs with a vertical dotted red line. While 
observations 
with the standard MWA correlator 
have an almost 50\% chance of detecting the predicted emission from GWs based on
the largely uncertain magnetar parameters, the VCS mode would either make a detection or rule out the jet-ISM interaction model undoubtedly.}

\textbf{The proximity of GW events also means less dispersion delay, which 
means we may require a faster response time than what is currently possible with MWA.
As discussed in \citet{Anderson20}, the MWA may only be on-target fast enough for frequencies $\lesssim130$\,MHz, the lower end of the MWA observing band. However, a strategy proposed by \citet{James19} for triggering MWA on `negative latency' aLIGO/Virgo alerts of BNS mergers (generated by detections of GWs during the inspiral phase before the merger) could help alleviate the requirement on response times, allowing the MWA enough time to capture any associated prompt, coherent radio emission. For further details on MWA follow-up strategies of GWs, see \citet{Kaplan16}, \citet{James19}, and \citet{Anderson20}.}

\subsubsection{Prospects for SKA-Low}

The low-frequency component of the Square Kilometre Array (SKA-Low; \citealt{Dewdney09}) is a radio telescope covering the frequency range 50\textendash350\,MHz and of an unprecedented sensitivity, resulting from the enormous number of dual-polarisation antennas \citep{Sokolowski21}. Given the model predictions shown in Section~\ref{sec:Fermi}, the far greater sensitivity of the SKA-Low will rigorously test these models. Therefore, it is plausible to implement our rapid-response system on the SKA-Low to search for coherent radio emission associated with GRBs. A future plan to enhance the SKA-Low system with the triggering capability to react to external transient alerts has been proposed by \citet{Sokolowski21}.

\section{conclusions}\label{sec:conc}

In this paper, we have searched for coherent radio emission from nine short GRBs in the frequency range of 170 and 200\,MHz using MWA rapid-response observations. 
These observations began within 30\,s to 10\,m postburst (7 of the 9 events within $\sim1$\,min post-burst), integrating for a maximum of 30\,min.
We have inspected the images of these nine GRBs that were made on timescales of 30\,min, 2\,min, 30\,s, and 5\,s but found no associated transient or variable emission within the GRB positional error regions, quoting flux density upper limits in Table~\ref{limit}.
We have also performed a de-dispersion search for transients using 0.5\,s\,/\,1.28\,MHz sub-band images, but found no FRB-like signals associated with the GRBs, quoting a range of fluence upper limits in Tables~\ref{dedispersion_Fermi} and \ref{dedispersion_Swift}.
Our fluence and flux density limits on transient emission associated with our nine short GRBs were compared to model predictions of coherent prompt and persistent emission applicable to BNS mergers. As a result of this work, we come to the following main conclusions:
\begin{enumerate}
\item The MWA rapid-response observations of \textit{Swift} GRB 190627A provides the most constraining upper limits on  coherent emission associated with short GRBs in our sample.
By fitting the stable magnetar model to its X-ray light curve (see Figure~\ref{X-ray}), we were able to acquire the magnetar remnant parameters (see Table~\ref{magnetar}). 
As GRB 190627A is the only event in our sample with a known redshift, we were able to constrain key parameters of the emission models described in Section~\ref{sec:models}, including
the radio emission efficiency of the nearly merged NSs ($\epsilon_r\lesssim10^{-4}$), the magnetic energy fraction in the GRB jet ($\epsilon_B\lesssim2\times10^{-4}$), and the radio emission efficiency of the magnetar remnant ($\epsilon_r\lesssim10^{-3}$, see Figures~\ref{GRB190627A_1}, \ref{GRB190627A_2} and \ref{GRB190627A_3}).

\item The fluence upper limits derived from the de-dispersion analysis of our MWA standard observations were sensitive enough to detect the predicted prompt radio emission produced by the GRB jet-ISM interaction (see Section~\ref{sec:model_ism}) up to a redshift of $z\sim0.5$ for a subset of typical magnetar parameters (see Figure~\ref{GRB190420.98_2}). While it is unlikely that the majority of short GRBs are at $z>0.5$ \citep{Gompertz20}, our sensitivity is not very constraining when considering the full regime of predictions (shaded region in Figure~\ref{GRB190420.98_2}) so non-detections may not be unexpected.
However, future MWA rapid-response observations using the VCS will be more sensitive to narrow, FRB-like signals, enabling us to probe this emission up to a much higher redshift for a larger range of potential magnetar parameters.

\item The flux density upper limits derived from the 30\,min observations of a sample of short GRBs suggest that we should be able to detect the persistent radio emission produced by a typical magnetar remnant (see Section~\ref{sec:model_pers}) up to a redshift of $z\sim0.6$ (see Figure~\ref{GRB190420.98_3}). 
Given that $\sim2/3$ of short GRBs have redshifts of $z \leq 0.6$ \citep{Gompertz20},  
our non-detection of persistent radio emission from this sample of short GRBs implies one or more of the following: that some GRBs are not genuinely short, no magnetar remnant was formed, the magnetar remnants do not have typical properties defined by \citet{Rowlinson19}, their radiation beams are pointing away from us, or the model is too simplistic.

\end{enumerate}

The MWA, with its large field of view and rapid-response triggering mode, is currently one of the most competitive radio telescopes for performing rapid follow-up observations of short GRBs in search of coherent prompt or persistent emission associated with BNS mergers. In the future, we plan to employ the VCS mode to trigger on GRBs, which is more sensitive to prompt, FRB-like signals.
Furthermore, our experiment with the MWA demonstrates the importance of incorporating rapid response capabilities into other low frequency facilities to enable programs that search for prompt radio emission associated with transients, particularly the SKA-Low, which will have superior instantaneous sensitivity on shorter timescales.

\begin{acknowledgements}
GEA is the recipient of an Australian Research Council Discovery Early Career Researcher Award (project number DE180100346), and JCAM-J is the recipient of an Australian Research Council Future Fellowship (project number FT140101082) funded by the Australian Government. DK was supported by NSF grant AST-1816492.
TM acknowledges the support of the Australian Research Council through grants FT150100099 and DP190100561.

This scientific work makes use of the Murchison Radio-astronomy Observatory, operated by CSIRO. We acknowledge the Wajarri Yamatji people as the traditional owners of the Observatory site. Support for the operation of the MWA is provided by the Australian Government (NCRIS), under a contract to Curtin University administered by Astronomy Australia Limited.

Parts of this research were conducted by the Australian Research Council Centre of Excellence for Gravitational Wave Discovery (OzGrav), project number CE170100004.
This work made use of data supplied by the UK {\it Swift} Science Data Centre at the University of Leicester and the {\it Swift} satellite. {\it Swift}, launched in November 2004, is a NASA mission in partnership with the Italian Space Agency and the UK Space Agency. {\it Swift} is managed by NASA Goddard. Penn State University controls science and flight operations from the Mission Operations Center in University Park, Pennsylvania. Los Alamos National Laboratory provides gamma-ray imaging analysis. 

This work was supported by resources provided by the Pawsey Supercomputing Centre with funding from the Australian Government and the Government of Western Australia.

The following software and packages were used to support this work: {\sc mwa\_trigger} \citep{Hancock19}\footnote{https://github.com/MWATelescope/mwa\_trigger/}, 
GLEAM-X pipeline\footnote{https://github.com/nhurleywalker/GLEAM-X-pipeline}, 
MWA-fast-image-transients\footnote{https://github.com/PaulHancock/MWA-fast-image-transients},
{\sc flux\_warp}\footnote{https://gitlab.com/Sunmish/flux\_warp/},
{\sc comet} \citep{Swinbank14}, 
{\sc voevent-parse} \citep{Staley16}, 
\Robbie{} \citep{Hancock19b}\footnote{https://github.com/PaulHancock/Robbie}, 
AegeanTools \citep{Hancock18}\footnote{https://github.com/PaulHancock/Aegean}, 
{\sc fits\_warp} \citep{Hurley18}, 
TOPCAT/stilts \citep{taylor05}, 
{\sc Astropy} \citep{TheAstropyCollaboration2013,TheAstropyCollaboration2018}, 
{\sc numpy} \citep{vanderWalt_numpy_2011}, 
{\sc scipy} \citep{Jones_scipy_2001}, 
{\sc matplotlib} \citep{hunter07},   docker\footnote{\href{https://www.docker.com/}{https://www.docker.com/}}, 
singularity \citep{kurtzer_singularity_2017}, Ned Wright's Cosmology Calculator\footnote{http:/www.astro.ucla.edu/$\sim$wright/CosmoCalc.html}~\citep{Wright06}.
This research has made use of NASA's Astrophysics Data System. 
This research has made use of SAOImage DS9, developed by Smithsonian Astrophysical Observatory.
This research has made use of the VizieR catalogue access tool \citep{ochsenbein00} and the SIMBAD database \citep{wenger00}, operated at CDS, Strasbourg, France.
\end{acknowledgements}

\bibliographystyle{pasa-mnras}
\bibliography{bib}

\begin{thebibliography}{}
\makeatletter
\relax
\def\mn@urlcharsother{\let\do\@makeother \do\$\do\&\do\#\do\^\do\_\do\%\do\~}
\definecolor{darkblue}{rgb}{0,0,0.597656}
\def\mndoi{\begingroup\mn@urlcharsother \@ifnextchar [ {\mndoi@} {\mndoi@[]}}
\def\mndoi@[#1]#2{\def\@tempa{#1}\ifx\@tempa\@empty \href
  {http://dx.doi.org/#2} {\textcolor{darkblue}{doi:#2}}\else \href
  {http://dx.doi.org/#2} {\textcolor{darkblue}{#1}}\fi \endgroup}
\def\mn@eprint#1#2{\mn@eprint@#1:#2::\@nil}
\def\mn@eprint@arXiv#1{\href {http://arxiv.org/abs/#1} {{\tt arXiv:#1}}}
\def\mn@eprint@dblp#1{\href {http://dblp.uni-trier.de/rec/bibtex/#1.xml}
  {dblp:#1}}
\def\mn@eprint@#1:#2:#3:#4\@nil{\def\@tempa {#1}\def\@tempb {#2}\def\@tempc
  {#3}\ifx \@tempc \@empty \let \@tempc \@tempb \let \@tempb \@tempa \fi \ifx
  \@tempb \@empty \def\@tempb {arXiv}\fi \@ifundefined
  {mn@eprint@\@tempb}{\@tempb:\@tempc}{\expandafter \expandafter \csname
  mn@eprint@\@tempb\endcsname \expandafter{\@tempc}}}

\bibitem[\protect\citeauthoryear{{Abbott} et~al.,}{{Abbott}
  et~al.}{2017}]{Abbott17}
{Abbott} B.~P.,  et~al., 2017, \mndoi [\prl] {10.1103/PhysRevLett.119.161101},
  \href {https://ui.adsabs.harvard.edu/abs/2017PhRvL.119p1101A} {119, 161101}

\bibitem[\protect\citeauthoryear{{Abbott} et~al.,}{{Abbott}
  et~al.}{2020}]{Abbott20}
{Abbott} B.~P.,  et~al., 2020, \mndoi [Living Reviews in Relativity]
  {10.1007/s41114-020-00026-9}, \href
  {https://ui.adsabs.harvard.edu/abs/2020LRR....23....3A} {23, 3}

\bibitem[\protect\citeauthoryear{{Ackermann} et~al.,}{{Ackermann}
  et~al.}{2010}]{Ackermann10}
{Ackermann} M.,  et~al., 2010, \mndoi [\apj] {10.1088/0004-637X/716/2/1178},
  \href {https://ui.adsabs.harvard.edu/abs/2010ApJ...716.1178A} {716, 1178}

\bibitem[\protect\citeauthoryear{{Ahumada} et~al.,}{{Ahumada}
  et~al.}{2021}]{Ahumada21}
{Ahumada} T.,  et~al., 2021, arXiv e-prints, \href
  {https://ui.adsabs.harvard.edu/abs/2021arXiv210505067A} {p. arXiv:2105.05067}

\bibitem[\protect\citeauthoryear{{Anderson} et~al.,}{{Anderson}
  et~al.}{2018a}]{Anderson18}
{Anderson} G.~E.,  et~al., 2018a, \mndoi [\mnras] {10.1093/mnras/stx2407},
  \href {https://ui.adsabs.harvard.edu/abs/2018MNRAS.473.1512A} {473, 1512}

\bibitem[\protect\citeauthoryear{{Anderson} et~al.,}{{Anderson}
  et~al.}{2018b}]{AndersonM18}
{Anderson} M.~M.,  et~al., 2018b, \mndoi [\apj] {10.3847/1538-4357/aad2d7},
  \href {https://ui.adsabs.harvard.edu/abs/2018ApJ...864...22A} {864, 22}

\bibitem[\protect\citeauthoryear{{Anderson} et~al.,}{{Anderson}
  et~al.}{2021a}]{Anderson20}
{Anderson} G.~E.,  et~al., 2021a, \mndoi [\pasa] {10.1017/pasa.2021.15}, \href
  {https://ui.adsabs.harvard.edu/abs/2021PASA...38...26A} {38, e026}

\bibitem[\protect\citeauthoryear{{Anderson} et~al.,}{{Anderson}
  et~al.}{2021b}]{Anderson21b}
{Anderson} G.~E.,  et~al., 2021b, \mndoi [\mnras] {10.1093/mnras/stab727},
  \href {https://ui.adsabs.harvard.edu/abs/2021MNRAS.503.4372A} {503, 4372}

\bibitem[\protect\citeauthoryear{{Astropy Collaboration} et~al.,}{{Astropy
  Collaboration} et~al.}{2018}]{TheAstropyCollaboration2018}
{Astropy Collaboration} et~al., 2018, \mndoi [The Astronomical Journal]
  {10.3847/1538-3881/aabc4f}, 156, 123

\bibitem[\protect\citeauthoryear{{Baird} et~al.,}{{Baird}
  et~al.}{1975}]{Baird75}
{Baird} G.~A.,  et~al., 1975, \mndoi [\apjl] {10.1086/181732}, \href
  {https://ui.adsabs.harvard.edu/abs/1975ApJ...196L..11B} {196, L11}

\bibitem[\protect\citeauthoryear{{Bannister}, {Murphy}, {Gaensler}  \&
  {Reynolds}}{{Bannister} et~al.}{2012}]{Bannister12}
{Bannister} K.~W.,  {Murphy} T.,  {Gaensler} B.~M.,   {Reynolds} J.~E.,  2012,
  \mndoi [\apj] {10.1088/0004-637X/757/1/38}, \href
  {https://ui.adsabs.harvard.edu/abs/2012ApJ...757...38B} {757, 38}

\bibitem[\protect\citeauthoryear{{Barthelmy} et~al.,}{{Barthelmy}
  et~al.}{2005}]{Barthelmy05}
{Barthelmy} S.~D.,  et~al., 2005, \mndoi [\ssr] {10.1007/s11214-005-5096-3},
  \href {https://ui.adsabs.harvard.edu/abs/2005SSRv..120..143B} {120, 143}

\bibitem[\protect\citeauthoryear{{Barthelmy} et~al.,}{{Barthelmy}
  et~al.}{2019}]{Barthelmy19}
{Barthelmy} S.~D.,  et~al., 2019, GRB Coordinates Network, \href
  {https://ui.adsabs.harvard.edu/abs/2019GCN.24899....1B} {24899, 1}

\bibitem[\protect\citeauthoryear{{B{\'e}gu{\'e}}, {Burgess}  \&
  {Greiner}}{{B{\'e}gu{\'e}} et~al.}{2017}]{Begue17}
{B{\'e}gu{\'e}} D.,  {Burgess} J.~M.,   {Greiner} J.,  2017, \mndoi [\apjl]
  {10.3847/2041-8213/aa9d85}, \href
  {https://ui.adsabs.harvard.edu/abs/2017ApJ...851L..19B} {851, L19}

\bibitem[\protect\citeauthoryear{{Belczynski}, {Holz}, {Fryer}, {Berger},
  {Hartmann}  \& {O'Shea}}{{Belczynski} et~al.}{2010}]{Belczynski10}
{Belczynski} K.,  {Holz} D.~E.,  {Fryer} C.~L.,  {Berger} E.,  {Hartmann}
  D.~H.,   {O'Shea} B.,  2010, \mndoi [\apj] {10.1088/0004-637X/708/1/117},
  \href {https://ui.adsabs.harvard.edu/abs/2010ApJ...708..117B} {708, 117}

\bibitem[\protect\citeauthoryear{{Bell} et~al.,}{{Bell} et~al.}{2019}]{Bell19}
{Bell} M.~E.,  et~al., 2019, \mndoi [\mnras] {10.1093/mnras/sty2801}, \href
  {https://ui.adsabs.harvard.edu/abs/2019MNRAS.482.2484B} {482, 2484}

\bibitem[\protect\citeauthoryear{{Berger}}{{Berger}}{2014}]{Berger14}
{Berger} E.,  2014, \mndoi [\araa] {10.1146/annurev-astro-081913-035926}, \href
  {https://ui.adsabs.harvard.edu/abs/2014ARA&A..52...43B} {52, 43}

\bibitem[\protect\citeauthoryear{{Bloom}, {Frail}  \& {Sari}}{{Bloom}
  et~al.}{2001}]{Bloom01}
{Bloom} J.~S.,  {Frail} D.~A.,   {Sari} R.,  2001, \mndoi [\aj]
  {10.1086/321093}, \href
  {https://ui.adsabs.harvard.edu/abs/2001AJ....121.2879B} {121, 2879}

\bibitem[\protect\citeauthoryear{{Burrows} et~al.,}{{Burrows}
  et~al.}{2005}]{Burrows05}
{Burrows} D.~N.,  et~al., 2005, \mndoi [\ssr] {10.1007/s11214-005-5097-2},
  \href {https://ui.adsabs.harvard.edu/abs/2005SSRv..120..165B} {120, 165}

\bibitem[\protect\citeauthoryear{{Burrows} et~al.,}{{Burrows}
  et~al.}{2006}]{Burrows06}
{Burrows} D.~N.,  et~al., 2006, \mndoi [\apj] {10.1086/508740}, \href
  {https://ui.adsabs.harvard.edu/abs/2006ApJ...653..468B} {653, 468}

\bibitem[\protect\citeauthoryear{{Cavallo} \& {Rees}}{{Cavallo} \&
  {Rees}}{1978}]{Cavallo78}
{Cavallo} G.,  {Rees} M.~J.,  1978, \mndoi [\mnras] {10.1093/mnras/183.3.359},
  \href {https://ui.adsabs.harvard.edu/abs/1978MNRAS.183..359C} {183, 359}

\bibitem[\protect\citeauthoryear{{Cenko} et~al.,}{{Cenko}
  et~al.}{2019a}]{Cenko19b}
{Cenko} S.~B.,  et~al., 2019a, GRB Coordinates Network, \href
  {https://ui.adsabs.harvard.edu/abs/2019GCN.25945....1C} {25945, 1}

\bibitem[\protect\citeauthoryear{{Cenko} et~al.,}{{Cenko}
  et~al.}{2019b}]{Cenko19}
{Cenko} S.~B.,  et~al., 2019b, GRB Coordinates Network, \href
  {https://ui.adsabs.harvard.edu/abs/2019GCN.25945....1C} {25945, 1}

\bibitem[\protect\citeauthoryear{{Chu}, {Howell}, {Rowlinson}, {Gao}, {Zhang},
  {Tingay}, {Bo{\"e}r}  \& {Wen}}{{Chu} et~al.}{2016}]{Chu16}
{Chu} Q.,  {Howell} E.~J.,  {Rowlinson} A.,  {Gao} H.,  {Zhang} B.,  {Tingay}
  S.~J.,  {Bo{\"e}r} M.,   {Wen} L.,  2016, \mndoi [\mnras]
  {10.1093/mnras/stw576}, \href
  {https://ui.adsabs.harvard.edu/abs/2016MNRAS.459..121C} {459, 121}

\bibitem[\protect\citeauthoryear{{Cordes} \& {Lazio}}{{Cordes} \&
  {Lazio}}{2002}]{Cordes02}
{Cordes} J.~M.,  {Lazio} T.~J.~W.,  2002, arXiv e-prints, \href
  {https://ui.adsabs.harvard.edu/abs/2002astro.ph..7156C} {pp
  astro--ph/0207156}

\bibitem[\protect\citeauthoryear{{DeLaunay}, {Tohuvavohu}  \&
  {Kennea}}{{DeLaunay} et~al.}{2020a}]{DeLaunay20}
{DeLaunay} J.,  {Tohuvavohu} A.,   {Kennea} J.,  2020a, GRB Coordinates
  Network, \href {https://ui.adsabs.harvard.edu/abs/2020GCN.27444....1D}
  {27444, 1}

\bibitem[\protect\citeauthoryear{{DeLaunay}, {Tohuvavohu}  \&
  {Kennea}}{{DeLaunay} et~al.}{2020b}]{James20}
{DeLaunay} J.,  {Tohuvavohu} A.,   {Kennea} J.,  2020b, GRB Coordinates
  Network, \href {https://ui.adsabs.harvard.edu/abs/2020GCN.27444....1D}
  {27444, 1}

\bibitem[\protect\citeauthoryear{{Dessenne} et~al.,}{{Dessenne}
  et~al.}{1996}]{Dessenne96}
{Dessenne} C.~A.~C.,  et~al., 1996, \mndoi [\mnras] {10.1093/mnras/281.3.977},
  \href {https://ui.adsabs.harvard.edu/abs/1996MNRAS.281..977D} {281, 977}

\bibitem[\protect\citeauthoryear{{Dewdney}, {Hall}, {Schilizzi}  \&
  {Lazio}}{{Dewdney} et~al.}{2009}]{Dewdney09}
{Dewdney} P.~E.,  {Hall} P.~J.,  {Schilizzi} R.~T.,   {Lazio} T.~J.~L.~W.,
  2009, \mndoi [IEEE Proceedings] {10.1109/JPROC.2009.2021005}, \href
  {https://ui.adsabs.harvard.edu/abs/2009IEEEP..97.1482D} {97, 1482}

\bibitem[\protect\citeauthoryear{{Duchesne}, {Johnston-Hollitt}, {Zhu}, {Wayth}
   \& {Line}}{{Duchesne} et~al.}{2020}]{Duchesne20}
{Duchesne} S.~W.,  {Johnston-Hollitt} M.,  {Zhu} Z.,  {Wayth} R.~B.,   {Line}
  J.~L.~B.,  2020, \mndoi [\pasa] {10.1017/pasa.2020.29}, \href
  {https://ui.adsabs.harvard.edu/abs/2020PASA...37...37D} {37, e037}

\bibitem[\protect\citeauthoryear{{Duncan} \& {Thompson}}{{Duncan} \&
  {Thompson}}{1992}]{Duncan92}
{Duncan} R.~C.,  {Thompson} C.,  1992, \mndoi [\apjl] {10.1086/186413}, \href
  {https://ui.adsabs.harvard.edu/abs/1992ApJ...392L...9D} {392, L9}

\bibitem[\protect\citeauthoryear{{Evans} et~al.,}{{Evans}
  et~al.}{2009}]{Evans09}
{Evans} P.~A.,  et~al., 2009, \mndoi [\mnras]
  {10.1111/j.1365-2966.2009.14913.x}, \href
  {https://ui.adsabs.harvard.edu/abs/2009MNRAS.397.1177E} {397, 1177}

\bibitem[\protect\citeauthoryear{{Evans} et~al.,}{{Evans}
  et~al.}{2010}]{Evans10}
{Evans} P.~A.,  et~al., 2010, \mndoi [\aap] {10.1051/0004-6361/201014819},
  \href {https://ui.adsabs.harvard.edu/abs/2010A&A...519A.102E} {519, A102}

\bibitem[\protect\citeauthoryear{{Farah} et~al.,}{{Farah}
  et~al.}{2019}]{Farah19}
{Farah} W.,  et~al., 2019, \mndoi [\mnras] {10.1093/mnras/stz1748}, \href
  {https://ui.adsabs.harvard.edu/abs/2019MNRAS.488.2989F} {488, 2989}

\bibitem[\protect\citeauthoryear{{Fong} \& {Berger}}{{Fong} \&
  {Berger}}{2013}]{Fong13}
{Fong} W.,  {Berger} E.,  2013, \mndoi [\apj] {10.1088/0004-637X/776/1/18},
  \href {https://ui.adsabs.harvard.edu/abs/2013ApJ...776...18F} {776, 18}

\bibitem[\protect\citeauthoryear{{Fong}, {Berger}, {Margutti}  \&
  {Zauderer}}{{Fong} et~al.}{2015}]{Fong15}
{Fong} W.,  {Berger} E.,  {Margutti} R.,   {Zauderer} B.~A.,  2015, \mndoi
  [\apj] {10.1088/0004-637X/815/2/102}, \href
  {https://ui.adsabs.harvard.edu/abs/2015ApJ...815..102F} {815, 102}

\bibitem[\protect\citeauthoryear{{Fong} et~al.,}{{Fong} et~al.}{2021}]{Fong21}
{Fong} W.,  et~al., 2021, \mndoi [\apj] {10.3847/1538-4357/abc74a}, \href
  {https://ui.adsabs.harvard.edu/abs/2021ApJ...906..127F} {906, 127}

\bibitem[\protect\citeauthoryear{{Gehrels} et~al.,}{{Gehrels}
  et~al.}{2004}]{Gehrels04}
{Gehrels} N.,  et~al., 2004, \mndoi [\apj] {10.1086/422091}, \href
  {https://ui.adsabs.harvard.edu/abs/2004ApJ...611.1005G} {611, 1005}

\bibitem[\protect\citeauthoryear{{Ghumatkar}, {Sharma}, {Bhattacharya},
  {Khanam}, {Vibhute}, {Vadawale}  \& {AstroSat CZTI
  Collaboration}}{{Ghumatkar} et~al.}{2019}]{Ghumatkar19}
{Ghumatkar} P.,  {Sharma} V.,  {Bhattacharya} D.,  {Khanam} T.,  {Vibhute} A.,
  {Vadawale} S.,   {AstroSat CZTI Collaboration} 2019, GRB Coordinates Network,
  \href {https://ui.adsabs.harvard.edu/abs/2019GCN.24846....1G} {25289, 1}

\bibitem[\protect\citeauthoryear{{Goldstein} et~al.,}{{Goldstein}
  et~al.}{2017}]{Goldstein17}
{Goldstein} A.,  et~al., 2017, \mndoi [\apjl] {10.3847/2041-8213/aa8f41}, \href
  {https://ui.adsabs.harvard.edu/abs/2017ApJ...848L..14G} {848, L14}

\bibitem[\protect\citeauthoryear{{Gompertz}, {Levan}  \& {Tanvir}}{{Gompertz}
  et~al.}{2020}]{Gompertz20}
{Gompertz} B.~P.,  {Levan} A.~J.,   {Tanvir} N.~R.,  2020, \mndoi [\apj]
  {10.3847/1538-4357/ab8d24}, \href
  {https://ui.adsabs.harvard.edu/abs/2020ApJ...895...58G} {895, 58}

\bibitem[\protect\citeauthoryear{{Gourdji}, {Rowlinson}, {Wijers}  \&
  {Goldstein}}{{Gourdji} et~al.}{2020}]{gourdji20}
{Gourdji} K.,  {Rowlinson} A.,  {Wijers} R.~A.~M.~J.,   {Goldstein} A.,  2020,
  \mndoi [\mnras] {10.1093/mnras/staa2128}, \href
  {https://ui.adsabs.harvard.edu/abs/2020MNRAS.497.3131G} {497, 3131}

\bibitem[\protect\citeauthoryear{{Gupta} et~al.,}{{Gupta}
  et~al.}{2021}]{Gupta21}
{Gupta} V.,  et~al., 2021, \mndoi [\mnras] {10.1093/mnras/staa3683}, \href
  {https://ui.adsabs.harvard.edu/abs/2021MNRAS.501.2316G} {501, 2316}

\bibitem[\protect\citeauthoryear{{Hancock}, {Trott}  \&
  {Hurley-Walker}}{{Hancock} et~al.}{2018}]{Hancock18}
{Hancock} P.~J.,  {Trott} C.~M.,   {Hurley-Walker} N.,  2018, \mndoi [\pasa]
  {10.1017/pasa.2018.3}, \href
  {https://ui.adsabs.harvard.edu/abs/2018PASA...35...11H} {35, e011}

\bibitem[\protect\citeauthoryear{{Hancock}, {Hurley-Walker}  \&
  {White}}{{Hancock} et~al.}{2019a}]{Hancock19b}
{Hancock} P.~J.,  {Hurley-Walker} N.,   {White} T.~E.,  2019a, \mndoi
  [Astronomy and Computing] {10.1016/j.ascom.2019.02.004}, \href
  {https://ui.adsabs.harvard.edu/abs/2019A&C....27...23H} {27, 23}

\bibitem[\protect\citeauthoryear{{Hancock} et~al.,}{{Hancock}
  et~al.}{2019b}]{Hancock19}
{Hancock} P.~J.,  et~al., 2019b, \mndoi [\pasa] {10.1017/pasa.2019.40}, \href
  {https://ui.adsabs.harvard.edu/abs/2019PASA...36...46H} {36, e046}

\bibitem[\protect\citeauthoryear{Hunter}{Hunter}{2007}]{hunter07}
Hunter J.~D.,  2007, \mndoi [Computing In Science \& Engineering]
  {10.1109/MCSE.2007.55}, 9, 90

\bibitem[\protect\citeauthoryear{{Hurley-Walker} \& {Hancock}}{{Hurley-Walker}
  \& {Hancock}}{2018}]{Hurley18}
{Hurley-Walker} N.,  {Hancock} P.~J.,  2018, \mndoi [Astronomy and Computing]
  {10.1016/j.ascom.2018.08.006}, \href
  {https://ui.adsabs.harvard.edu/abs/2018A&C....25...94H} {25, 94}

\bibitem[\protect\citeauthoryear{{Hurley-Walker} et~al.,}{{Hurley-Walker}
  et~al.}{2017}]{Hurley17}
{Hurley-Walker} N.,  et~al., 2017, \mndoi [\mnras] {10.1093/mnras/stw2337},
  \href {https://ui.adsabs.harvard.edu/abs/2017MNRAS.464.1146H} {464, 1146}

\bibitem[\protect\citeauthoryear{{Hurley} et~al.,}{{Hurley}
  et~al.}{2013}]{Hurley13}
{Hurley} K.,  et~al., 2013, \mndoi [\apjs] {10.1088/0067-0049/207/2/39}, \href
  {https://ui.adsabs.harvard.edu/abs/2013ApJS..207...39H} {207, 39}

\bibitem[\protect\citeauthoryear{{Inoue}, {Asano}  \& {Ioka}}{{Inoue}
  et~al.}{2011}]{Inoue11}
{Inoue} T.,  {Asano} K.,   {Ioka} K.,  2011, \mndoi [\apj]
  {10.1088/0004-637X/734/2/77}, \href
  {https://ui.adsabs.harvard.edu/abs/2011ApJ...734...77I} {734, 77}

\bibitem[\protect\citeauthoryear{{James}, {Anderson}, {Wen}, {Bosveld}, {Chu},
  {Kovalam}, {Slaven-Blair}  \& {Williams}}{{James} et~al.}{2019}]{James19}
{James} C.~W.,  {Anderson} G.~E.,  {Wen} L.,  {Bosveld} J.,  {Chu} Q.,
  {Kovalam} M.,  {Slaven-Blair} T.~J.,   {Williams} A.,  2019, \mndoi [\mnras]
  {10.1093/mnrasl/slz129}, \href
  {https://ui.adsabs.harvard.edu/abs/2019MNRAS.489L..75J} {489, L75}

\bibitem[\protect\citeauthoryear{{Japelj} et~al.,}{{Japelj}
  et~al.}{2019}]{Japelj19}
{Japelj} J.,  et~al., 2019, GRB Coordinates Network, \href
  {https://ui.adsabs.harvard.edu/abs/2019GCN.24916....1J} {24916, 1}

\bibitem[\protect\citeauthoryear{Jones, Oliphant, Peterson  \& Others}{Jones
  et~al.}{2001}]{Jones_scipy_2001}
Jones E.,  Oliphant T.,  Peterson P.,   Others 2001, {SciPy: Open source
  scientific tools for Python}, \url {https://www.scipy.org/}

\bibitem[\protect\citeauthoryear{{Kann}}{{Kann}}{2013}]{Kann13}
{Kann} D.~A.,  2013, in {Castro-Tirado} A.~J.,  {Gorosabel} J.,   {Park} I.~H.,
   eds,  EAS Publications Series Vol. 61, EAS Publications Series. pp 309--317
  (\mn@eprint {arXiv} {1212.0040}), \mndoi{10.1051/eas/1361049}

\bibitem[\protect\citeauthoryear{{Kaplan} et~al.,}{{Kaplan}
  et~al.}{2015}]{Kaplan15}
{Kaplan} D.~L.,  et~al., 2015, \mndoi [\apjl] {10.1088/2041-8205/814/2/L25},
  \href {https://ui.adsabs.harvard.edu/abs/2015ApJ...814L..25K} {814, L25}

\bibitem[\protect\citeauthoryear{{Kaplan}, {Murphy}, {Rowlinson}, {Croft},
  {Wayth}  \& {Trott}}{{Kaplan} et~al.}{2016}]{Kaplan16}
{Kaplan} D.~L.,  {Murphy} T.,  {Rowlinson} A.,  {Croft} S.~D.,  {Wayth} R.~B.,
   {Trott} C.~M.,  2016, \mndoi [\pasa] {10.1017/pasa.2016.43}, \href
  {https://ui.adsabs.harvard.edu/abs/2016PASA...33...50K} {33, e050}

\bibitem[\protect\citeauthoryear{{Katz}}{{Katz}}{1997}]{Katz97}
{Katz} J.~I.,  1997, \mndoi [\apj] {10.1086/304896}, \href
  {https://ui.adsabs.harvard.edu/abs/1997ApJ...490..633K} {490, 633}

\bibitem[\protect\citeauthoryear{{Kouveliotou}, {Meegan}, {Fishman}, {Bhat},
  {Briggs}, {Koshut}, {Paciesas}  \& {Pendleton}}{{Kouveliotou}
  et~al.}{1993}]{Kouveliotou93}
{Kouveliotou} C.,  {Meegan} C.~A.,  {Fishman} G.~J.,  {Bhat} N.~P.,  {Briggs}
  M.~S.,  {Koshut} T.~M.,  {Paciesas} W.~S.,   {Pendleton} G.~N.,  1993, \mndoi
  [\apjl] {10.1086/186969}, \href
  {https://ui.adsabs.harvard.edu/abs/1993ApJ...413L.101K} {413, L101}

\bibitem[\protect\citeauthoryear{Kurtzer, Sochat  \& Bauer}{Kurtzer
  et~al.}{2017}]{kurtzer_singularity_2017}
Kurtzer G.~M.,  Sochat V.,   Bauer M.~W.,  2017, \mndoi [PLOS ONE]
  {10.1371/journal.pone.0177459}, 12, e0177459

\bibitem[\protect\citeauthoryear{{Lasky}, {Haskell}, {Ravi}, {Howell}  \&
  {Coward}}{{Lasky} et~al.}{2014}]{Lasky14}
{Lasky} P.~D.,  {Haskell} B.,  {Ravi} V.,  {Howell} E.~J.,   {Coward} D.~M.,
  2014, \mndoi [\prd] {10.1103/PhysRevD.89.047302}, \href
  {https://ui.adsabs.harvard.edu/abs/2014PhRvD..89d7302L} {89, 047302}

\bibitem[\protect\citeauthoryear{{Lattimer}}{{Lattimer}}{2012}]{Lattimer12}
{Lattimer} J.~M.,  2012, \mndoi [Annual Review of Nuclear and Particle Science]
  {10.1146/annurev-nucl-102711-095018}, \href
  {https://ui.adsabs.harvard.edu/abs/2012ARNPS..62..485L} {62, 485}

\bibitem[\protect\citeauthoryear{{Lien} et~al.,}{{Lien} et~al.}{2016}]{Lien16}
{Lien} A.,  et~al., 2016, \mndoi [\apj] {10.3847/0004-637X/829/1/7}, \href
  {https://ui.adsabs.harvard.edu/abs/2016ApJ...829....7L} {829, 7}

\bibitem[\protect\citeauthoryear{Lipunov \& Panchenko}{Lipunov \&
  Panchenko}{1996}]{Lipunov96}
Lipunov V.~M.,  Panchenko I.~E.,  1996, Astronomy \& Astrophysics, 312, 937

\bibitem[\protect\citeauthoryear{{L{\"u}}, {Liang}, {Zhang}  \&
  {Zhang}}{{L{\"u}} et~al.}{2010}]{Lu10}
{L{\"u}} H.-J.,  {Liang} E.-W.,  {Zhang} B.-B.,   {Zhang} B.,  2010, \mndoi
  [\apj] {10.1088/0004-637X/725/2/1965}, \href
  {https://ui.adsabs.harvard.edu/abs/2010ApJ...725.1965L} {725, 1965}

\bibitem[\protect\citeauthoryear{{Lucek} \& {Bell}}{{Lucek} \&
  {Bell}}{2000}]{Lucek00}
{Lucek} S.~G.,  {Bell} A.~R.,  2000, \mndoi [\mnras]
  {10.1046/j.1365-8711.2000.03363.x}, \href
  {https://ui.adsabs.harvard.edu/abs/2000MNRAS.314...65L} {314, 65}

\bibitem[\protect\citeauthoryear{{Lyutikov}}{{Lyutikov}}{2013}]{Lyutikov13}
{Lyutikov} M.,  2013, \mndoi [\apj] {10.1088/0004-637X/768/1/63}, \href
  {https://ui.adsabs.harvard.edu/abs/2013ApJ...768...63L} {768, 63}

\bibitem[\protect\citeauthoryear{{Macquart} et~al.,}{{Macquart}
  et~al.}{2020}]{JP20}
{Macquart} J.~P.,  et~al., 2020, \mndoi [\nat] {10.1038/s41586-020-2300-2},
  \href {https://ui.adsabs.harvard.edu/abs/2020Natur.581..391M} {581, 391}

\bibitem[\protect\citeauthoryear{{Mailyan} \& {Meegan}}{{Mailyan} \&
  {Meegan}}{2019}]{Mailyan19}
{Mailyan} B.,  {Meegan} C.,  2019, GRB Coordinates Network, \href
  {https://ui.adsabs.harvard.edu/abs/2019GCN.25636....1M} {25636, 1}

\bibitem[\protect\citeauthoryear{{Medvedev}, {Fiore}, {Fonseca}, {Silva}  \&
  {Mori}}{{Medvedev} et~al.}{2005}]{Medvedev05}
{Medvedev} M.~V.,  {Fiore} M.,  {Fonseca} R.~A.,  {Silva} L.~O.,   {Mori}
  W.~B.,  2005, \mndoi [\apjl] {10.1086/427921}, \href
  {https://ui.adsabs.harvard.edu/abs/2005ApJ...618L..75M} {618, L75}

\bibitem[\protect\citeauthoryear{{Meegan} et~al.,}{{Meegan}
  et~al.}{2009}]{Meegan09}
{Meegan} C.,  et~al., 2009, \mndoi [\apj] {10.1088/0004-637X/702/1/791}, \href
  {https://ui.adsabs.harvard.edu/abs/2009ApJ...702..791M} {702, 791}

\bibitem[\protect\citeauthoryear{Metzger \& Zivancev}{Metzger \&
  Zivancev}{2016}]{Metzger16}
Metzger B.~D.,  Zivancev C.,  2016, Monthly Notices of the Royal Astronomical
  Society, 461, 4435

\bibitem[\protect\citeauthoryear{{Metzger}, {Berger}  \& {Margalit}}{{Metzger}
  et~al.}{2017}]{Metzger17}
{Metzger} B.~D.,  {Berger} E.,   {Margalit} B.,  2017, \mndoi [\apj]
  {10.3847/1538-4357/aa633d}, \href
  {https://ui.adsabs.harvard.edu/abs/2017ApJ...841...14M} {841, 14}

\bibitem[\protect\citeauthoryear{{Milosavljevi{\'c}} \&
  {Nakar}}{{Milosavljevi{\'c}} \& {Nakar}}{2006}]{Milosavljevi06}
{Milosavljevi{\'c}} M.,  {Nakar} E.,  2006, \mndoi [\apj] {10.1086/507975},
  \href {https://ui.adsabs.harvard.edu/abs/2006ApJ...651..979M} {651, 979}

\bibitem[\protect\citeauthoryear{{Mizuno}, {Pohl}, {Niemiec}, {Zhang},
  {Nishikawa}  \& {Hardee}}{{Mizuno} et~al.}{2011}]{Mizuno11}
{Mizuno} Y.,  {Pohl} M.,  {Niemiec} J.,  {Zhang} B.,  {Nishikawa} K.-I.,
  {Hardee} P.~E.,  2011, \mndoi [\apj] {10.1088/0004-637X/726/2/62}, \href
  {https://ui.adsabs.harvard.edu/abs/2011ApJ...726...62M} {726, 62}

\bibitem[\protect\citeauthoryear{{Narayan}, {Paczynski}  \& {Piran}}{{Narayan}
  et~al.}{1992}]{Narayan92}
{Narayan} R.,  {Paczynski} B.,   {Piran} T.,  1992, \mndoi [\apjl]
  {10.1086/186493}, \href
  {https://ui.adsabs.harvard.edu/abs/1992ApJ...395L..83N} {395, L83}

\bibitem[\protect\citeauthoryear{{Narayana Bhat} et~al.,}{{Narayana Bhat}
  et~al.}{2016}]{Narayana16}
{Narayana Bhat} P.,  et~al., 2016, \mndoi [\apjs] {10.3847/0067-0049/223/2/28},
  \href {https://ui.adsabs.harvard.edu/abs/2016ApJS..223...28N} {223, 28}

\bibitem[\protect\citeauthoryear{{Obenberger} et~al.,}{{Obenberger}
  et~al.}{2014}]{Obenberger14}
{Obenberger} K.~S.,  et~al., 2014, \mndoi [\apj] {10.1088/0004-637X/785/1/27},
  \href {https://ui.adsabs.harvard.edu/abs/2014ApJ...785...27O} {785, 27}

\bibitem[\protect\citeauthoryear{{Ochsenbein}, {Bauer}  \&
  {Marcout}}{{Ochsenbein} et~al.}{2000}]{ochsenbein00}
{Ochsenbein} F.,  {Bauer} P.,   {Marcout} J.,  2000, \mndoi [\aaps]
  {10.1051/aas:2000169}, \href
  {https://ui.adsabs.harvard.edu/abs/2000A&AS..143...23O} {143, 23}

\bibitem[\protect\citeauthoryear{{Offringa} \& {Smirnov}}{{Offringa} \&
  {Smirnov}}{2017}]{Offringa17}
{Offringa} A.~R.,  {Smirnov} O.,  2017, \mndoi [\mnras]
  {10.1093/mnras/stx1547}, \href
  {https://ui.adsabs.harvard.edu/abs/2017MNRAS.471..301O} {471, 301}

\bibitem[\protect\citeauthoryear{{Offringa} et~al.,}{{Offringa}
  et~al.}{2014}]{Offringa14}
{Offringa} A.~R.,  et~al., 2014, \mndoi [\mnras] {10.1093/mnras/stu1368}, \href
  {https://ui.adsabs.harvard.edu/abs/2014MNRAS.444..606O} {444, 606}

\bibitem[\protect\citeauthoryear{{Palaniswamy}, {Wayth}, {Trott}, {McCallum},
  {Tingay}  \& {Reynolds}}{{Palaniswamy} et~al.}{2014}]{Palaniswamy14}
{Palaniswamy} D.,  {Wayth} R.~B.,  {Trott} C.~M.,  {McCallum} J.~N.,  {Tingay}
  S.~J.,   {Reynolds} C.,  2014, \mndoi [\apj] {10.1088/0004-637X/790/1/63},
  \href {https://ui.adsabs.harvard.edu/abs/2014ApJ...790...63P} {790, 63}

\bibitem[\protect\citeauthoryear{{Panaitescu}}{{Panaitescu}}{2005}]{Panaitescu05}
{Panaitescu} A.,  2005, \mndoi [\mnras] {10.1111/j.1365-2966.2005.09532.x},
  \href {https://ui.adsabs.harvard.edu/abs/2005MNRAS.363.1409P} {363, 1409}

\bibitem[\protect\citeauthoryear{{Panaitescu} \& {Kumar}}{{Panaitescu} \&
  {Kumar}}{2002}]{Panaitescu02}
{Panaitescu} A.,  {Kumar} P.,  2002, \mndoi [\apj] {10.1086/340094}, \href
  {https://ui.adsabs.harvard.edu/abs/2002ApJ...571..779P} {571, 779}

\bibitem[\protect\citeauthoryear{{Ravi} \& {Lasky}}{{Ravi} \&
  {Lasky}}{2014}]{Ravi14}
{Ravi} V.,  {Lasky} P.~D.,  2014, \mndoi [\mnras] {10.1093/mnras/stu720}, \href
  {https://ui.adsabs.harvard.edu/abs/2014MNRAS.441.2433R} {441, 2433}

\bibitem[\protect\citeauthoryear{{Rees} \& {Meszaros}}{{Rees} \&
  {Meszaros}}{1992}]{Rees92}
{Rees} M.~J.,  {Meszaros} P.,  1992, \mndoi [\mnras] {10.1093/mnras/258.1.41P},
  \href {https://ui.adsabs.harvard.edu/abs/1992MNRAS.258P..41R} {258, 41}

\bibitem[\protect\citeauthoryear{{Rezzolla}, {Giacomazzo}, {Baiotti}, {Granot},
  {Kouveliotou}  \& {Aloy}}{{Rezzolla} et~al.}{2011}]{Rezzolla11}
{Rezzolla} L.,  {Giacomazzo} B.,  {Baiotti} L.,  {Granot} J.,  {Kouveliotou}
  C.,   {Aloy} M.~A.,  2011, \mndoi [\apjl] {10.1088/2041-8205/732/1/L6}, \href
  {https://ui.adsabs.harvard.edu/abs/2011ApJ...732L...6R} {732, L6}

\bibitem[\protect\citeauthoryear{{Rowlinson} \& {Anderson}}{{Rowlinson} \&
  {Anderson}}{2019}]{Rowlinson19}
{Rowlinson} A.,  {Anderson} G.~E.,  2019, \mndoi [\mnras]
  {10.1093/mnras/stz2295}, \href
  {https://ui.adsabs.harvard.edu/abs/2019MNRAS.489.3316R} {489, 3316}

\bibitem[\protect\citeauthoryear{{Rowlinson} et~al.,}{{Rowlinson}
  et~al.}{2010}]{Rowlinson10}
{Rowlinson} A.,  et~al., 2010, \mndoi [\mnras]
  {10.1111/j.1365-2966.2010.17354.x}, \href
  {https://ui.adsabs.harvard.edu/abs/2010MNRAS.409..531R} {409, 531}

\bibitem[\protect\citeauthoryear{{Rowlinson}, {O'Brien}, {Metzger}, {Tanvir}
  \& {Levan}}{{Rowlinson} et~al.}{2013}]{Rowlinson13}
{Rowlinson} A.,  {O'Brien} P.~T.,  {Metzger} B.~D.,  {Tanvir} N.~R.,   {Levan}
  A.~J.,  2013, \mndoi [\mnras] {10.1093/mnras/sts683}, \href
  {https://ui.adsabs.harvard.edu/abs/2013MNRAS.430.1061R} {430, 1061}

\bibitem[\protect\citeauthoryear{{Rowlinson} et~al.,}{{Rowlinson}
  et~al.}{2019}]{Rowlinson19b}
{Rowlinson} A.,  et~al., 2019, \mndoi [\mnras] {10.1093/mnras/stz2866}, \href
  {https://ui.adsabs.harvard.edu/abs/2019MNRAS.490.3483R} {490, 3483}

\bibitem[\protect\citeauthoryear{{Rowlinson} et~al.,}{{Rowlinson}
  et~al.}{2020}]{Rowlinson20}
{Rowlinson} A.,  et~al., 2020, arXiv e-prints, \href
  {https://ui.adsabs.harvard.edu/abs/2020arXiv200812657R} {p. arXiv:2008.12657}

\bibitem[\protect\citeauthoryear{{Rowlinson} et~al.,}{{Rowlinson}
  et~al.}{2021}]{Rowlinson21}
{Rowlinson} A.,  et~al., 2021, \mndoi [\mnras] {10.1093/mnras/stab2060}, \href
  {https://ui.adsabs.harvard.edu/abs/2021MNRAS.506.5268R} {506, 5268}

\bibitem[\protect\citeauthoryear{{Sakamoto} et~al.,}{{Sakamoto}
  et~al.}{2019}]{Sakamoto19}
{Sakamoto} T.,  et~al., 2019, GRB Coordinates Network, \href
  {https://ui.adsabs.harvard.edu/abs/2019GCN.24899....1B} {25953, 1}

\bibitem[\protect\citeauthoryear{{Santana}, {Barniol Duran}  \&
  {Kumar}}{{Santana} et~al.}{2014}]{Santana14}
{Santana} R.,  {Barniol Duran} R.,   {Kumar} P.,  2014, \mndoi [\apj]
  {10.1088/0004-637X/785/1/29}, \href
  {https://ui.adsabs.harvard.edu/abs/2014ApJ...785...29S} {785, 29}

\bibitem[\protect\citeauthoryear{{Sarin}, {Lasky}  \& {Ashton}}{{Sarin}
  et~al.}{2020}]{Sarin20}
{Sarin} N.,  {Lasky} P.~D.,   {Ashton} G.,  2020, \mndoi [\mnras]
  {10.1093/mnras/staa3090}, \href
  {https://ui.adsabs.harvard.edu/abs/2020MNRAS.499.5986S} {499, 5986}

\bibitem[\protect\citeauthoryear{{Seaman} et~al.,}{{Seaman}
  et~al.}{2011}]{Seaman11}
{Seaman} R.,  et~al., 2011, {Sky Event Reporting Metadata Version 2.0}, IVOA
  Recommendation 11 July 2011 (\mn@eprint {arXiv} {1110.0523}),
  \mndoi{10.5479/ADS/bib/2011ivoa.spec.0711S}

\bibitem[\protect\citeauthoryear{{Soderberg} et~al.,}{{Soderberg}
  et~al.}{2006}]{Soderberg06}
{Soderberg} A.~M.,  et~al., 2006, \mndoi [\apj] {10.1086/506429}, \href
  {https://ui.adsabs.harvard.edu/abs/2006ApJ...650..261S} {650, 261}

\bibitem[\protect\citeauthoryear{{Sokolowski} et~al.,}{{Sokolowski}
  et~al.}{2017}]{Sokolowski17}
{Sokolowski} M.,  et~al., 2017, \mndoi [\pasa] {10.1017/pasa.2017.54}, \href
  {https://ui.adsabs.harvard.edu/abs/2017PASA...34...62S} {34, e062}

\bibitem[\protect\citeauthoryear{{Sokolowski} et~al.,}{{Sokolowski}
  et~al.}{2021}]{Sokolowski21}
{Sokolowski} M.,  et~al., 2021, \mndoi [\pasa] {10.1017/pasa.2021.16}, \href
  {https://ui.adsabs.harvard.edu/abs/2021PASA...38...23S} {38, e023}

\bibitem[\protect\citeauthoryear{{Sonbas} et~al.,}{{Sonbas}
  et~al.}{2019a}]{Sonbas19b}
{Sonbas} E.,  et~al., 2019a, GRB Coordinates Network, \href
  {https://ui.adsabs.harvard.edu/abs/2019GCN.24888....1S} {24888, 1}

\bibitem[\protect\citeauthoryear{{Sonbas} et~al.,}{{Sonbas}
  et~al.}{2019b}]{Sonbas19}
{Sonbas} E.,  et~al., 2019b, GRB Coordinates Network, \href
  {https://ui.adsabs.harvard.edu/abs/2019GCN.24888....1S} {24888, 1}

\bibitem[\protect\citeauthoryear{{Spergel} et~al.,}{{Spergel}
  et~al.}{2003}]{Spergel03}
{Spergel} D.~N.,  et~al., 2003, \mndoi [\apjs] {10.1086/377226}, \href
  {https://ui.adsabs.harvard.edu/abs/2003ApJS..148..175S} {148, 175}

\bibitem[\protect\citeauthoryear{{Staley} \& {Fender}}{{Staley} \&
  {Fender}}{2016}]{Staley16}
{Staley} T.~D.,  {Fender} R.,  2016, arXiv e-prints, \href
  {https://ui.adsabs.harvard.edu/abs/2016arXiv160603735S} {p. arXiv:1606.03735}

\bibitem[\protect\citeauthoryear{{Svinkin} et~al.,}{{Svinkin}
  et~al.}{2017}]{Svinkin17}
{Svinkin} D.,  et~al., 2017, GRB Coordinates Network, \href
  {https://ui.adsabs.harvard.edu/abs/2017GCN.21735....1S} {21735, 1}

\bibitem[\protect\citeauthoryear{{Swinbank}}{{Swinbank}}{2014}]{Swinbank14}
{Swinbank} J.,  2014, \mndoi [Astronomy and Computing]
  {10.1016/j.ascom.2014.09.001}, \href
  {https://ui.adsabs.harvard.edu/abs/2014A&C.....7...12S} {7, 12}

\bibitem[\protect\citeauthoryear{{Taylor}}{{Taylor}}{2005}]{taylor05}
{Taylor} M.~B.,  2005, in {Shopbell} P.,  {Britton} M.,   {Ebert} R.,  eds,
  Astronomical Society of the Pacific Conference Series Vol. 347, Astronomical
  Data Analysis Software and Systems XIV. p.~29

\bibitem[\protect\citeauthoryear{{Taylor}, {Manchester}  \& {Lyne}}{{Taylor}
  et~al.}{1993}]{Taylor93}
{Taylor} J.~H.,  {Manchester} R.~N.,   {Lyne} A.~G.,  1993, \mndoi [\apjs]
  {10.1086/191832}, \href
  {https://ui.adsabs.harvard.edu/abs/1993ApJS...88..529T} {88, 529}

\bibitem[\protect\citeauthoryear{{The Astropy Collaboration} et~al.,}{{The
  Astropy Collaboration} et~al.}{2013}]{TheAstropyCollaboration2013}
{The Astropy Collaboration} et~al., 2013, \mndoi [Astronomy {\&} Astrophysics]
  {10.1051/0004-6361/201322068}, 558, 9

\bibitem[\protect\citeauthoryear{{Thompson}}{{Thompson}}{1994}]{Thompson94}
{Thompson} C.,  1994, \mndoi [\mnras] {10.1093/mnras/270.3.480}, \href
  {https://ui.adsabs.harvard.edu/abs/1994MNRAS.270..480T} {270, 480}

\bibitem[\protect\citeauthoryear{{Tingay} et~al.,}{{Tingay}
  et~al.}{2013}]{Tingay13}
{Tingay} S.~J.,  et~al., 2013, \mndoi [\pasa] {10.1017/pasa.2012.007}, \href
  {https://ui.adsabs.harvard.edu/abs/2013PASA...30....7T} {30, e007}

\bibitem[\protect\citeauthoryear{{Totani}}{{Totani}}{2013}]{Totani13}
{Totani} T.,  2013, \mndoi [\pasj] {10.1093/pasj/65.5.L12}, \href
  {https://ui.adsabs.harvard.edu/abs/2013PASJ...65L..12T} {65, L12}

\bibitem[\protect\citeauthoryear{{Tremblay} et~al.,}{{Tremblay}
  et~al.}{2015}]{Tremblay15}
{Tremblay} S.~E.,  et~al., 2015, \mndoi [\pasa] {10.1017/pasa.2015.6}, \href
  {https://ui.adsabs.harvard.edu/abs/2015PASA...32....5T} {32, e005}

\bibitem[\protect\citeauthoryear{{Usov}}{{Usov}}{1992}]{Usov92}
{Usov} V.~V.,  1992, \mndoi [\nat] {10.1038/357472a0}, \href
  {https://ui.adsabs.harvard.edu/abs/1992Natur.357..472U} {357, 472}

\bibitem[\protect\citeauthoryear{{Usov}}{{Usov}}{1994}]{Usov94}
{Usov} V.~V.,  1994, \mndoi [\mnras] {10.1093/mnras/267.4.1035}, \href
  {https://ui.adsabs.harvard.edu/abs/1994MNRAS.267.1035U} {267, 1035}

\bibitem[\protect\citeauthoryear{{Usov} \& {Katz}}{{Usov} \&
  {Katz}}{2000}]{Usov00}
{Usov} V.~V.,  {Katz} J.~I.,  2000, \aap, \href
  {https://ui.adsabs.harvard.edu/abs/2000A&A...364..655U} {364, 655}

\bibitem[\protect\citeauthoryear{{Veres}, {Hui}, {Meegan}, {Wood}  \& {Fermi
  GBM Team}}{{Veres} et~al.}{2020a}]{Veres20}
{Veres} P.,  {Hui} C.~M.,  {Meegan} C.,  {Wood} J.,   {Fermi GBM Team} 2020a,
  GRB Coordinates Network, \href
  {https://ui.adsabs.harvard.edu/abs/2020GCN.27453....1V} {27453, 1}

\bibitem[\protect\citeauthoryear{{Veres}, {Meegan}  \& {Fermi GBM
  Team}}{{Veres} et~al.}{2020b}]{Veres20b}
{Veres} P.,  {Meegan} C.,   {Fermi GBM Team} 2020b, GRB Coordinates Network,
  \href {https://ui.adsabs.harvard.edu/abs/2020GCN.27466....1V} {27466, 1}

\bibitem[\protect\citeauthoryear{{Wayth} et~al.,}{{Wayth}
  et~al.}{2018}]{Wayth18}
{Wayth} R.~B.,  et~al., 2018, \mndoi [\pasa] {10.1017/pasa.2018.37}, \href
  {https://ui.adsabs.harvard.edu/abs/2018PASA...35...33W} {35, 33}

\bibitem[\protect\citeauthoryear{{Wenger} et~al.,}{{Wenger}
  et~al.}{2000}]{wenger00}
{Wenger} M.,  et~al., 2000, \mndoi [\aaps] {10.1051/aas:2000332}, \href
  {https://ui.adsabs.harvard.edu/abs/2000A&AS..143....9W} {143, 9}

\bibitem[\protect\citeauthoryear{{Wright}}{{Wright}}{2006}]{Wright06}
{Wright} E.~L.,  2006, \mndoi [\pasp] {10.1086/510102}, \href
  {https://ui.adsabs.harvard.edu/abs/2006PASP..118.1711W} {118, 1711}

\bibitem[\protect\citeauthoryear{{Yao}, {Manchester}  \& {Wang}}{{Yao}
  et~al.}{2017}]{YMW16}
{Yao} J.~M.,  {Manchester} R.~N.,   {Wang} N.,  2017, \mndoi [\apj]
  {10.3847/1538-4357/835/1/29}, \href
  {https://ui.adsabs.harvard.edu/abs/2017ApJ...835...29Y} {835, 29}

\bibitem[\protect\citeauthoryear{{Yost}, {Harrison}, {Sari}  \& {Frail}}{{Yost}
  et~al.}{2003}]{Yost03}
{Yost} S.~A.,  {Harrison} F.~A.,  {Sari} R.,   {Frail} D.~A.,  2003, \mndoi
  [\apj] {10.1086/378288}, \href
  {https://ui.adsabs.harvard.edu/abs/2003ApJ...597..459Y} {597, 459}

\bibitem[\protect\citeauthoryear{{Zhang}}{{Zhang}}{2014}]{Zhang14}
{Zhang} B.,  2014, \mndoi [\apjl] {10.1088/2041-8205/780/2/L21}, \href
  {https://ui.adsabs.harvard.edu/abs/2014ApJ...780L..21Z} {780, L21}

\bibitem[\protect\citeauthoryear{{Zhang} \& {M{\'e}sz{\'a}ros}}{{Zhang} \&
  {M{\'e}sz{\'a}ros}}{2001}]{Zhang01}
{Zhang} B.,  {M{\'e}sz{\'a}ros} P.,  2001, \mndoi [\apjl] {10.1086/320255},
  \href {https://ui.adsabs.harvard.edu/abs/2001ApJ...552L..35Z} {552, L35}

\bibitem[\protect\citeauthoryear{{Zhang} et~al.,}{{Zhang}
  et~al.}{2009}]{Zhang09}
{Zhang} B.,  et~al., 2009, \mndoi [\apj] {10.1088/0004-637X/703/2/1696}, \href
  {https://ui.adsabs.harvard.edu/abs/2009ApJ...703.1696Z} {703, 1696}

\bibitem[\protect\citeauthoryear{{van Haarlem} et~al.,}{{van Haarlem}
  et~al.}{2013}]{Haarlem13}
{van Haarlem} M.~P.,  et~al., 2013, \mndoi [\aap]
  {10.1051/0004-6361/201220873}, \href
  {https://ui.adsabs.harvard.edu/abs/2013A&A...556A...2V} {556, A2}

\bibitem[\protect\citeauthoryear{{van der Horst} et~al.,}{{van der Horst}
  et~al.}{2008}]{Horst08}
{van der Horst} A.~J.,  et~al., 2008, \mndoi [\aap]
  {10.1051/0004-6361:20078051}, \href
  {https://ui.adsabs.harvard.edu/abs/2008A&A...480...35V} {480, 35}

\bibitem[\protect\citeauthoryear{van~der Walt, Colbert  \& Varoquaux}{van~der
  Walt et~al.}{2011}]{vanderWalt_numpy_2011}
van~der Walt S.,  Colbert S.~C.,   Varoquaux G.,  2011, \mndoi [Computing in
  Science {\&} Engineering] {10.1109/MCSE.2011.37}, 13, 22

\makeatother
\end{thebibliography}

\appendix
\section{MWA images}\label{appendix:images}
\begin{figure*}
\centering
\includegraphics[width=\textwidth]{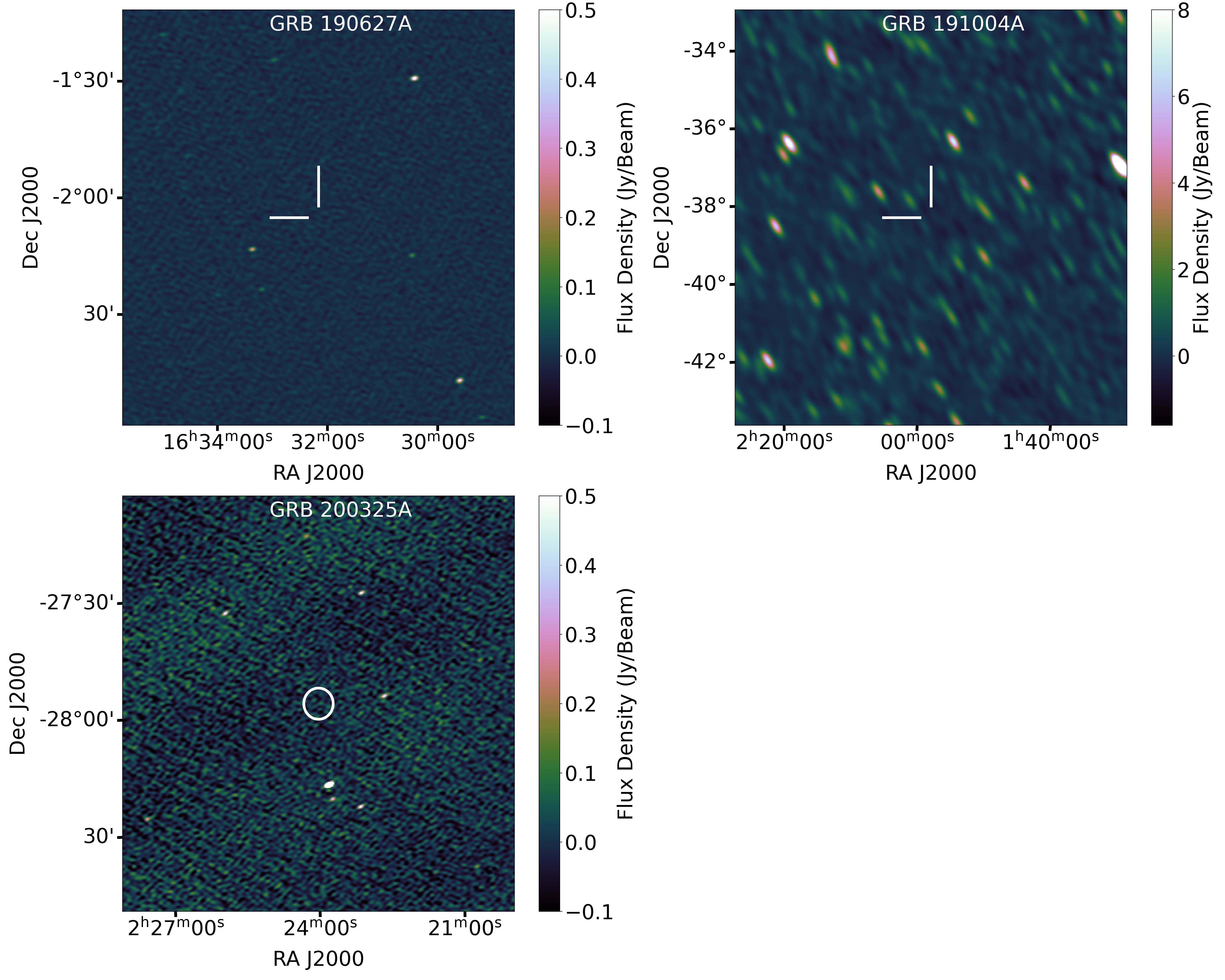}
\caption{The first 2\,min snapshots showing the regions surrounding the three GRBs localised by \textit{Swift}. The white lines in the top 2 panels 
point to the positions of GRB 190627A and 191004A localised by \textit{Swift}\textendash XRT to within a synthesised beam of the MWA, and the white circle in the bottom panel
indicates the position of GRB 200325A localised by \textit{Swift}\textendash BAT to within $50$ synthesised beams.} 
\label{images_swift}
\end{figure*}

\begin{figure*}
\centering
\includegraphics[width=\textwidth]{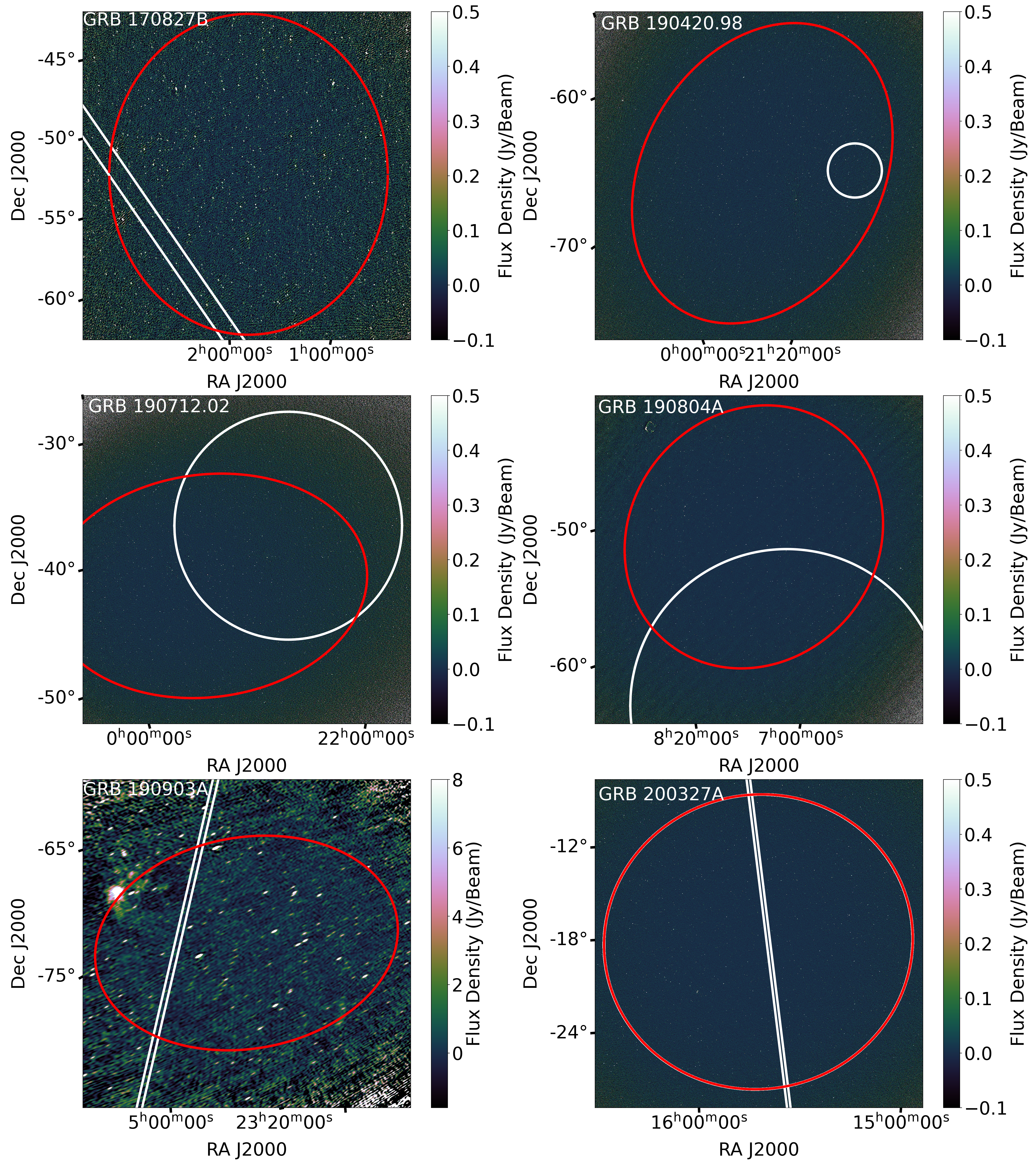}
\caption{Similar to Figure~\ref{images_swift}, here we present the first 2\,min snapshots of field covering the \textit{Fermi} GRBs in our sample. The white lines show the boundaries of the GRB localisation by \textit{Fermi} or the IPN, the red lines show the boundaries of the MWA primary beam 50\% response, and their overlaps show the ROIs within which we searched for transients and variables (see Section~\ref{sec:tran_var_search}).} 
\label{images_fermi}
\end{figure*}

In Figures \ref{images_swift} and \ref{images_fermi}, we show the first 2\,min snapshots of the nine short GRBs in our sample. These images were visually inspected to judge the data quality for our sample selection (see Section~\ref{sec:sample}). The positions of those GRBs localised by \textit{Swift} are indicated by white lines or a circle in Figure~\ref{images_swift}, and the ROIs of those GRBs localised by \textit{Fermi} or the IPN are indicated by the overlap between the GRB error positions (white curves) and the MWA primary beams 50\% response (red curves) in Figure~\ref{images_fermi} (see Section~\ref{sec:tran_var_search}). Note that the observations of GRB 190903A and 191004A were taken in the MWA compact configuration so their images 
have a larger resolution (synthesised beam) than those taken during Phase I or in the extended configuration.

\section{Transient and variable candidates for Fermi GRBs} \label{sec:appendix}

In Table \ref{var}, we provide the light curve  variability parameters derived via priorized fitting  at the position of the two \textit{Swift} GRBs on timescales of 2\,min, 30\,s and 5\,s, i.e. the modulation index ($m$), the de-biased modulation index ($m_d$; which takes into account the errors on each data-point), and the probability of observing such variability in a non-variable (steady) source (p\_val; see Section \ref{sec:tran_var_search} for details).

In Table \ref{transient} of this appendix, we 
report the number of transient and variable candidates that were detected by {\sc Robbie} within each \textit{Fermi} GRB ROI in our sample. The expected number of false positive transient and variable candidates assuming Gaussian statistics are also provided (see Section~\ref{sec:tran_var_sel} for details).

\begin{table*}
\begin{threeparttable}
\centering
\resizebox{.8\columnwidth}{!}{\hspace{-0.0cm}\begin{tabular}{c c c c c}
\hline
GRB & Timescale & \multicolumn{3}{c}{Variability statistics} \\
&  & $m$ & $m_d$ & p\_val \\
\hline
190627A & 2min & 6.47 & -1.6 & 0.7 \\
 & 30s & 5.82 & 2.53 & 0.75 \\
 & 5s & -20.04 & -6.24 & 0.53 \\
\hline
191004A & 2min & -0.62 & -0.54 & 0.05 \\
 & 30s & -0.78 & -0.4 & 0.02 \\
 & 5s & -0.42 & 0.51 & 0.41 \\
\hline
\end{tabular}}
\caption{Light curve variability statistics derived via priorized fitting at the positions of the \textit{Swift} GRBs for different monitoring timescales. Quoted are the modulation index ($m$), the de-biased modulation index (\textit{$\text{m}_d$}), and the probability of being a non-variable source (p\_val; see Section~\ref{sec:tran_var_search}).}
\label{var}
\end{threeparttable}
\end{table*}

\begin{table*}
\begin{threeparttable}
\hspace{-1cm}\resizebox{2.3\columnwidth}{!}{\begin{tabular}{c c c c c c c c}
\hline
GRB & ROI\tnote{1} & Timescale & Expected false transients & Expected false transients & Detected transients & Expected false variables & Detected variables \\
 & & & above $6\sigma$\tnote{2} & in GRB ROI\tnote{3} & in GRB ROI\tnote{4} & in GRB ROI\tnote{5} & in GRB ROI\tnote{6} \\
\hline
170827B & $\sim2\times10^4$ & 2min &  $3\times10^{-4}$ & 1.44 & 0 & 0 & 0 \\
 & &30s &  $10^{-3}$ & 0.66 & 0 & 1.92 & 0\\
 & &5s & $7\times10^{-3}$ & 1.68 & 0 & 4.14 & 0\\
\hline
190420.98 & $\sim2\times10^5$ &2min & $2\times10^{-3}$ & 0 & 0 & 0 & 0\\
 & &30s & $8\times10^{-3}$ & 0 & 0 & 0.97 & 1\\
 & &5s & $5\times10^{-2}$ & 0.04 & 1 & 2.23 & 1 \\
\hline
190712.02 & $\sim4\times10^6$ &2min & $6\times10^{-2}$ & 0 & 0 & 0 & 0 \\
 & &30s & $0.24$ & 0 & 0 & 0.55 & 0\\
 & &5s & $1.44$ & 1.1 & 1 & 6.6 & 5\\
\hline
190804A & $\sim3\times10^6$ &2min & $3\times10^{-2}$ & 0.31 & 0 & 0.31 & 0 \\
 & &30s & $0.12$ & 0.31 & 0 & 0 & 0\\
 & &5s & $0.72$ & 1.55 & 2 & 4.03 & 4\\
\hline
190903A & $\sim10^3$ &2min & $2\times10^{-5}$ & 0 & 0 & 0.05 & 0 \\
 & &30s & $8\times10^{-5}$ & 0 & 0 & 0.07 & 0\\
 & &5s & $5\times10^{-4}$ & 0.01 & 0 & 0.06 & 0\\
\hline
200325A & $\sim50$ &2min & $7\times10^{-7}$ & 0 & 0 & 0 & 0\\
 & &30s & $3\times10^{-6}$ &  0 & 0 & 0 & 0 \\
 & &5s & $2\times10^{-5}$ &  0  & 0 & 0 & 0 \\
\hline
200327A & $\sim10^4$ &2min & $2\times10^{-4}$   & 0.06 & 0 & 0 & 0 \\
 & &30s & $8\times10^{-4}$ &   0.42  & 0 & 0.12 & 0\\
 & &5s & $5\times10^{-3}$ & 1.67     & 0 & 0.72 & 0\\
\hline
\end{tabular}}
\end{threeparttable}
\caption{The number of detected transients and variables within each \textit{Fermi} GRB ROI on different timescales as output by {\sc Robbie}, which are compared to the
expected false positive transient and variable rates as defined in Section~\ref{sec:tran_var_sel}.\\
1: The size of the ROI is in units of synthesised beams;\\
2: The expected number of false positive transient candidates above $6\sigma$ in the ROI that is calculated assuming Gaussian statistics;\\
3: The expected transient false positive rate in the ROI based on the number of false positive candidates in the MWA 50\% primary beam;\\
4: The detected number of transient candidates in the ROI by {\sc Robbie};\\
5: The expected variable false positive rate in the ROI based on the number of false positive candidates in the MWA 50\% primary beam;\\
6: The detected number of variable candidates in the ROI by {\sc Robbie}.
}
\label{transient}
\end{table*}

\end{document}